%% file: fjvt_paper.tex
\newcommand*{\ATLASLATEXPATH}{latex/}
\documentclass[cernpreprint,texlive=2016,txfonts,UKenglish]{\ATLASLATEXPATH atlasdoc}
 \pdfoutput=1
\PreprintIdNumber{CERN-EP-2017-055}

\usepackage[]{\ATLASLATEXPATH atlaspackage}
\usepackage[]{\ATLASLATEXPATH atlascover}
\usepackage{\ATLASLATEXPATH atlasbiblatex}
\usepackage{subfig}

\usepackage{\ATLASLATEXPATH atlascontribute}

\usepackage{\ATLASLATEXPATH atlasphysics}

\graphicspath{{logos/}{figures/}}

\input{fjvt_paper-metadata}

\hypersetup{pdftitle={Forward JVT},pdfauthor={M. Klein, F. Rubbo, A. Schwartzman}}

\begin{document}

\maketitle

\tableofcontents

\input{defines.tex}

\section{Introduction}
\label{sec:intro}

In order to enhance the capability of the experiments to discover physics beyond the Standard Model, 
the Large Hadron Collider (LHC)
operates at the conditions yielding the highest integrated luminosity achievable.
Therefore, the collisions of proton bunches result
not only in large transverse-momentum transfer proton--proton (\pp) interactions, but also in additional
collisions within the same bunch crossing, primarily consisting of low-energy quantum chromodynamics
(QCD) processes. 
Such additional \pp collisions are referred to as \textit{in-time} \textit{pile-up} interactions.
In addition to in-time pile-up, \textit{out-of-time} pile-up refers to the energy deposits 
in the ATLAS calorimeter from previous and following bunch crossings
with respect to the triggered event.
In this paper, in-time and out-of-time pile-up are referred collectively as pile-up (PU).

\begin{figure}[htbp]
\begin{center}
\subfloat[][]{\includegraphics[width=0.475\textwidth]{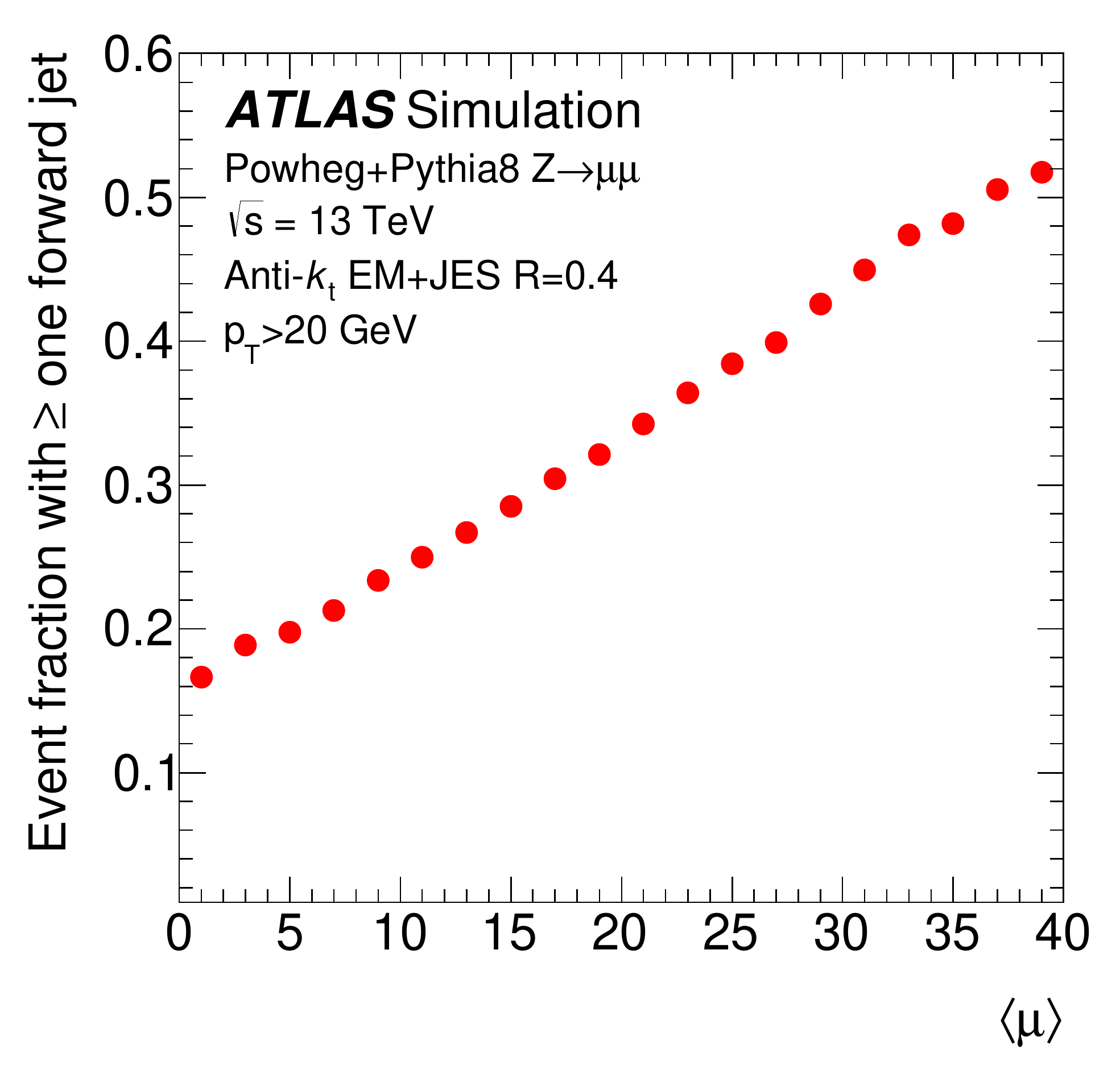}\label{appfrac}}
\subfloat[][]{\includegraphics[width=0.475\textwidth]{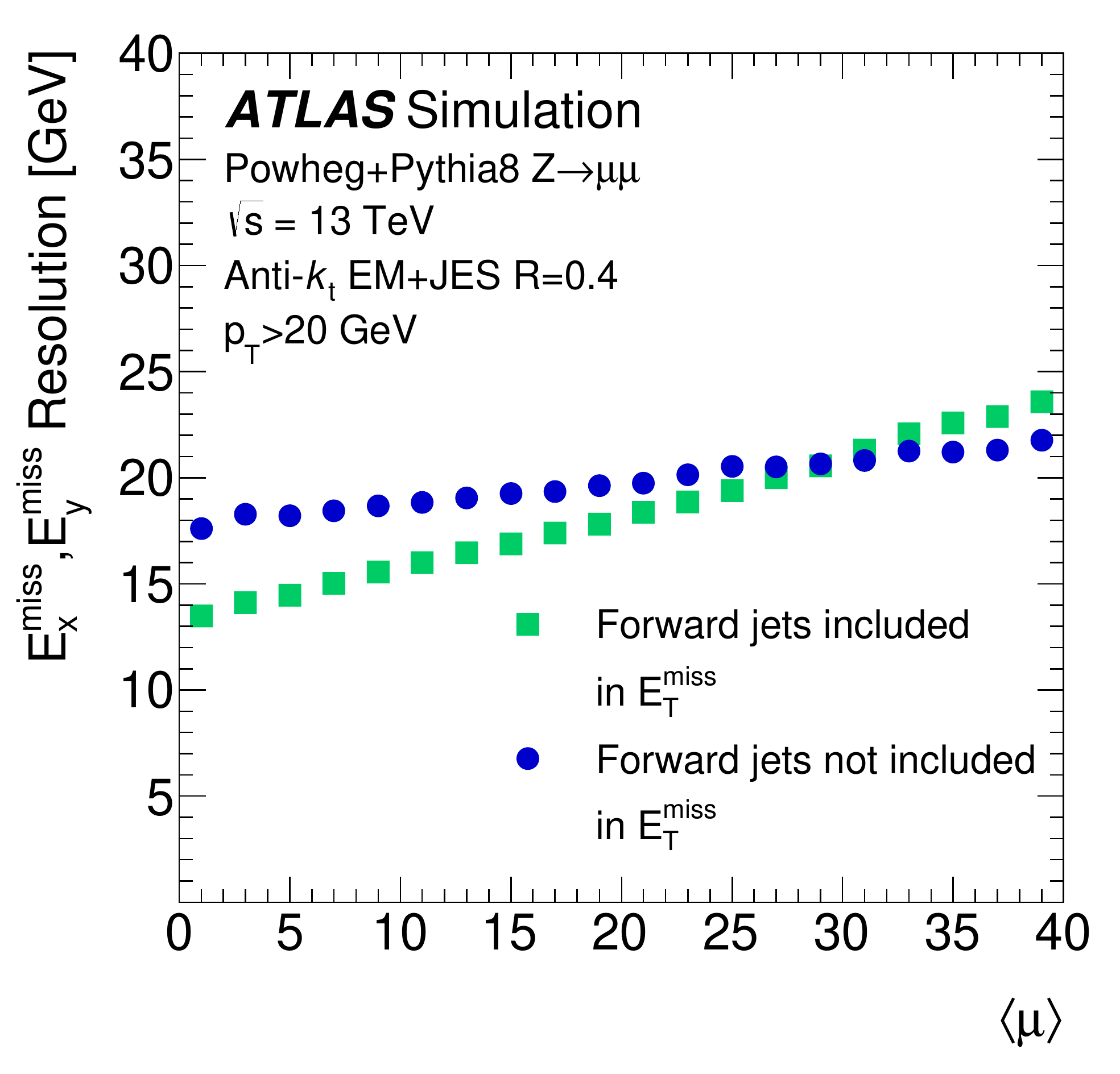}\label{appmet}}
\end{center}
\caption{\protect\subref{appfrac} Fraction of simulated $\Zboson+$jets events with at least one forward jet  
and \protect\subref{appmet} the resolution of the \met components \metx and \mety  
as a function of $\langle\mu\rangle$.
Jets and \met definitions are described in Section~\ref{sec:objects}.
}
\label{fig:fJVTapp}
\end{figure}

In Ref.~\cite{Aad:2015ina} it was shown that pile-up jets can be 
effectively removed using track and vertex information with the jet-vertex-tagger (\JVT) 
technique. The CMS Collaboration employs a pile-up mitigation strategy based on tracks and 
jet shapes~\cite{CMS-PAS-JME-13-005}.
A limitation of the \JVT discriminant used by the ATLAS Collaboration is that it can only be used for jets within the 
coverage\footnote{\ATLAS uses a right-handed coordinate system with 
its origin at the nominal interaction point (IP) in the centre of the detector 
and the $z$-axis along the beam pipe. The $x$-axis points from the IP to the 
centre of the LHC ring, and the $y$-axis points upward. Cylindrical coordinates 
$(r, \phi)$ are used in the transverse plane, $\phi$ being the azimuthal angle 
around the beam pipe. 
The pseudorapidity is defined in terms of the polar angle $\theta$ as 
$\eta = - \ln \tan(\theta/2)$.
}  of the tracking detector, $|\eta|<2.5$. However, in the ATLAS detector, jets are reconstructed 
in the range $|\eta|<4.5$.
The rejection of pile-up jets in the \textit{forward} region, here defined as $2.5<|\eta|<4.5$, 
is crucial to enhance the sensitivity of key analyses such as
the measurement of Higgs boson production in the vector-boson fusion (VBF) process. 
Figure~\ref{fig:fJVTapp}(a) shows how the fraction of $\Zboson+$jets events with 
at least one forward jet\footnote{The jet reconstruction is described in Section~\ref{sec:objects}.} with $\pt>20\GeV$, 
an important background for VBF analyses, 
rises quickly with busier pile-up conditions, quantified by the average number of interactions per bunch crossing ($\langle\mu\rangle$).
Likewise, the resolution of the missing transverse momentum (\met) components \metx and \mety
in $\Zboson+$jets events is also affected by the presence of forward pile-up jets.
The inclusion of forward jets allows a more precise \met calculation
but a more pronounced pile-up dependence, as shown in Figure~\ref{fig:fJVTapp}(b).
At higher $\langle\mu\rangle$, improving the \met resolution depends on rejecting all forward jets, 
unless the impact of pile-up jets specifically can be mitigated.

In this paper, the phenomenology of pile-up jets with $|\eta|>2.5$ is investigated
in detail, and techniques to identify and reject them are presented.
The paper is organized as follows. Section~\ref{sec:objects} briefly describes the ATLAS
detector, the event reconstruction and selection.
The physical origin and classification of pile-up jets are described in Section~\ref{sec:pileup}. 
Section~\ref{sec:width} describes the use of jet shape variables for the 
identification and rejection of forward pile-up jets.
The \textit{forward \JVT} (\fJVT) technique is presented
in Section~\ref{sec:fjvt} along with its performance and efficiency measurements.
The usage of jet shape variables in improving \fJVT performance is presented in Section~\ref{sec:comb}, 
while the application of forward pile-up jet rejection in a VBF analysis is discussed in Section~\ref{sec:vbf}.
The conclusions are presented in Section~\ref{sec:conclusion}.

\section{Experimental setup}
\label{sec:objects}

\subsection{ATLAS detector}
\label{sec:atlas}

The \ATLAS detector is a general-purpose particle detector covering almost $4\pi$ 
in solid angle and consisting of a tracking system called the inner detector (ID), a calorimeter system, 
and a muon spectrometer (MS). The details of the detector are given in 
Ref.~\cite{ATLAS:JINST,ATLAS-TDR-19}.

The ID consists of silicon pixel and microstrip tracking detectors covering
the pseudorapidity range of $|\eta| < 2.5$ and
a straw-tube tracker covering $|\eta| < 2.0$.
These components are immersed in an axial 2\,T 
magnetic field provided by a superconducting solenoid.

The electromagnetic (EM) and hadronic calorimeters are composed 
of multiple subdetectors covering the range $|\eta|<4.9$, generally divided into barrel
($|\eta| < 1.4$), endcap ($1.4 < |\eta| < 3.2$) and forward ($3.2 < |\eta| < 4.9$) 
regions. 
The barrel and endcap sections of the EM calorimeter use liquid argon (LAr) as the active medium and lead absorbers. 
The hadronic endcap calorimeter ($1.5<|\eta|<3.2$) uses copper absorbers and LAr, 
while in the forward ($3.1<|\eta|<4.9$) region LAr, copper and tungsten are used.
The LAr calorimeter read-out~\cite{Abreu:2010zzc}, with a pulse length between 60 and 600 ns,
is sensitive to signals from the preceding 24 bunch crossings. It uses bipolar shaping with positive and
negative output, which ensures that the signal induced by out-of-time pile-up averages to zero.
In the region $|\eta|<1.7$, the hadronic (Tile) calorimeter is constructed from steel absorber and scintillator 
tiles and is separated into barrel ($|\eta|<1.0$) and extended barrel ($0.8<|\eta|<1.7$) sections. 
The fast response of the Tile calorimeter makes it less sensitive to out-of-time pile-up.

The MS forms the outer layer of the ATLAS detector and is dedicated to the detection and measurement
of high-energy muons in the region $|\eta|<2.7$.
A multi-level trigger system of dedicated hardware and software filters is used 
to select \pp collisions producing high-\pT particles.

\subsection{Data and MC samples}

The studies presented in this paper are performed using a data set 
of $pp$ collisions at $\sqrt{s}=13\TeV$, corresponding to an integrated
luminosity of $3.2$~\ifb, collected in 2015 during which the LHC operated 
with a bunch spacing of 25 ns.
There are on average 13.5 interactions per bunch crossing in the data sample used 
for the analysis.

Samples of simulated events used for comparisons with data are reweighted to match the 
distribution of the number of pile-up interactions observed in data. The average number of interactions 
per bunch crossing $\langle\mu\rangle$ in the data used as reference for the reweighting is divided by a 
scale factor of $1.16\pm0.07$.
This scale factor takes into account the fraction of visible cross-section
due to inelastic \pp collisions as measured in the data~\cite{Aad:2011eu}
and is required to obtain good agreement with the number of inelastic interactions reconstructed in the tracking detector
as predicted in the reweighted simulation.
In order to extend the study of the pile-up dependence, simulated samples with an
average of 22 interactions per bunch crossing are also used.
Dijet events are simulated with the \PYTHIAeight~\cite{Sjostrand:2007gs} event generator
using the NNPDF2.3LO~\cite{Carrazza:2013axa} set of parton distribution
functions (PDFs) and the parameter values set according to the A14 underlying-event 
tune~\cite{ATL-PHYS-PUB-2014-021}.
Simulated \ttbar events are generated with 
\Powheg~v2.0~\cite{Alioli:2010xd,Frixione:2007vw,Nason:2004rx} using
the CT10 PDF set~\cite{Lai:2010vv};
\PYTHIAsix~\cite{Sjostrand:2006za} is used for fragmentation and 
hadronization with the Perugia2012~\cite{Skands:2010ak}
tune that employs the CTEQ6L1~\cite{Pumplin:2002vw} PDF set.
A sample of leptonically decaying Z bosons produced with jets ($\Zboson(\to\ell\ell)$+jets) 
and VBF $H\to\tau\tau$ samples are generated with \Powheg v1.0 and 
\PYTHIAeight is used for fragmentation and hadronization with the AZNLO tune~\cite{ATL-PHYS-PUB-2013-017} 
and the CTEQ6L1 PDF set.
For all samples, the EvtGen v1.2.0 program~\cite{EvtGen} is used for properties 
of the bottom and charm hadron decays.
The effect of in-time as well as out-of-time pile-up 
is simulated using minimum-bias events generated with \PYTHIAeight to reflect the
pile-up conditions during the 2015 data-taking period, 
using the A2 tune~\cite{ATL-PHYS-PUB-2012-003} and the 
MSTW2008LO~\cite{Martin:2009iq} PDF set.
All generated events are processed with a detailed simulation of the \ATLAS detector 
response~\cite{Aad:2010ah} based on \GEANTfour~\cite{Agostinelli:2002hh}
and subsequently reconstructed and analysed in the same way as the data.

\subsection{Event reconstruction}
\label{sec:OS}
The raw data collected by the ATLAS detector is reconstructed in the form of particle candidates
and jets using various pattern recognition algorithms.
The reconstruction used in this analysis are detailed in
Ref.~\cite{Aad:2015ina}, while an overview is presented in this section.

\paragraph{Inputs to jet reconstruction}

Jets in ATLAS are reconstructed from clusters of energy deposits in the calorimeters.
Two methods of combining calorimeter cell information are considered in this paper: 
topological clusters and towers.

Topological clusters (topo-clusters)~\cite{Aad:2016upy} are built from neighbouring calorimeter cells. 
The algorithm uses as seeds calorimeter cells
with energy significance\footnote{The cell noise 
$\sigma_\mathrm{noise}$ is the sum in quadrature of the readout electronic noise and the cell noise due to
pile-up, estimated in simulation~\cite{Aad:2016upy,ATLASjets}.} $|E_\mathrm{cell}|/\sigma_\mathrm{noise}>4$,
combines all neighbouring cells with
$|E_\mathrm{cell}|/\sigma_\mathrm{noise}>2$ and finally adds neighbouring cells
without any significance requirement.
Topo-clusters are used as input for jet reconstruction.

Calorimeter towers are fixed-size objects ($\Delta\eta\times\Delta\phi=0.1\times0.1$)~\cite{cscbook}
that ensure a uniform segmentation of the calorimeter information.
Instead of building clusters, the cells are projected onto a fixed grid in $\eta$ and $\phi$ corresponding to 6400 towers.
Calorimeter cells which completely fit within a tower contribute their total energy
to the single tower.
Other cells extending beyond the tower boundary contribute to multiple
towers, depending on the overlap fraction of the cell area with the towers.
In the following, towers are matched geometrically to jets reconstructed using topo-clusters.

\paragraph{Vertices and tracks}

The event hard-scatter primary vertex is defined as the 
reconstructed primary vertex with the largest $\sum p_\mathrm{T}^2$ of constituent tracks.
When evaluating performance in simulation,
only events where the reconstructed hard-scatter primary vertex lies
$|\Delta z|<0.1$\,mm from the true hard-scatter interaction are considered.
For the physics processes considered, the reconstructed hard-scatter primary vertex
matches the true hard-scatter interaction more than 95\% of the time.
Tracks are required to have $\pT> 0.5\GeV$ and to satisfy quality 
criteria designed to reject poorly measured or fake tracks~\cite{ATL-PHYS-PUB-2015-051}. 
Tracks are assigned to primary vertices based on the track-to-vertex matching 
resulting from the vertex reconstruction.
Tracks not included in vertex reconstruction are assigned to the nearest vertex based on the distance 
$|\Delta z \times \sin \theta|$, 
up to a maximum distance of 3.0 mm. Tracks not matched to any vertex are not considered.
Tracks are then assigned to jets by adding them to the jet
clustering process with infinitesimal \pt{}, a procedure known as ghost-association~\cite{Cacciari:2008gn}.

\paragraph{Jets}
\label{sec:jets}

Jets are reconstructed from topo-clusters at the EM scale\footnote{The EM scale 
corresponds to the energy deposited in the calorimeter by electromagnetically interacting particles 
without any correction accounting for the loss of signal for hadrons.} 
using the anti-$k_t$~\cite{Cacciari:2008gp} algorithm, as implemented in 
\Fastjet~\cite{Cacciari:2011ma}, with a radius parameter $R=0.4$.  
After a jet-area-based subtraction of pile-up energy, a response correction is applied to each
jet reconstructed in the calorimeter to calibrate it to the particle-level jet energy scale~\cite{Aad:2015ina,ATLASjets,ATLAS-CONF-2015-002}.
Unless noted otherwise, jets are required to have $20\GeV < \pT < 50\GeV$. Higher-\pT forward jets are ignored due to
their negligible pile-up rate at the pile-up conditions considered in this paper.
\textit{Central} jets are required to be within $|\eta|$ of 2.5 so that most of their charged particles are within 
the tracking coverage of the inner detector. 
\textit{Forward} jets are those in the region $2.5<|\eta|<4.5$, and no tracks associated with 
their charged particles are measured beyond $|\eta|=2.5$.

Jets built from particles in the Monte Carlo generator's event record (``truth particles'') are also considered.
Truth-particle jets are reconstructed using the 
anti-$k_t$ algorithm with $R=0.4$ from stable\footnote{Truth particles are considered stable 
if their decay length $c \tau$ is greater than $1\unit{cm}$. 
A truth particle is considered to be interacting if it is expected to deposit most of 
its energy in the calorimeters; muons and neutrinos are considered to be 
non-interacting.}
final-state truth particles from the 
simulated hard-scatter (\textit{truth-particle hard-scatter jets}) or in-time pile-up 
(\textit{truth-particle pile-up jets}) interaction of choice. 
A third type of truth-particle jet (\textit{inclusive truth-particle jets}) is reconstructed by 
considering truth particles from all interactions 
simultaneously, in order to study the effects of pile-up interactions on truth-particle pile-up jets.

The simulation studies in this paper require a 
classification of the reconstructed jets into three categories: \textit{hard-scatter jets}, 
\textit{QCD pile-up jets}, and \textit{stochastic pile-up jets}.
Jets are thus truth-labelled based on a matching criterion to truth-particle jets.
Similarly to Ref.~\cite{Aad:2015ina}, jets are first classified as 
hard-scatter or pile-up jets.
Jets are labelled as hard-scatter jets if 
a truth-particle hard-scatter jet with $\pT > 10\GeV$
is found within $\Delta R = \sqrt{(\Delta \eta)^2 + (\Delta \phi)^2}$ of 0.3.
The $\pT>10\GeV$ requirement is used to avoid accidental matches of reconstructed jets 
with soft activity from the hard-scatter interaction.
In cases where more than one truth-particle jet is matched, 
$\pT^\mathrm{truth}$ is defined from the highest-\pT truth-particle hard-scatter jet 
within $\Delta R$ of 0.3.

Jets are labelled as pile-up jets if no truth-particle hard-scatter jet with $\pT > 4\GeV$ 
is found within $\Delta R$ of 0.6.
These pile-up jets are further classified as QCD pile-up if they are matched within 
$\Delta R<0.3$ to a truth-particle pile-up jet or as stochastic pile-up jets 
if there is no truth-particle pile-up jet within $\Delta R<0.6$, 
requiring that truth-particle pile-up jets have $\pT > 10\GeV$ in both cases.
Jets with $0.3<\Delta R<0.6$ relative to truth-particle hard-scatter jets with $\pT > 10\GeV$ 
or $\Delta R<0.3$ of truth-particle hard-scatter jets with $4\GeV < \pT < 10\GeV$ are not
labelled because their nature cannot be unambiguously determined.
These jets are therefore not used for performance based on simulation.

\paragraph{Jet Vertex Tagger}

The Jet Vertex Tagger (JVT) is built out of the combination of two jet variables, 
\cJVF and $\RpT^0$, that provide information to separate hard-scatter jets from pile-up jets.
The quantity \cJVF~\cite{Aad:2015ina} is defined for each jet as 
\begin{equation}
\cJVF = \frac{\sum{\pT^{\mathrm{trk}}(\mathrm{PV}_0)}}{\sum \pT^{\mathrm{trk}}(\mathrm{PV}_0) + \frac{\sum_{i\geq1}\sum \pT^{\mathrm{trk}}(\mathrm{PV}_i)}{ (k \cdot \nPUtrk)}},
\label{eq:corrJVF}
\end{equation}
where PV$_i$ denotes the reconstructed event vertices (PV$_0$ is 
the identified hard-scatter vertex and the PV$_i$ are sorted by decreasing $\sum p_\mathrm{T}^2$),
and $\sum{\pT^{\mathrm{trk}}(\mathrm{PV}_0)}$ is the scalar \pT sum 
of the tracks that are associated with the jet and originate from the hard-scatter 
vertex.
The term $\pT^{\mathrm{PU}}=\sum_{i\geq1}\sum \pT^{\mathrm{trk}}(\mathrm{PV}_i)$ denotes 
the scalar \pT sum of the tracks associated with the jet and originating from pile-up 
vertices. 
To correct for the linear increase of $\pT^{\mathrm{PU}}$ with the 
total number of pile-up tracks per event (\nPUtrk), $\pT^{\mathrm{PU}}$ is divided by 
$(k \cdot \nPUtrk)$ with the parameter $k$ set to 0.01~\cite{Aad:2015ina}.\footnote{The parameter 
$k$ does not affect performance
and is chosen to ensure that the \cJVF distribution stretches over the full range between 0 and 1.}

The variable $\RpT^0$ is defined as the scalar \pT sum of the tracks that are associated 
with the jet and 
originate from the hard-scatter vertex divided by the fully calibrated jet $\pT$, 
which includes pile-up subtraction:
\begin{equation}
\RpT^0 = \frac{\sum{\pT^{\mathrm{trk}}(\mathrm{PV}_0)}}{\pT^\mathrm{jet}}.
\label{eq:RpT}
\end{equation}
This observable tests the compatibility between the jet \pt and the 
total \pt of the hard-scatter charged particles within the jet. 
Its average value for hard-scatter jets is approximately 0.5, as the numerator
does not account for the neutral particles in the jet.
The \JVT discriminant is built by defining a two-dimensional likelihood 
based on a k-nearest neighbour (kNN) algorithm~\cite{Hocker:2007ht}.
An extension of the $\RpT^0$ variable computed with respect to any vertex $i$ in the event,
$\RpT^i=\sum_k{\pT^{\mathrm{trk}_k}(\mathrm{PV}_i)}/\pT^\mathrm{jet}$, is also used 
in this analysis.

\paragraph{Electrons and muons}
Electrons are built from EM clusters and associated ID tracks.
They are required to satisfy $|\eta|<2.47$ and $p_\mathrm{T}>10\GeV$, as well as
reconstruction quality and isolation criteria~\cite{Aad:2014fxa}.
Muons are built from an ID track (for $|\eta|<2.5$) and an MS track.
Muons are required to satisfy $p_\mathrm{T}>10\GeV$ as well as reconstruction quality 
and isolation criteria~\cite{Aad:2016jkr}.
Correction factors are applied to simulated events to account for mismodelling of lepton isolation, trigger efficiency, 
and quality selection variables.

\paragraph{\met}
The missing transverse momentum, \metvec, corresponds to the negative vector sum of the
transverse momenta
of selected electron, photon, and muon candidates, as well as jets and tracks not used in reconstruction~\cite{Aad:2016nrq}. 
The scalar magnitude \met represents the total transverse momentum imbalance in an event. 

\section{Origin and structure of pile-up jets}
\label{sec:pileup}

The additional transverse energy from pile-up interactions contributing to jets originating
from the hard-scatter (HS) interaction is subtracted on an event-by-event basis using the 
jet-area method~\cite{Aad:2015ina,Cacciari:2007fd}.
However, the jet-area subtraction assumes a uniform pile-up distribution across the calorimeter,
while local fluctuations of pile-up can cause additional jets to be reconstructed.
The additional jets can be classified into two categories:
\textit{QCD pile-up jets}, where the particles in the jet stem mostly from a single QCD process occuring 
in a single pile-up interaction, 
and \textit{stochastic jets}, which combine particles from different interactions.
Figure \ref{fig:eventDisplayCentral} shows an event with a hard-scatter jet, a QCD pile-up
jet and a stochastic pile-up jet.
Most of the particles associated with the hard-scatter jet originate from the primary interaction. 
Most of the particles associated with the QCD pile-up jet originate from
a single pile-up interaction. The stochastic pile-up jet includes
particles associated with both pile-up interactions in the event, without a single prevalent source.

\begin{figure}[htbp]
\begin{center}
\includegraphics[height=60mm]{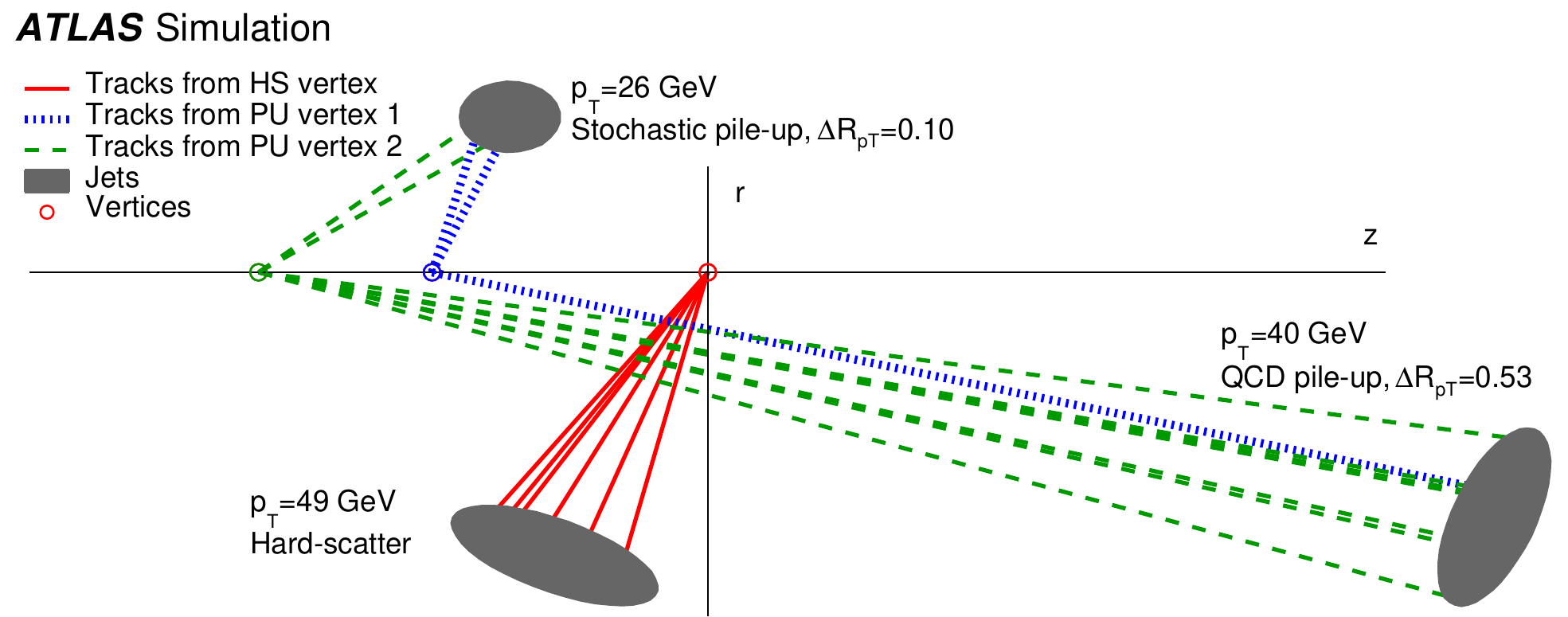}
\end{center}
\caption{Display of a simulated event in $r$--$z$ view containing a hard-scatter jet, 
  a QCD pile-up jet, and a stochastic pile-up jet.
 The \diffRpT values (defined in Section~\ref{sec:diffrpt}) are quoted for the two pile-up jets.}
\label{fig:eventDisplayCentral}
\end{figure}
While this binary classification is convenient for the purpose of description,
the boundary between the two categories is somewhat arbitrary.
This is particularly true in harsh pile-up conditions, with dozens of concurrent
$pp$ interactions, where every jet, including those originating primarily from the identified
hard-scatter interaction, also has contributions from multiple pile-up interactions.

In order to identify and reject forward pile-up jets, a twofold strategy is adopted.
Stochastic jets have intrinsic differences in shape with respect to hard-scatter and QCD pile-up jets, 
and this shape can be used for discrimination.
On the other hand, the calorimeter signature of QCD pile-up jets does not differ fundamentally from that of hard-scatter jets.
Therefore, QCD pile-up jets are identified by exploiting transverse momentum conservation 
in individual pile-up interactions.

The nature of pile-up jets can vary significantly whether
or not most of the jet energy originates from a single interaction.
Figure~\ref{fig:quotienttr} shows the fraction of QCD pile-up jets among all pile-up jets, 
when considering inclusive truth-particle jets.
The corresponding distributions for reconstructed jets are shown in Figure~\ref{fig:quotient}.
When considering only in-time pile-up contributions (Figure~\ref{fig:quotienttr}),
the fraction of QCD pile-up jets depends on the pseudorapidity and 
\pt of the jet and the average number of interactions per bunch crossing $\langle\mu\rangle$.
Stochastic jets are more likely at low \pt and $|\eta|$ and in harsher pile-up conditions.
However, the comparison between Figure~\ref{fig:quotienttr}, containing inclusive truth-particle jets, 
and Figure~\ref{fig:quotient}, containing reconstructed jets,
suggests that only a small fraction of stochastic jets are due to in-time pile-up.
Indeed, the fraction of QCD pile-up jets decreases significantly once out-of-time pile-up effects and 
detector noise and resolution are taken into account.
Even though the average amount of out-of-time energy is higher in the forward region,
topo-clustering results in a stronger suppression of this contribution in the forward region.
Therefore, the fraction of QCD pile-up jets increases in the forward region, and it
constitutes more than 80\% of pile-up jets with \pT > 30 \GeV overall.
The fraction of stochastic jets becomes more prominent at low \pt and it grows as the number 
of interactions increases.
The majority of pile-up jets in the forward region are QCD pile-up jets, although a sizeable
fraction of stochastic jets is present in both the central and forward regions.

\begin{figure}[htbp]
\begin{center}
\subfloat[][]{\includegraphics[width=0.475\textwidth]{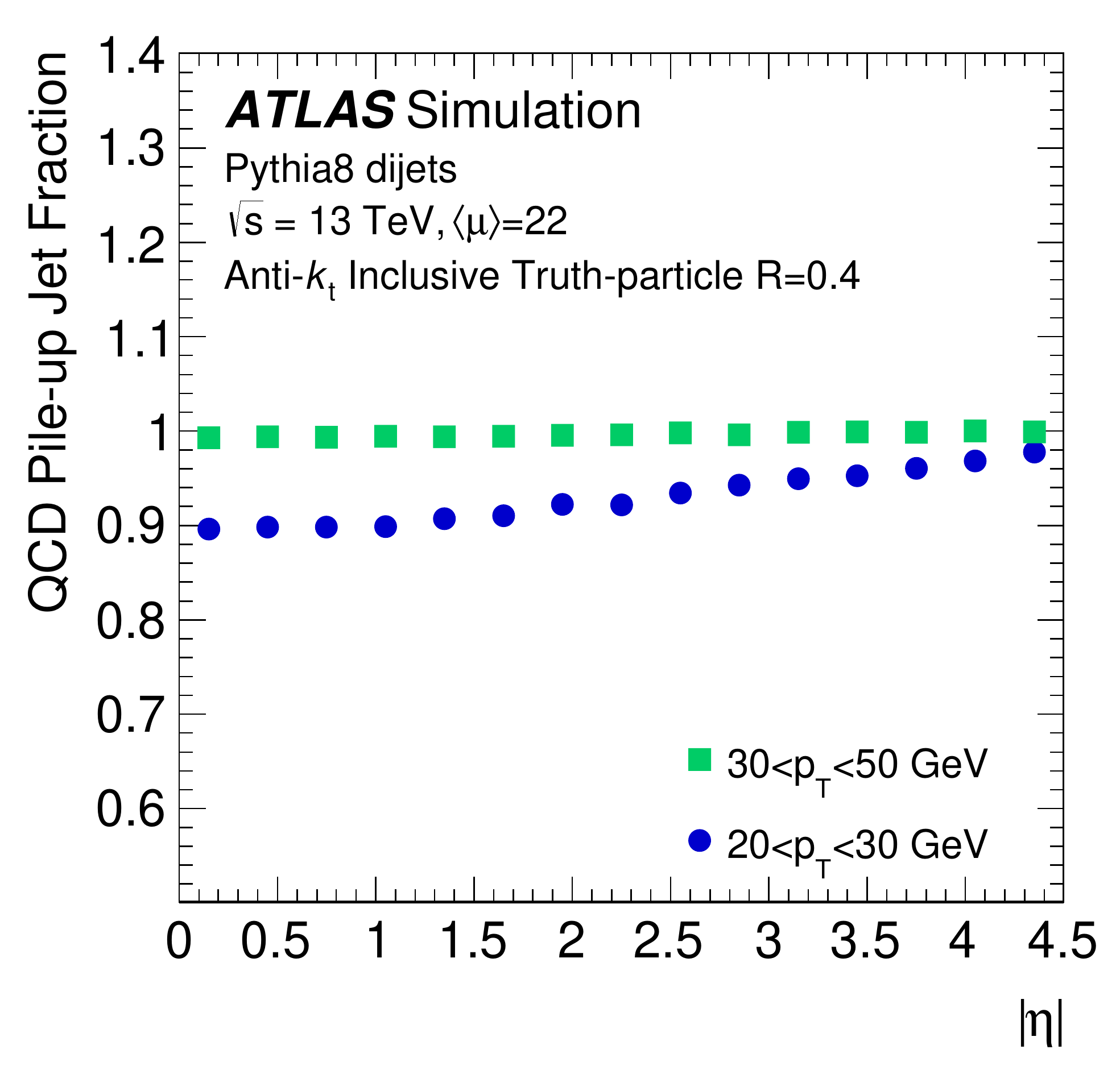}\label{qcdvseta}}\quad
\subfloat[][]{\includegraphics[width=0.475\textwidth]{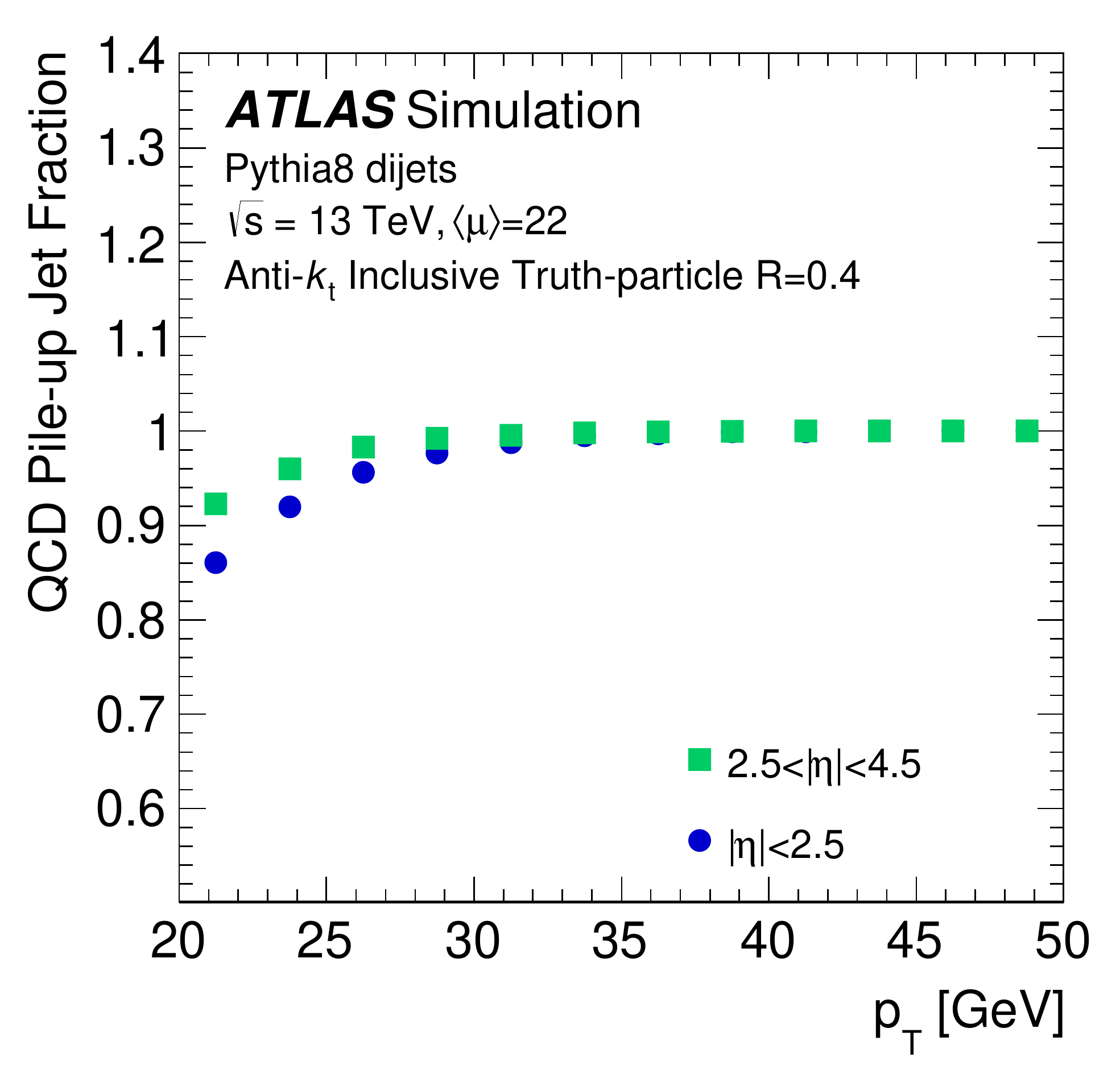}\label{qcdvspt}}\\
\subfloat[][]{\includegraphics[width=0.475\textwidth]{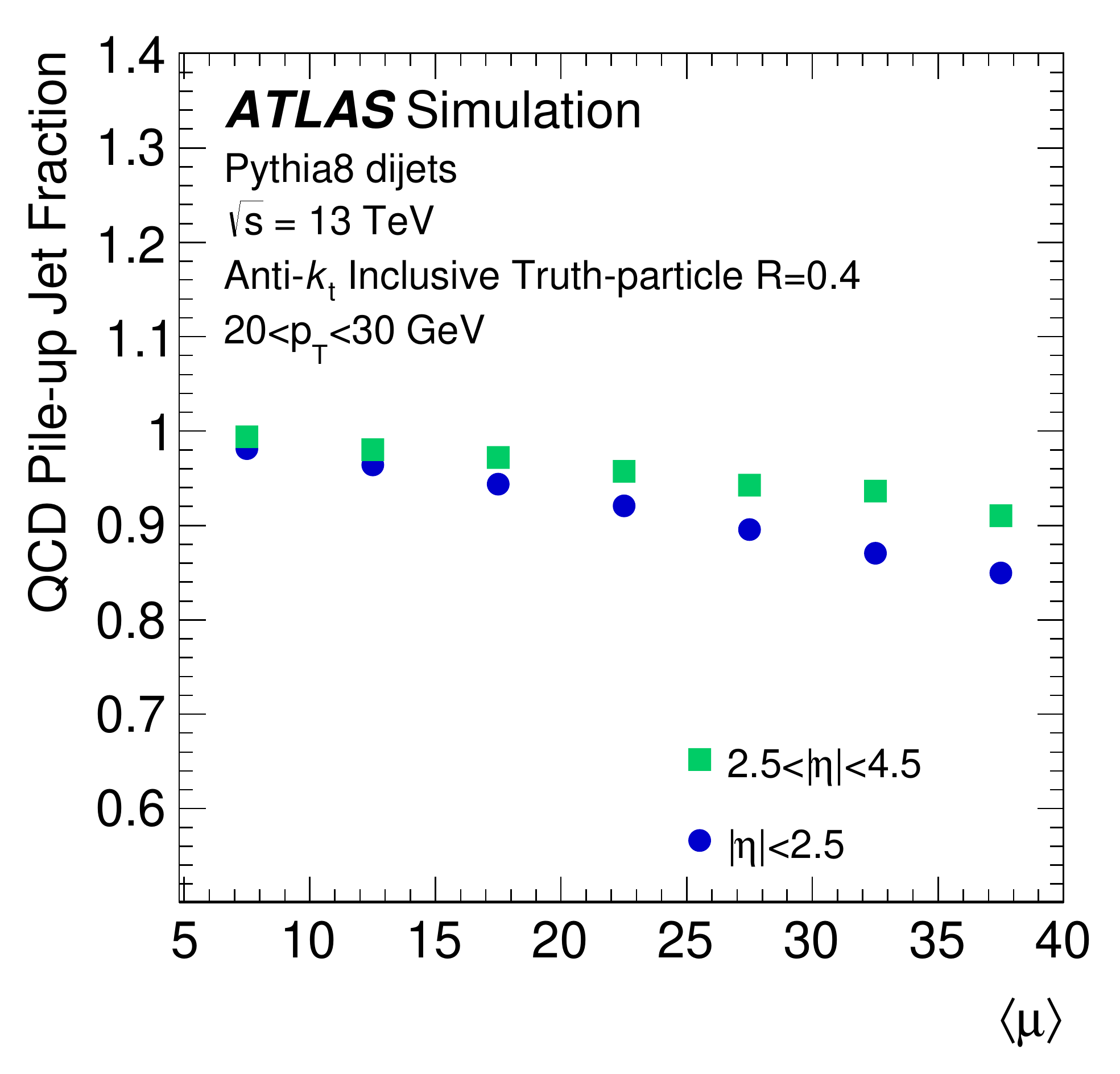}\label{qcdvsnpv20}}\quad
\subfloat[][]{\includegraphics[width=0.475\textwidth]{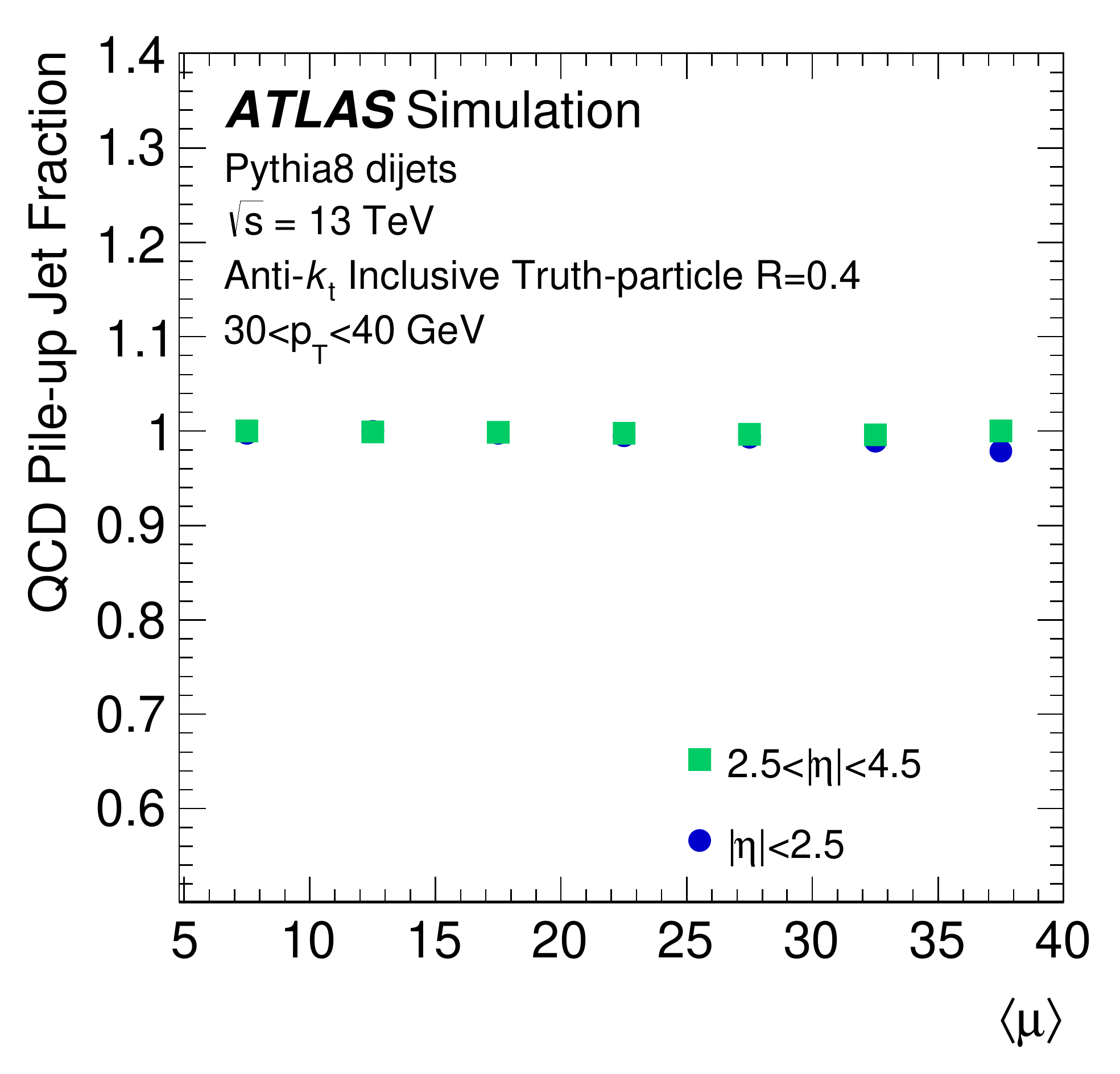}\label{qcdvsnpv30}}
\end{center}
\caption{Fraction of pile-up tagged inclusive truth-particle jets classified as QCD pile-up jets
as a function of \protect\subref{qcdvseta} $|\eta|$, \protect\subref{qcdvspt} \pt, 
and \protect\subref{qcdvsnpv20} $\langle\mu\rangle$ for jets with $20\GeV<\pT<30\GeV$ and 
\protect\subref{qcdvsnpv30} $30\GeV<\pT<40\GeV$, 
as estimated in dijet events with \PYTHIAeight pile-up simulation. The inclusive truth-particle jets are
reconstructed from truth particles originating from all in-time pile-up interactions.}
\label{fig:quotienttr}
\end{figure}

\begin{figure}[htbp]
\begin{center}
\subfloat[][]{\includegraphics[width=0.475\textwidth]{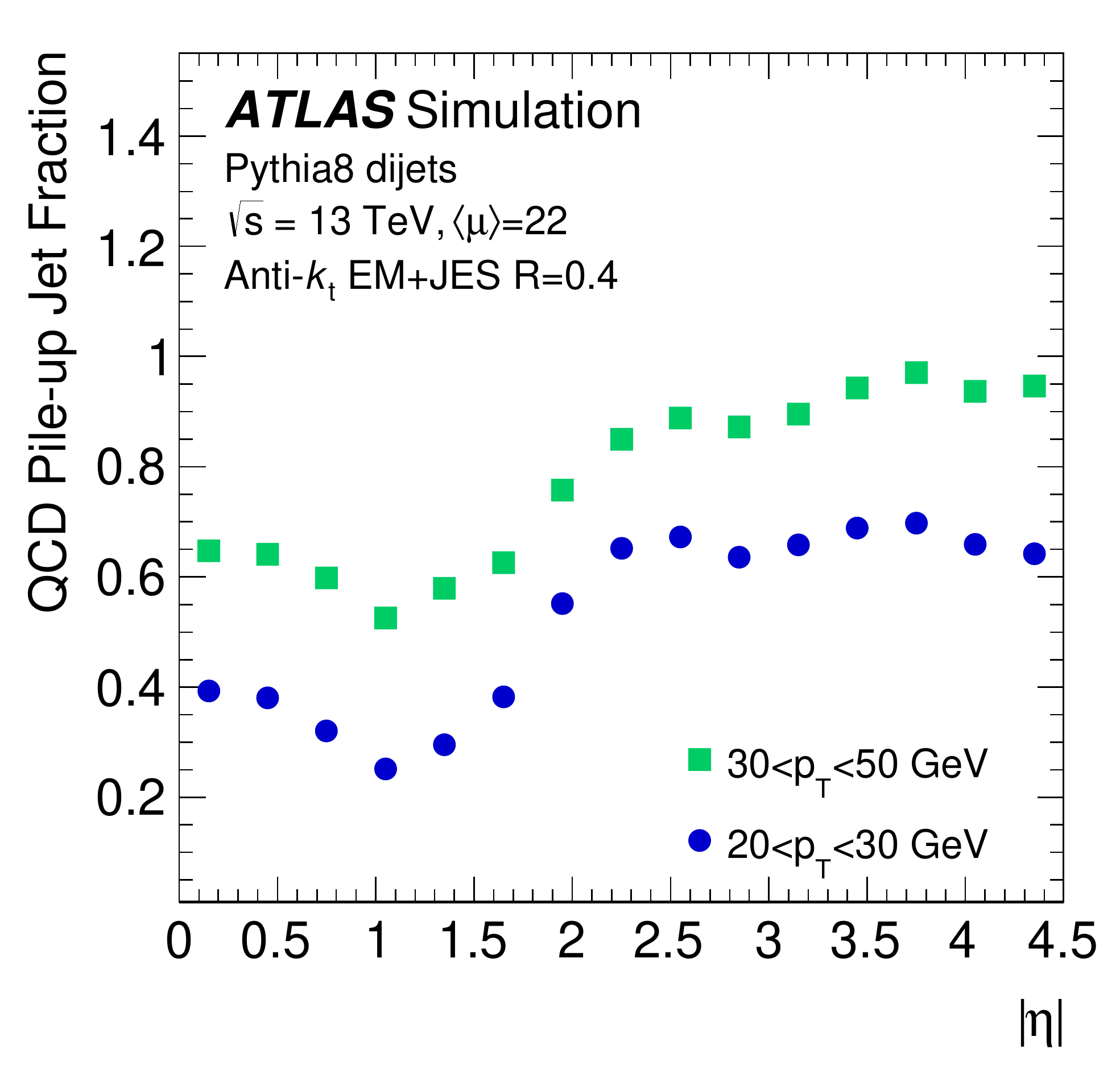}\label{qcdvseta}}\quad
\subfloat[][]{\includegraphics[width=0.475\textwidth]{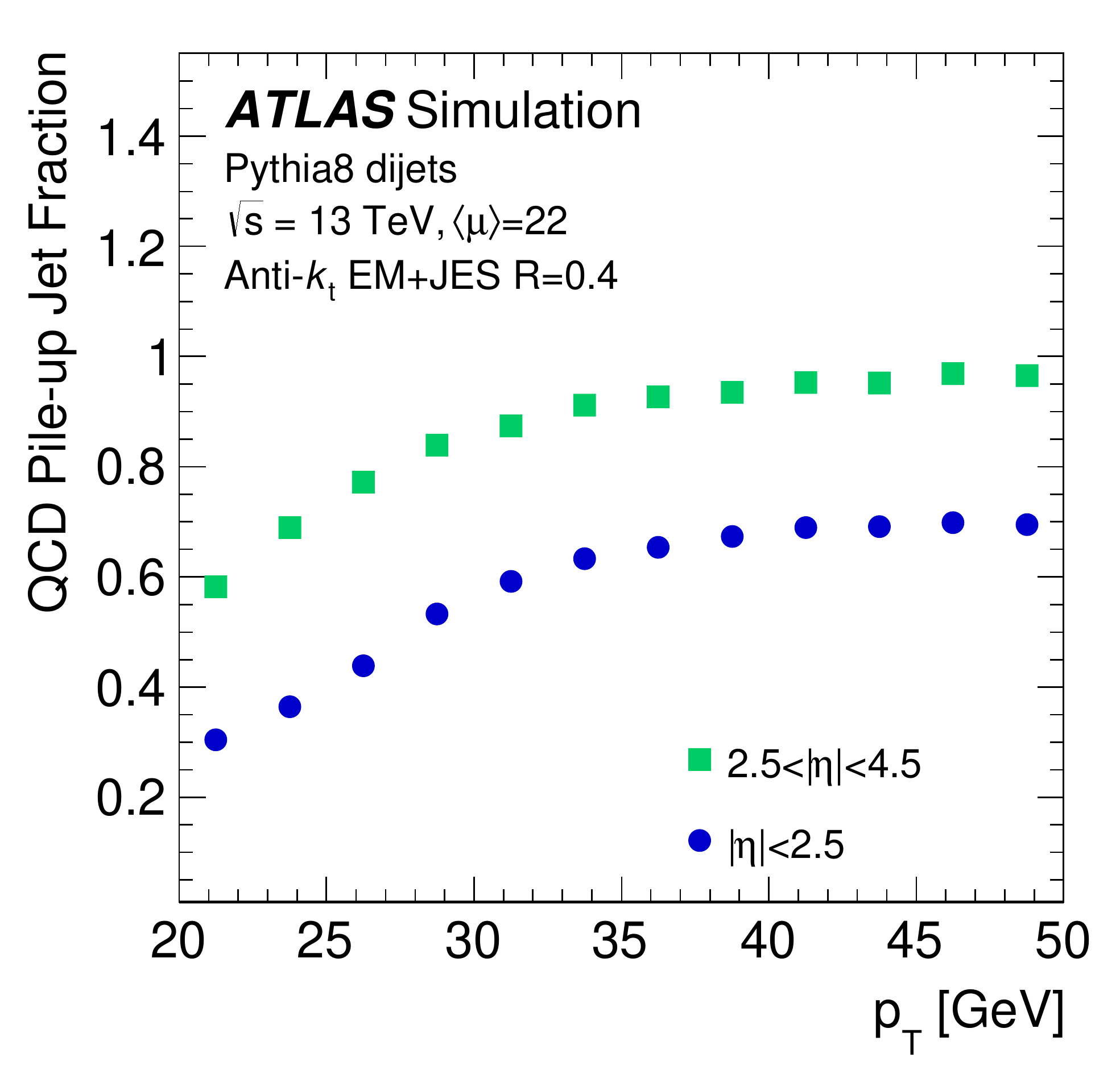}\label{qcdvspt}}\\
\subfloat[][]{\includegraphics[width=0.475\textwidth]{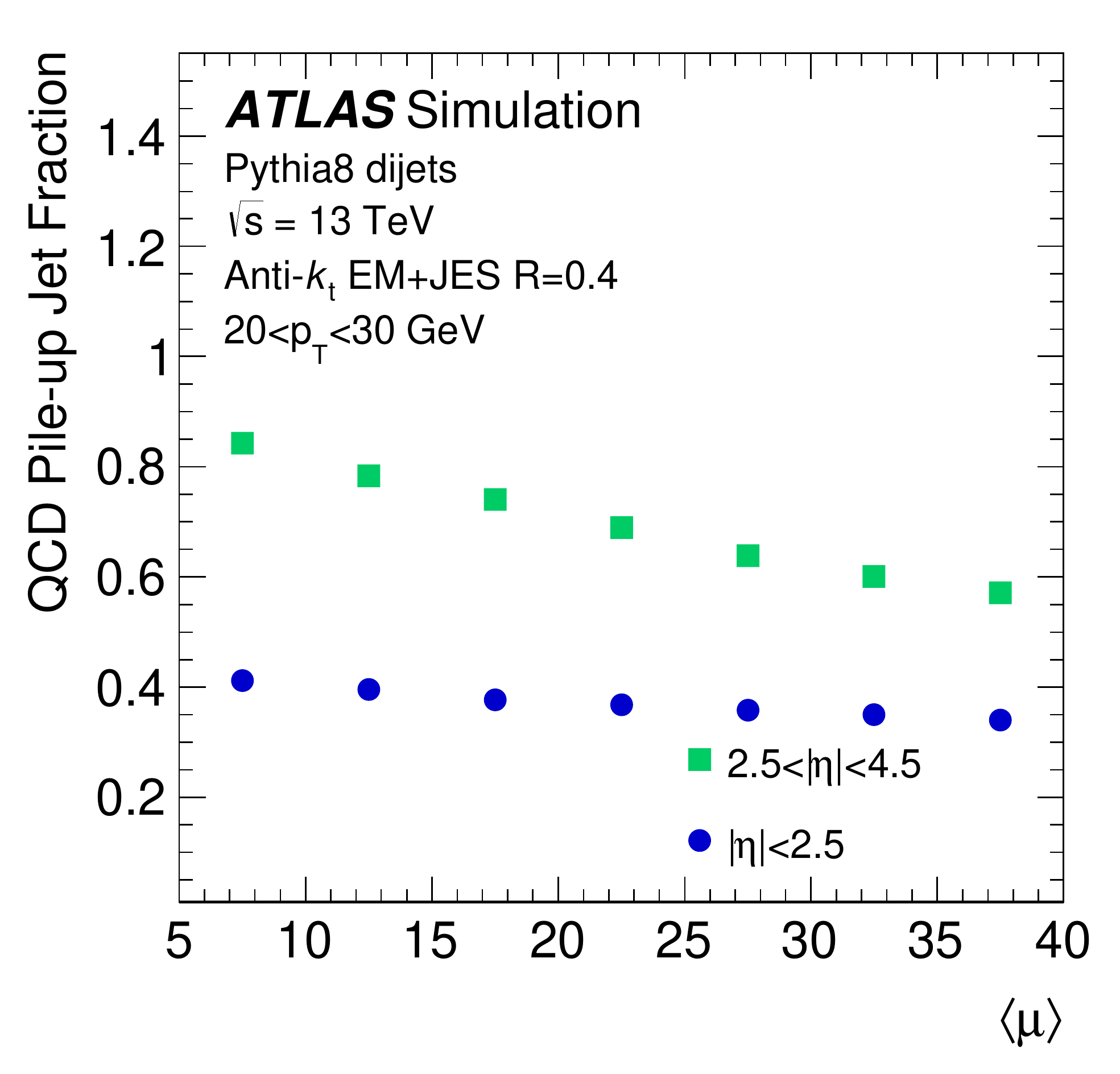}\label{qcdvsnpv20}}\quad
\subfloat[][]{\includegraphics[width=0.475\textwidth]{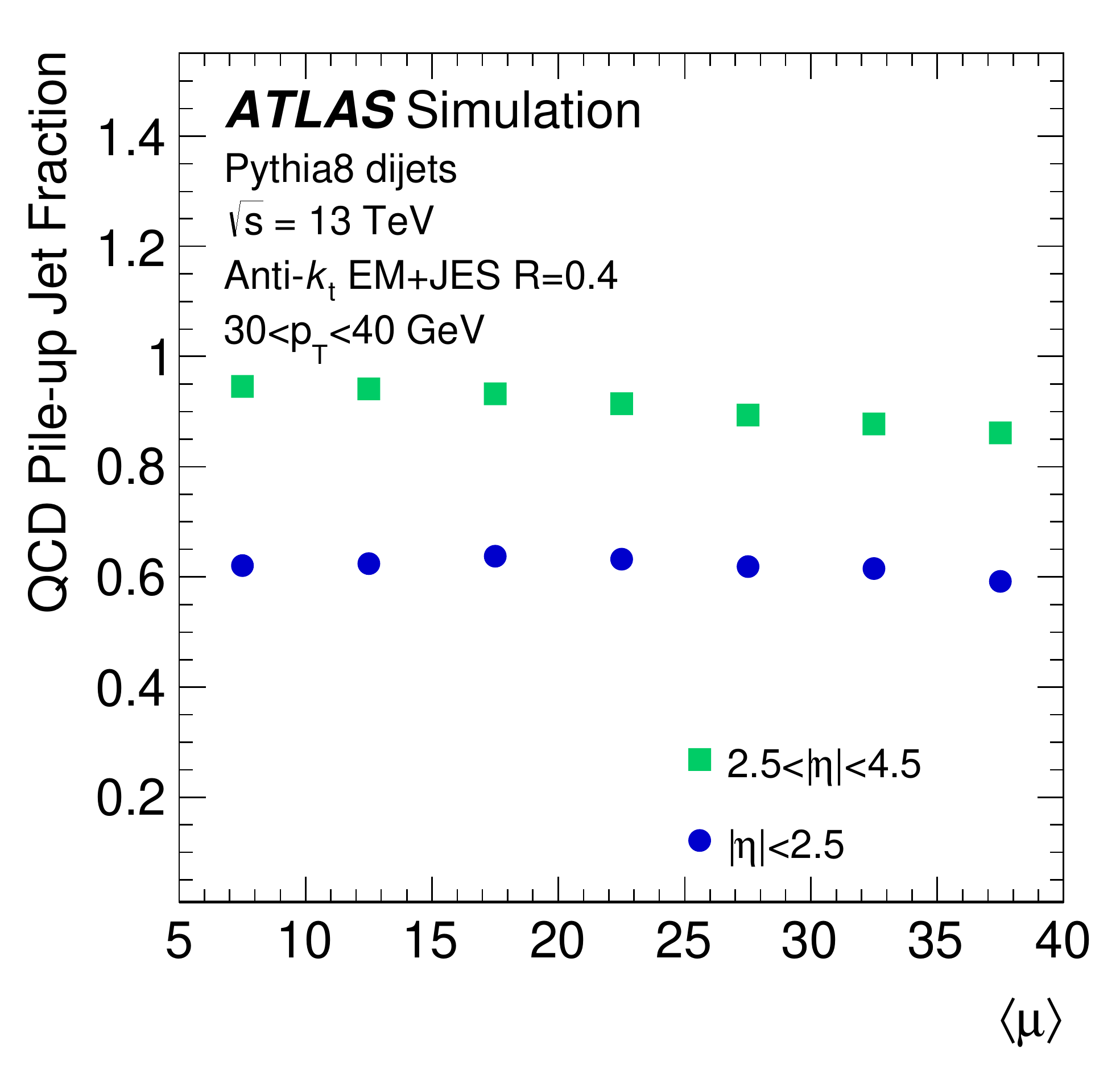}\label{qcdvsnpv30}}
\end{center}
\caption{Fraction of reconstructed pile-up jets classified as QCD pile-up jets,
as a function of \protect\subref{qcdvseta} $|\eta|$, \protect\subref{qcdvspt} \pt, 
and \protect\subref{qcdvsnpv20} $\langle\mu\rangle$ for jets with $20\GeV<\pT<30\GeV$
and \protect\subref{qcdvsnpv30} $30\GeV<\pT<40\GeV$, 
as estimated in dijet events with \PYTHIAeight pile-up simulation.}
\label{fig:quotient}
\end{figure}

In the following, each source of forward pile-up jets is addressed with algorithms
targeting its specific features.

\section{Stochastic pile-up jet tagging with time and shape information}
\label{sec:width}

Given the evidence presented in Section~\ref{sec:pileup} that out-of-time pile-up 
plays an important role for stochastic jets, a direct handle consists of the
timing information associated with the jet.
The jet timing $t_\mathrm{jet}$ is defined as the energy-weighted average of the timing of the 
constituent clusters.
In turn, the cluster timing is defined as the energy-weighted average of the timing of the 
constituent calorimeter cells.
The jet timing distribution, shown in Figure~\ref{fig:distTime}, is symmetric and centred at
$t_\mathrm{jet}=0$ for both the hard-scatter and pile-up jets.
However, the significantly wider distribution for stochastic jets reveals the large
out-of-time pile-up contribution.
For jets with $20<\pt{}<30$ \GeV{}, requiring $|t_\mathrm{jet}|<12$ ns ensures that 20\% of 
stochastic pile-up jets are rejected while keeping 99\% of hard-scatter jets.
In the following, this is always applied as a baseline requirement when identifying stochastic pile-up jets.

\begin{figure}[htbp]
\begin{center}
\subfloat[][]{\includegraphics[width=0.5\textwidth]{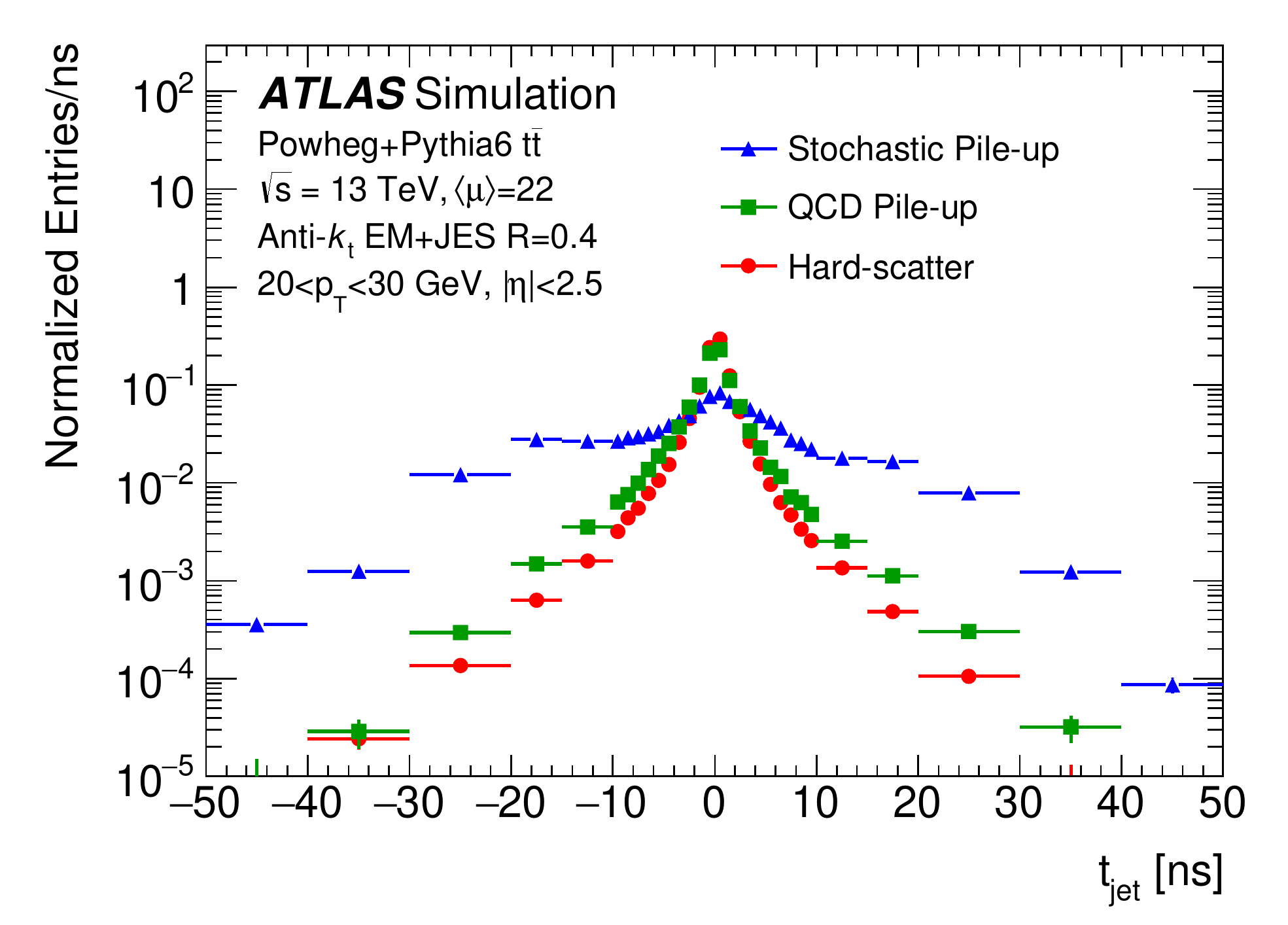}\label{centraltime}}
\subfloat[][]{\includegraphics[width=0.5\textwidth]{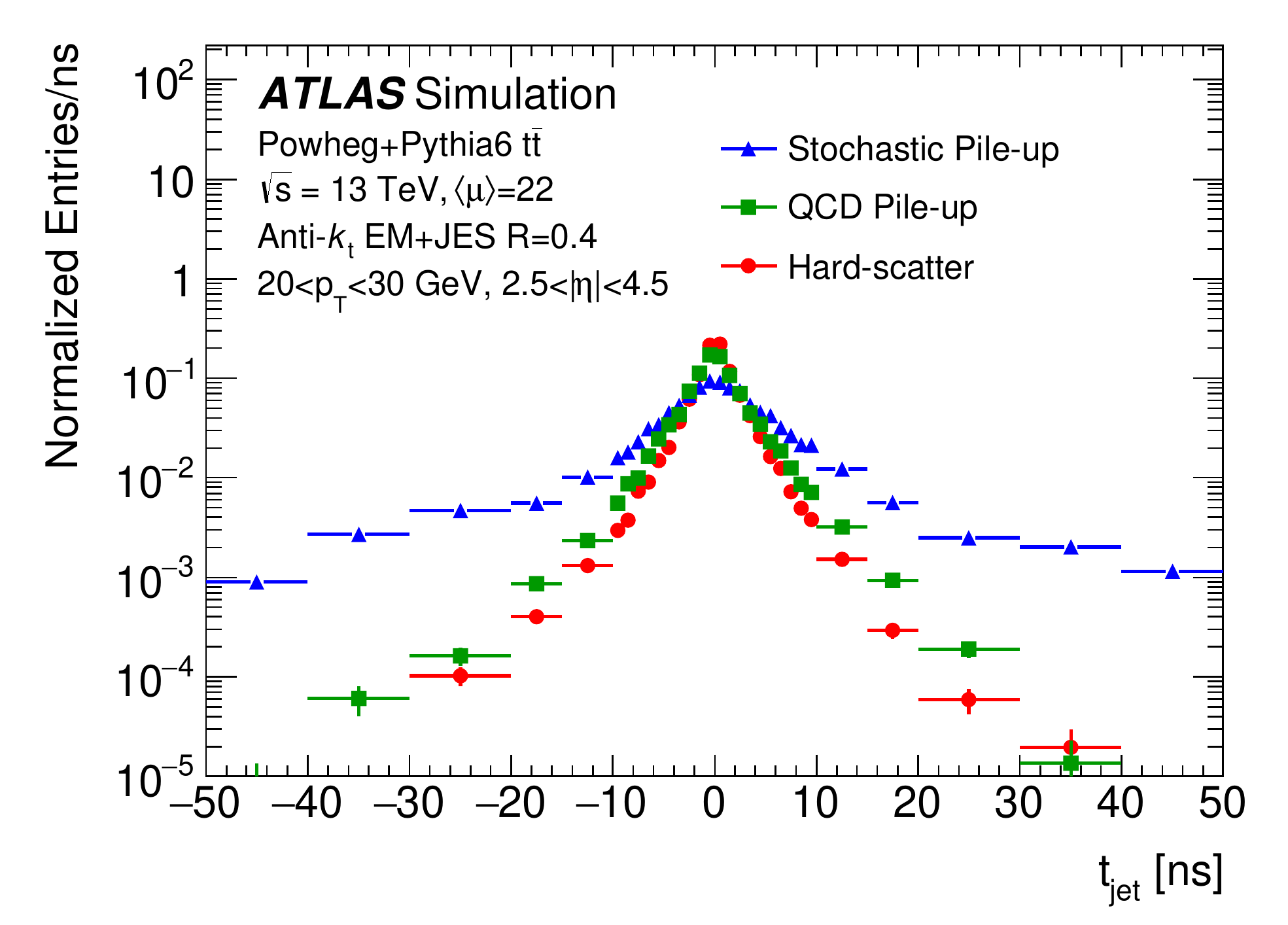}\label{forwardtime}}
\end{center}
\caption{Distribution of the jet timing $t_\mathrm{jet}$ for hard-scatter, QCD pile-up 
and stochastic pile-up jets in the \protect\subref{centraltime} central and \protect\subref{forwardtime} forward region.}
\label{fig:distTime}
\end{figure}

Stochastic jets can be further suppressed using shape information.
Being formed from a random collection of particles from different interactions,
stochastic jets lack the characteristic dense energy core of jets originating
from the showering and hadronization of a hard-scatter parton.
The energy is instead spread rather uniformly within the jet cone.
Therefore, pile-up mitigation techniques based on jet shapes have been shown to be
effective in suppressing stochastic pile-up jets \cite{CMS-PAS-JME-13-005}.
In this section, the challenges of this approach are presented, and 
different algorithms exploiting the jet shape information are described 
and characterized.

The jet width $w$ is a variable that characterizes the energy spread within a jet. 
It is defined as
\begin{equation}
w = \frac{\sum_k{\DeltaR(\mathrm{jet},k)\pTk}}{\sum_k{\pTk}},
\label{eq:clusterwidth}
\end{equation}
where the index $k$ runs over the jet constituents and $\DeltaR(\mathrm{jet},k)$ is
the angular distance between the jet constituent $k$ and the jet axis.
The jet width is a useful observable for identifying stochastic jets, as 
the average width is significantly larger for jets 
with a smaller fraction of energy originating from a single interaction.

In simulation the jet width can be computed using truth-particles (\textit{truth-particle width}),
as a reference point to benchmark the performance of the reconstructed observable.
At detector level, the jet constituents are calorimeter topo-clusters.
In general, topo-clustering compresses the calorimeter information while retaining its
fine granularity. Ideally, each cluster captures the energy shower from a single
incoming particle.
However, the cluster multiplicity in jets decreases quickly 
in the forward region, to the point where jets are formed by a single cluster and
the jet width can no longer be defined.
An alternative approach consists of using as constituents
the 11 by 11 grid of calorimeter towers centred around the jet axis.
The use of calorimeter towers ensures a fixed multiplicity given by the $0.1\times0.1$ granularity
so that the jet width always contains jet shape information.

\begin{figure}[htbp]
\begin{center}
\subfloat[][]{\includegraphics[width=0.475\textwidth]{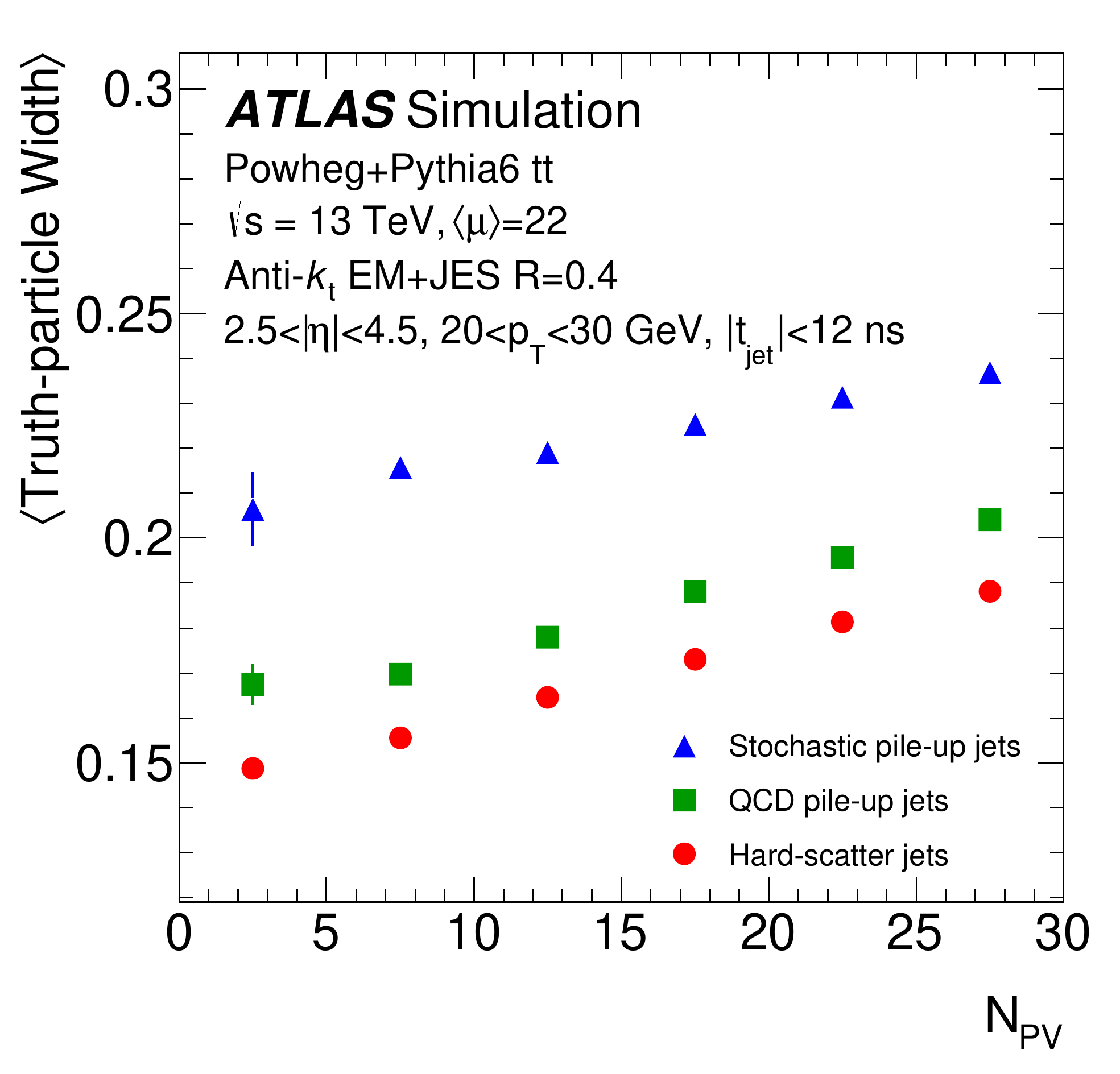}\label{vsnpvtruth}}\quad
\subfloat[][]{\includegraphics[width=0.475\textwidth]{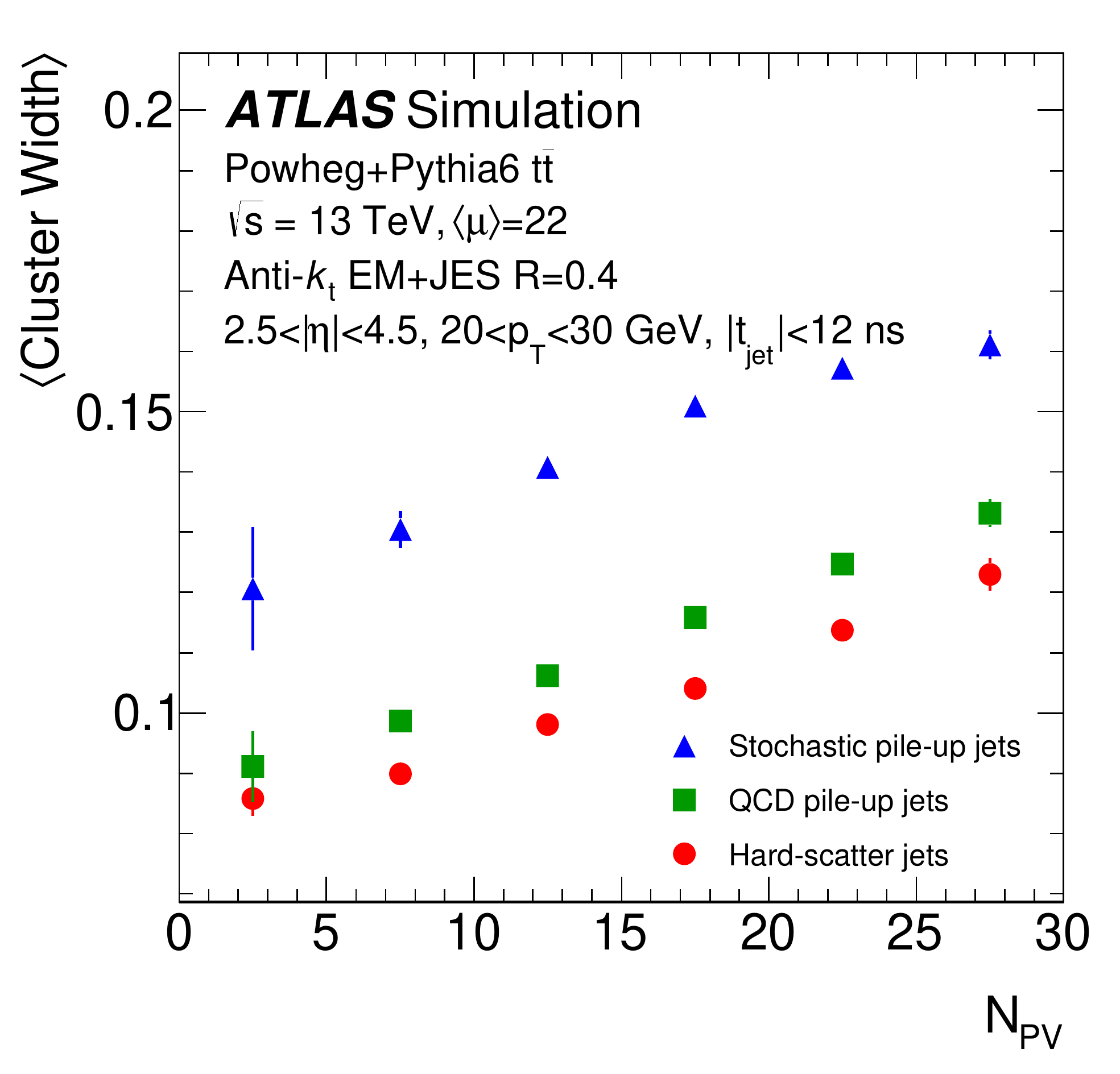}\label{vsnpvcluster}}\\
\subfloat[][]{\includegraphics[width=0.475\textwidth]{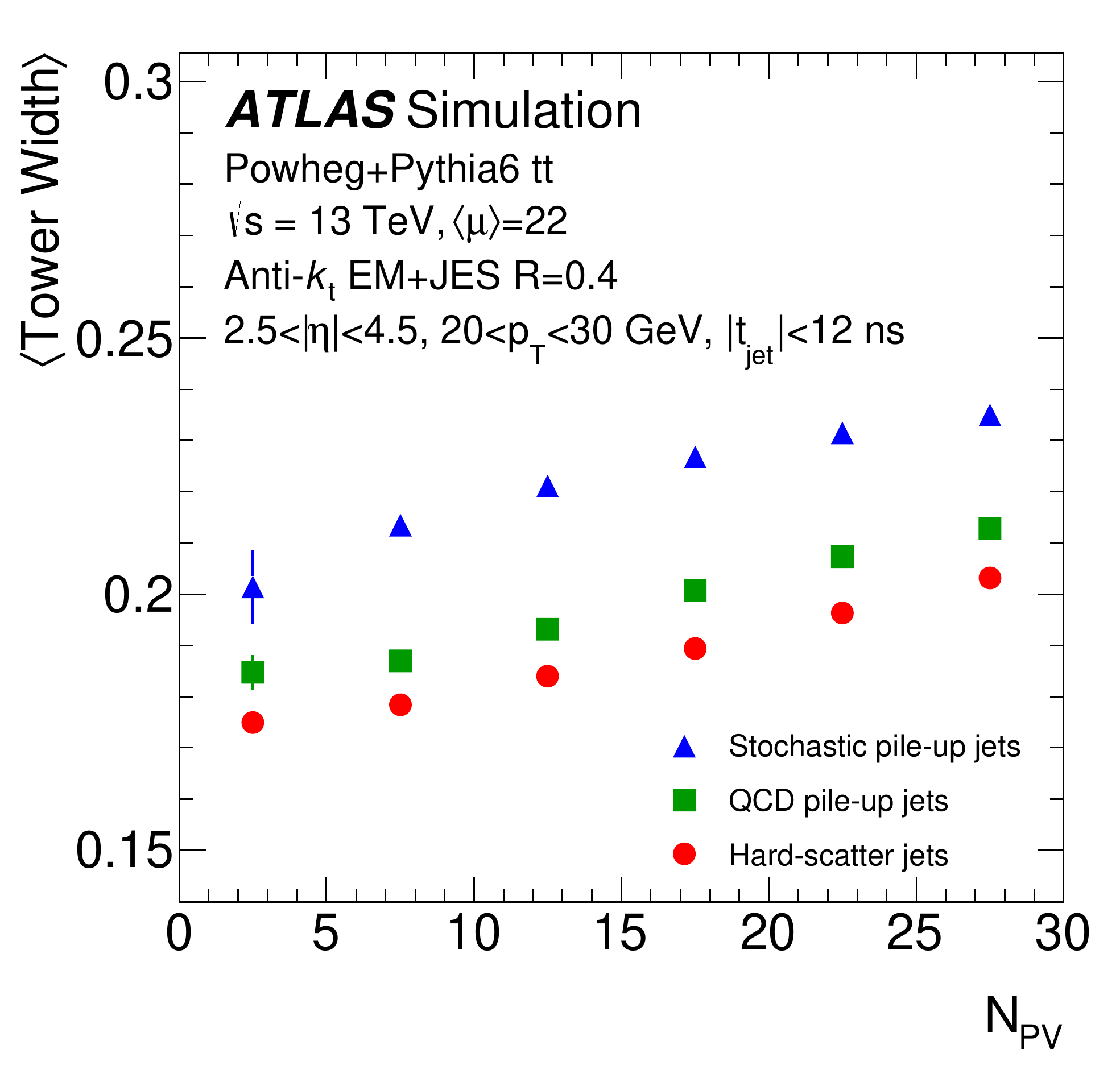}\label{vsnpvtower}}
\end{center}
\caption{Dependence of the average jet width on the number of reconstructed primary vertices (\NPV). 
The distributions are shown using \protect\subref{vsnpvtruth} hard-scatter and in-time pile-up truth-particles, 
\protect\subref{vsnpvcluster} clusters, or \protect\subref{vsnpvtower} towers as constituents.}
\label{fig:width_limits}
\end{figure}

As shown in Figure~\ref{fig:width_limits}, the average jet width depends on the pile-up conditions.
At higher pile-up values, a larger number of pile-up particles are likely
to contribute to a jet, thus broadening the energy distribution within the jet
itself. As a result, the width drifts towards higher values for hard-scatter, QCD pile-up, 
and stochastic jets. 
The difference in width between hard-scatter and QCD pile-up jets is due to the different underlying
\pt spectra. The spectrum of QCD pile-up jets is softer than that of the hard-scatter jets for the process considered (\ttbar);
therefore, a significant fraction of QCD pile-up jets are reconstructed with \pt between $20\GeV$
and $30 \GeV$ because the stochastic and out-of-time component is larger than in hard-scatter jets.

\begin{figure}[htbp]
\begin{center}
\includegraphics[height=70mm]{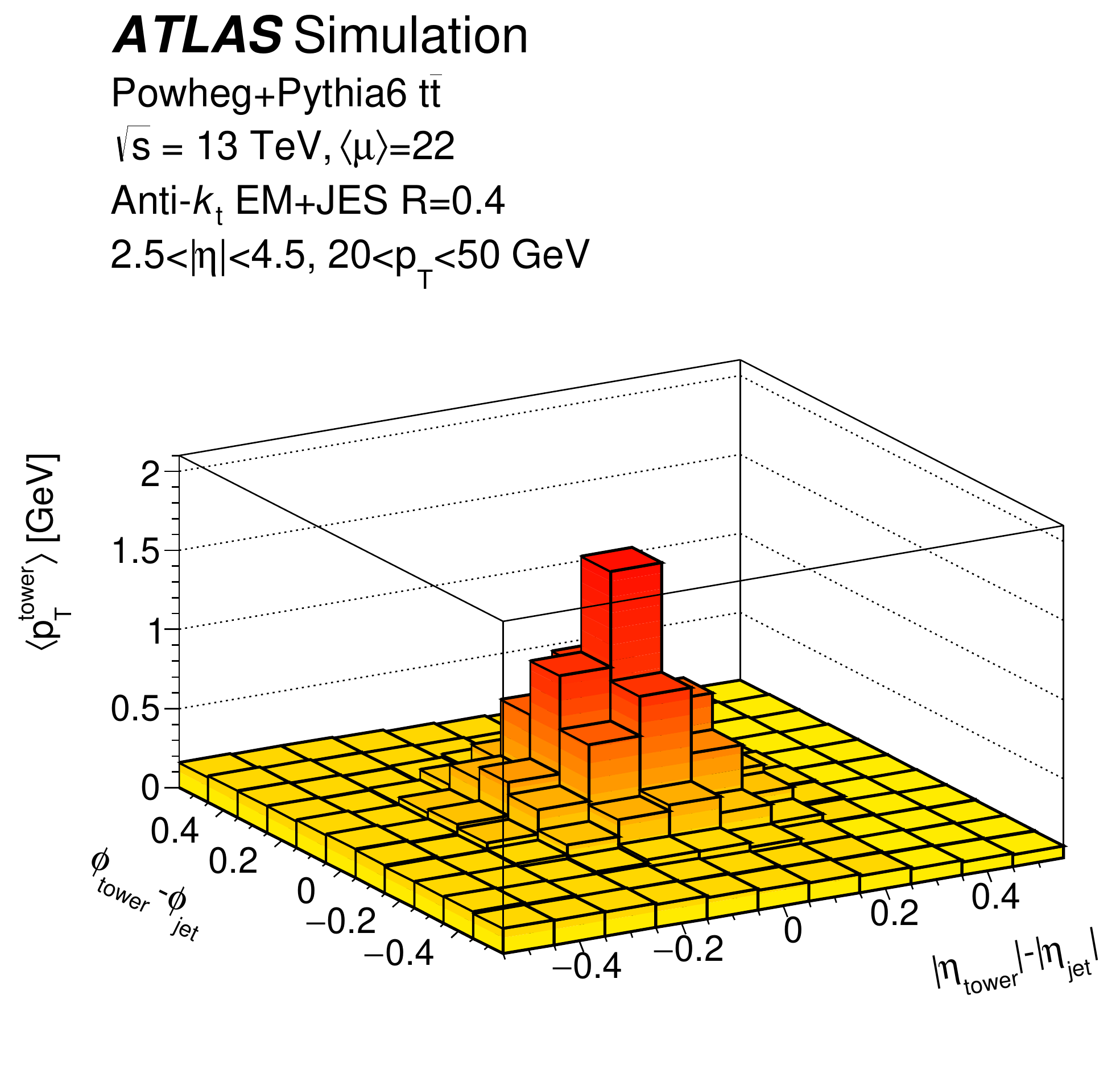}
\end{center}
\caption{
Distribution of the average tower \pt for hard-scatter jets as a function of the angular distance from the jet axis in $\eta$ 
and $\phi$ in simulated \ttbar events.}
\label{fig:towerDist}
\end{figure}

Using calorimeter towers as constituents, it is possible
to explore the \pt distribution within a jet with a fixed $\eta\times\phi$ granularity.
Figure \ref{fig:towerDist} shows the two-dimensional \pt distribution
around the jet axis for hard-scatter jets. The distribution is symmetric in $\phi$, while the pile-up pedestal
decreases with increasing $\eta$, as is expected in the forward region.
A new variable, designed to exploit the full information about tower constituents, 
is considered.
The two-dimensional\footnote{The simultaneous fit of both dimensions was found to perform better than the fit of a 1D projection.} 
\pt distribution in the \Deltaeta--\Deltaphibare plane centred 
around the jet axis is fitted with a function
\begin{equation}
f = \alpha+\beta\Delta\eta+\gamma \mathrm{e}^{-\frac{1}{2}\left(\frac{\Deltaeta}{0.1}\right)^2-\frac{1}{2}\left(\frac{\Deltaphibare}{0.1}\right)^2}.
\label{eq:gausfunc}
\end{equation}

Both the width of the Gaussian component of the fit and the range in which the fit is
performed are treated as jet-independent constants.
The fit range, an $11\times11$ tower grid, optimizes the balance between an improved 
constant ($\alpha$) and linear ($\beta$) term measurement by using a larger range and 
a decreased risk of including outside pile-up fluctuations by using a smaller range.
On average, the jet tower \pt distribution is symmetric with respect to $\Delta\phi$, 
and pile-up rejection at constant hard-scatter efficiency is improved by averaging the tower momenta at $\Deltaphi$ and 
$-\Deltaphi$ so that fluctuations are partially cancelled before performing the fit.

The constant ($\alpha$) and linear ($\beta$) terms in the fit capture the average stochastic pile-up 
contribution to the jet \pt distribution, while the Gaussian term describes
the \pt distribution from the underlying QCD jet.
The parameter $\gamma$ therefore represents a stochastic pile-up-subtracted estimate of the \pt
of such a QCD pile-up jet in a $\Delta R=0.1$ core assuming a Gaussian \pt distribution of its
constituent towers.
By definition, $\gamma$ does not depend on the amount of pile-up in the event,
but only on the nature of the jet as stochastic or QCD.
In order to make the fitting procedure more robust, the Gaussian width parameter is fixed.
While the width of a QCD pile-up jet is expected to depend on the truth-particle jet \pt and $\eta$,
such dependence is negligible in the \pt range relevant for these studies (20--50 \GeV).
Figure~\ref{fig:singleJets}, showing projections of the tower distribution with the fit function overlaid,
illustrates the characteristic peaking shape of pure QCD jets compared with the flatter distribution
in stochastic jets.
The hard-scatter jet distribution displays the expected, sharply peaked distribution, 
while the stochastic pile-up jet distribution is flat with various off-centre features, reflecting the randomness of the underlying processes.

\begin{figure}[htbp]
\begin{center}
\subfloat[][]{\includegraphics[width=0.475\textwidth]{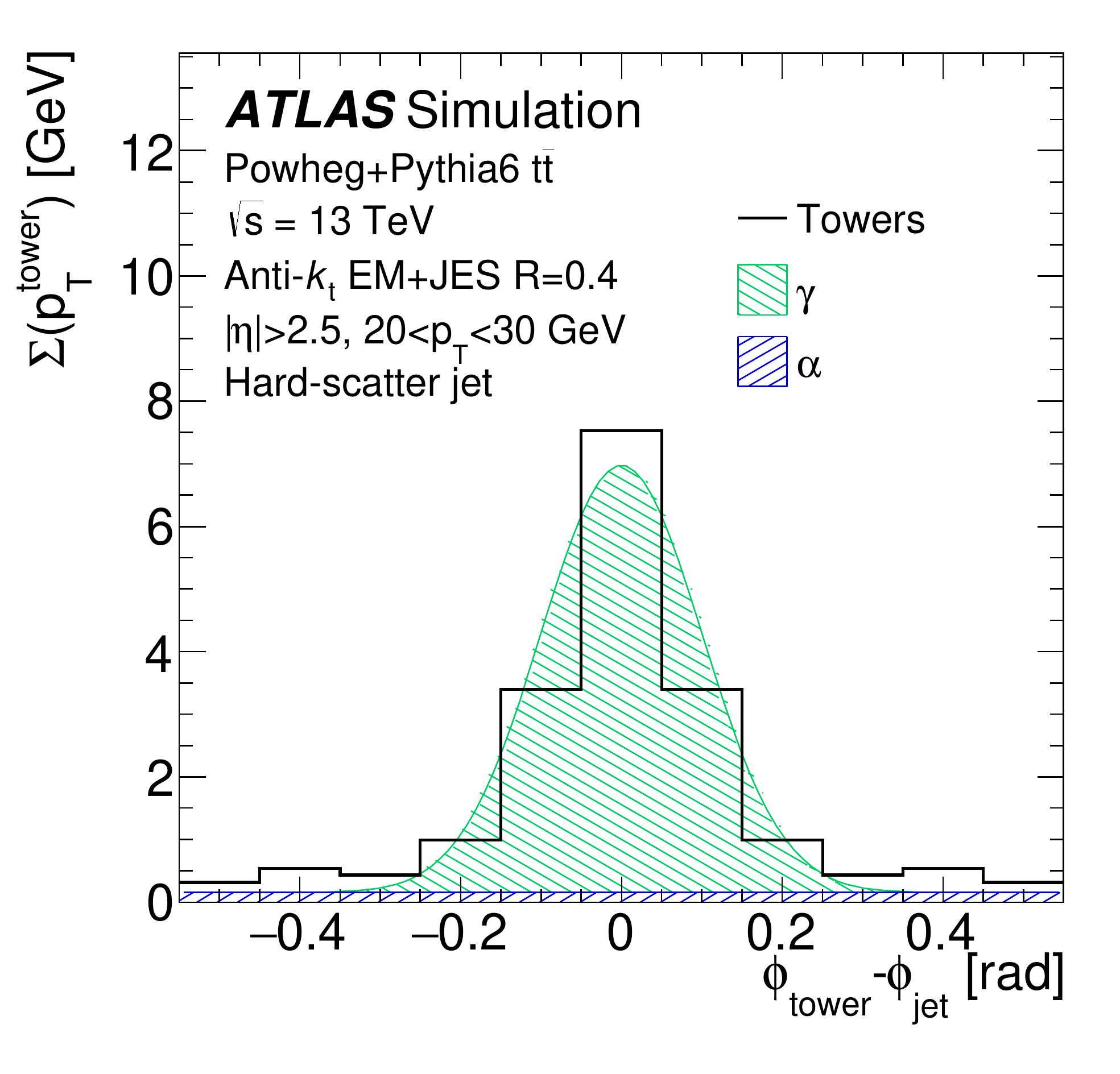}\label{jeths}}
\subfloat[][]{\includegraphics[width=0.475\textwidth]{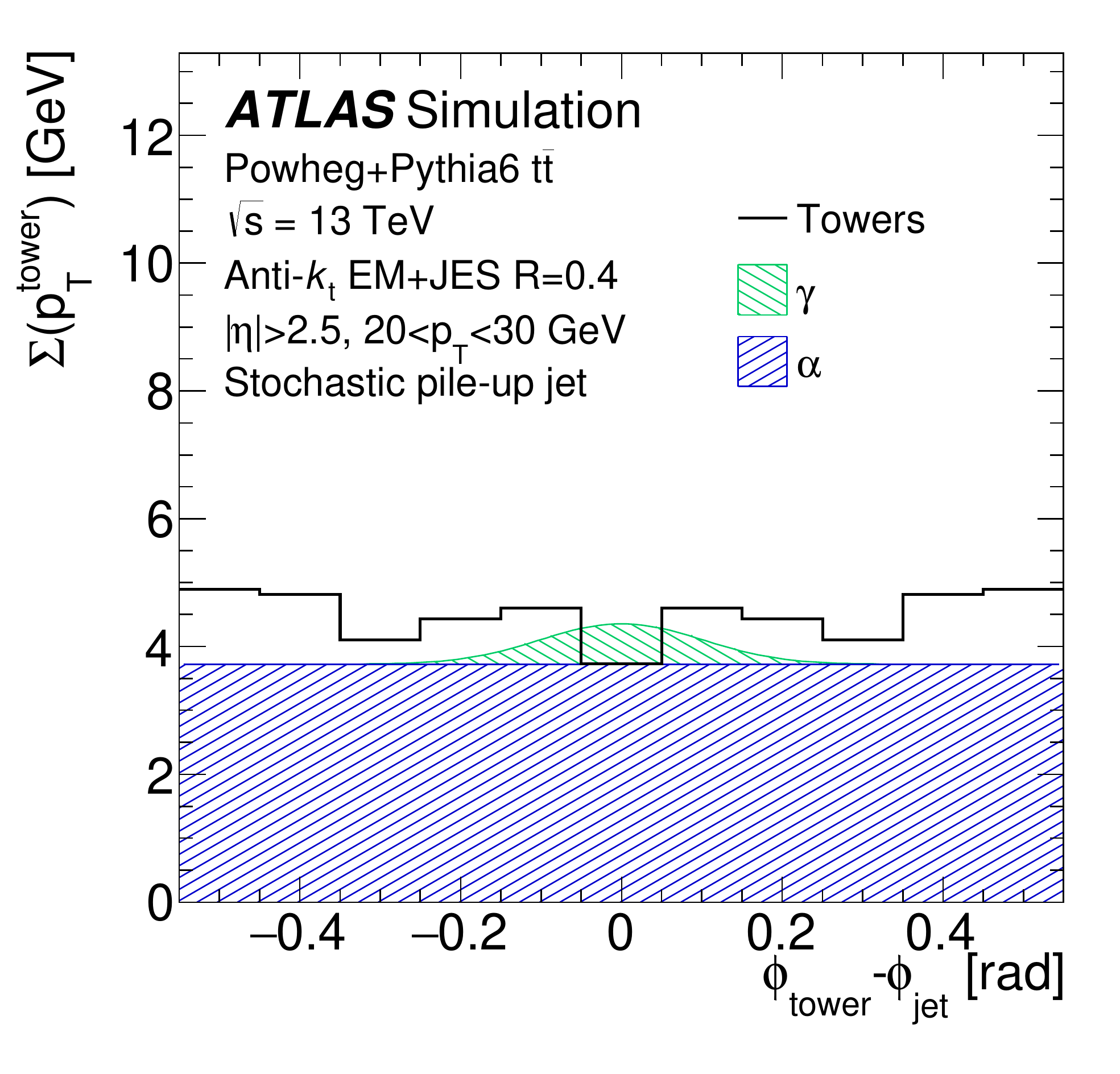}\label{jetsto}}
\end{center}
\caption{Symmetrized tower \pt distribution projections in $\phi$
for an example \protect\subref{jeths} hard-scatter jet and \protect\subref{jetsto} stochastic
pile-up jet in simulated \ttbar events.
The black histogram line corresponds to the projection of the 2D tower distribution.
The fit model closely follows the hard-scatter jet distribution, yielding a large Gaussian signal,
while stochastic pile-up jets feature multiple smaller signals, away from the jet core.}
\label{fig:singleJets}
\end{figure}

The performance of the $\gamma$ variable and of the cluster-based
and tower-based widths is compared in Figure~\ref{fig:roc_constCompare},
where the efficiency for stochastic pile-up jets is shown as a function of the
hard-scatter jet efficiency. Each curve is obtained by applying an upper or lower
bound on the jet width or $\gamma$, respectively, in order to select hard-scatter jets.
The tower-based width outperforms the cluster-based width over the whole efficiency 
range, while the $\gamma$ variable performs similarly to the tower-based width.
The hard-scatter efficiency and pile-up efficiency dependence on the number of reconstructed vertices
in the event (\NPV) and $\eta$ is shown in Figure~\ref{fig:eff_constCompare};
the requirement for each discriminant is tuned so that an overall efficiency of 90\%
is achieved for hard-scatter jets. 
By construction, the performance of the $\gamma$ variable is less affected by the 
pile-up conditions than the two width variables.

\begin{figure}[htbp]
\begin{center}
\subfloat[][]{\includegraphics[width=0.475\textwidth]{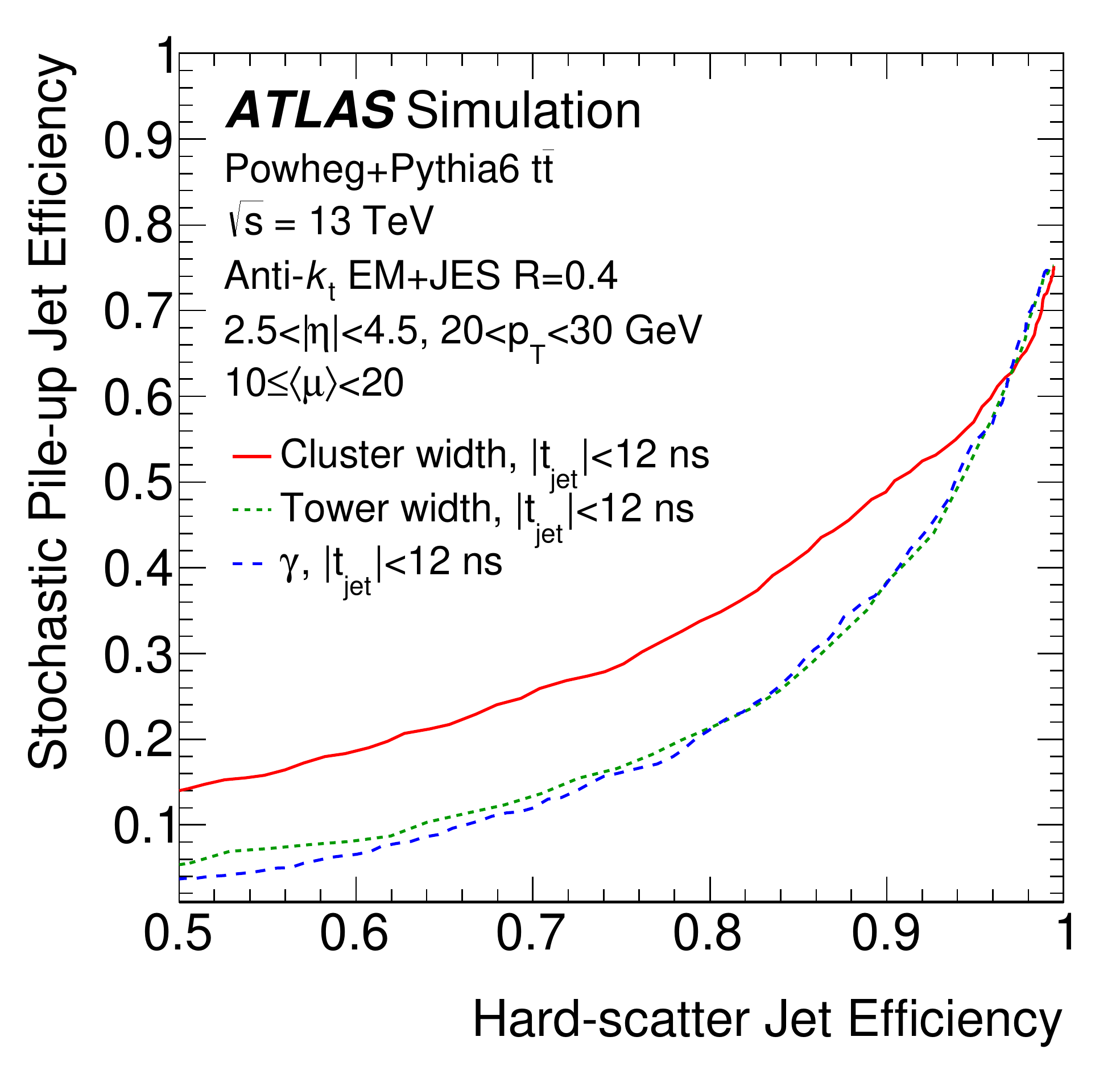}\label{rocconst10}}
\subfloat[][]{\includegraphics[width=0.475\textwidth]{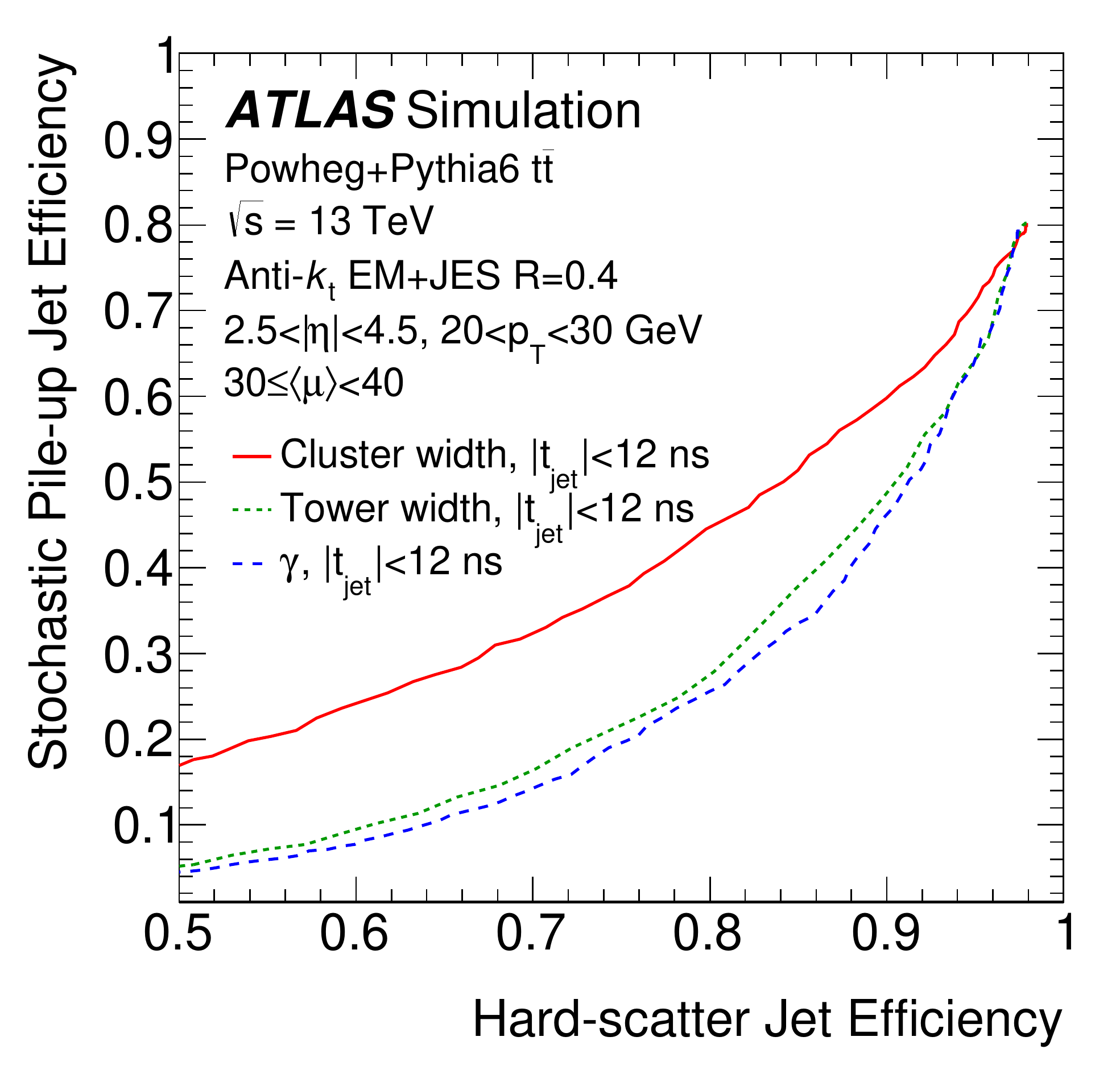}\label{rocconst30}}
\end{center}
\caption{Efficiency for stochastic pile-up jets  
as a function of the efficiency for hard-scatter jets
using different shape-based discriminants: \protect\subref{rocconst10} $10\leq\langle\mu\rangle<20$ and 
\protect\subref{rocconst30} $30\leq\langle\mu\rangle<40$ in simulated \ttbar events.}
\label{fig:roc_constCompare}
\end{figure}

\begin{figure}[htbp]
\begin{center}
\subfloat[][]{\includegraphics[width=0.475\textwidth]{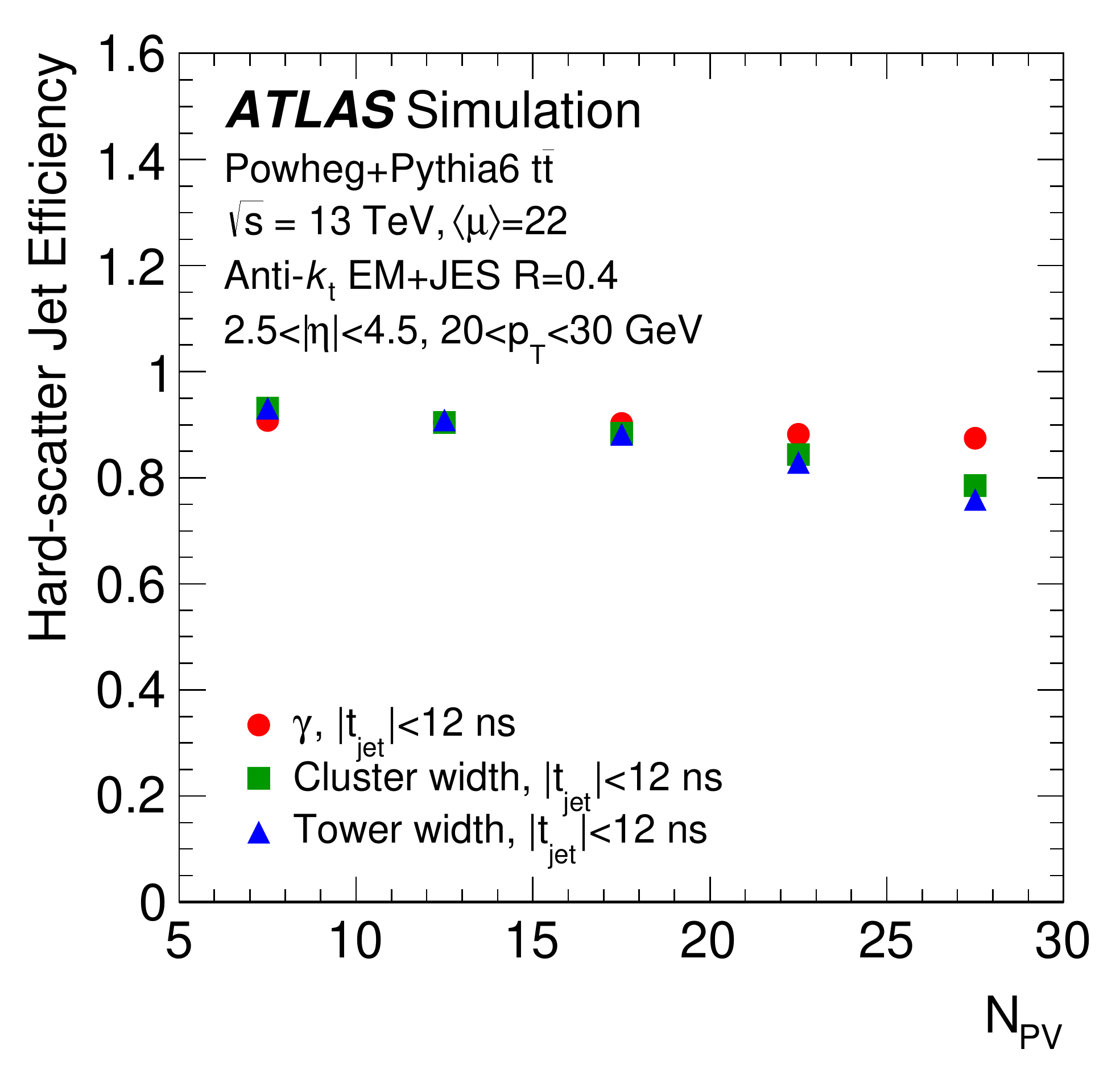}\label{effnpv}}\quad
\subfloat[][]{\includegraphics[width=0.475\textwidth]{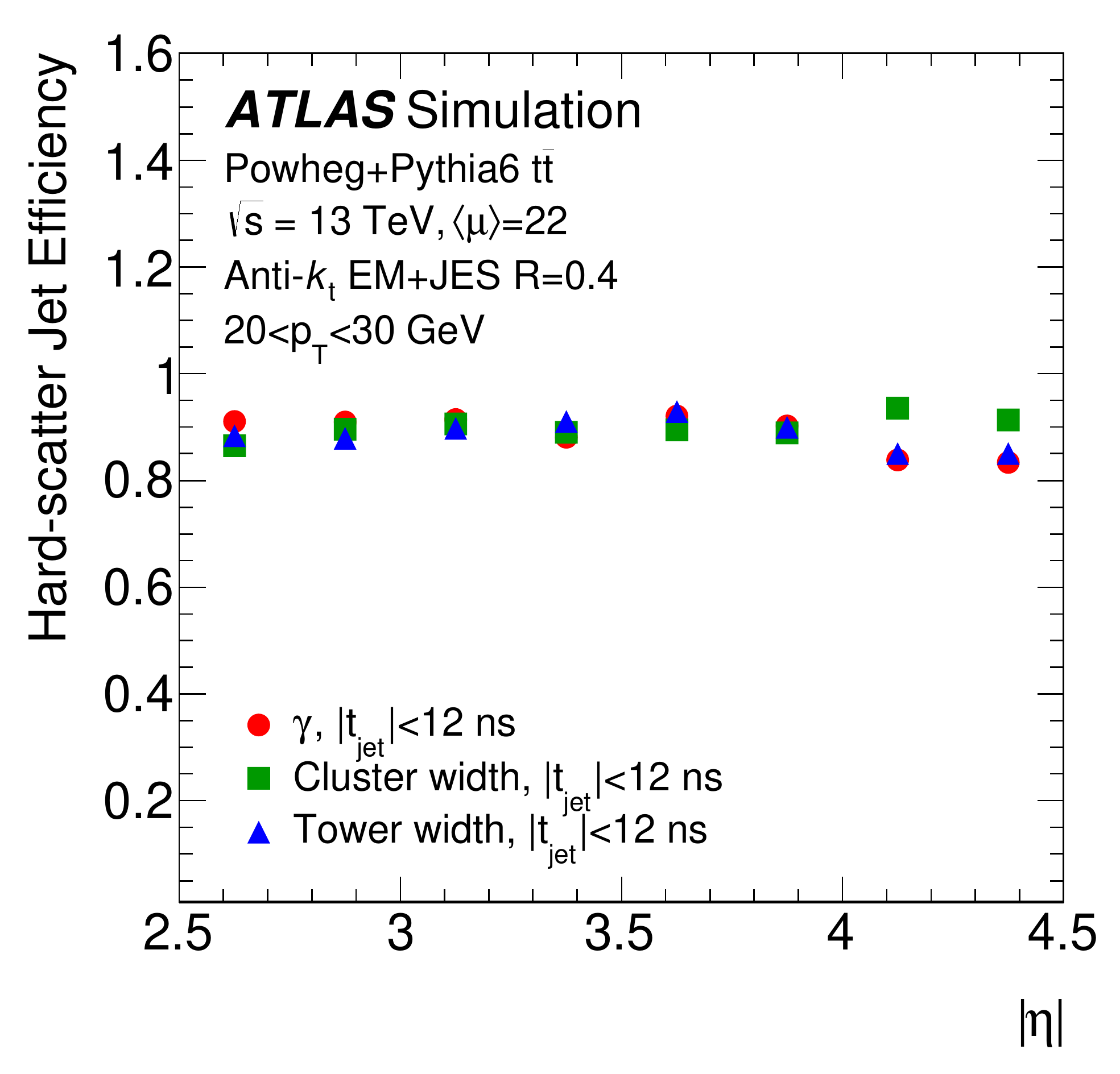}\label{effeta}}\\
\subfloat[][]{\includegraphics[width=0.475\textwidth]{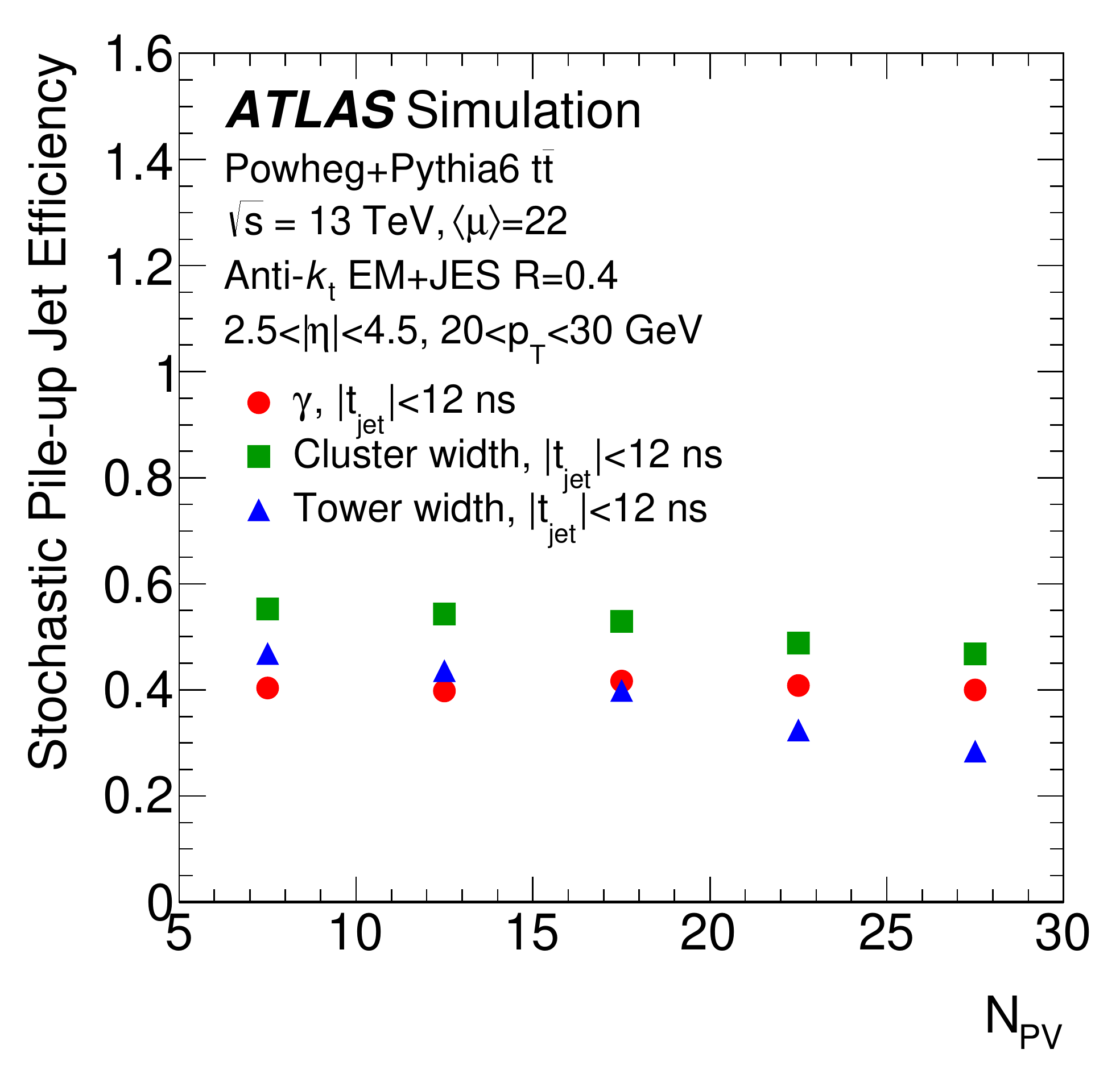}\label{fakenpv}}\quad
\subfloat[][]{\includegraphics[width=0.475\textwidth]{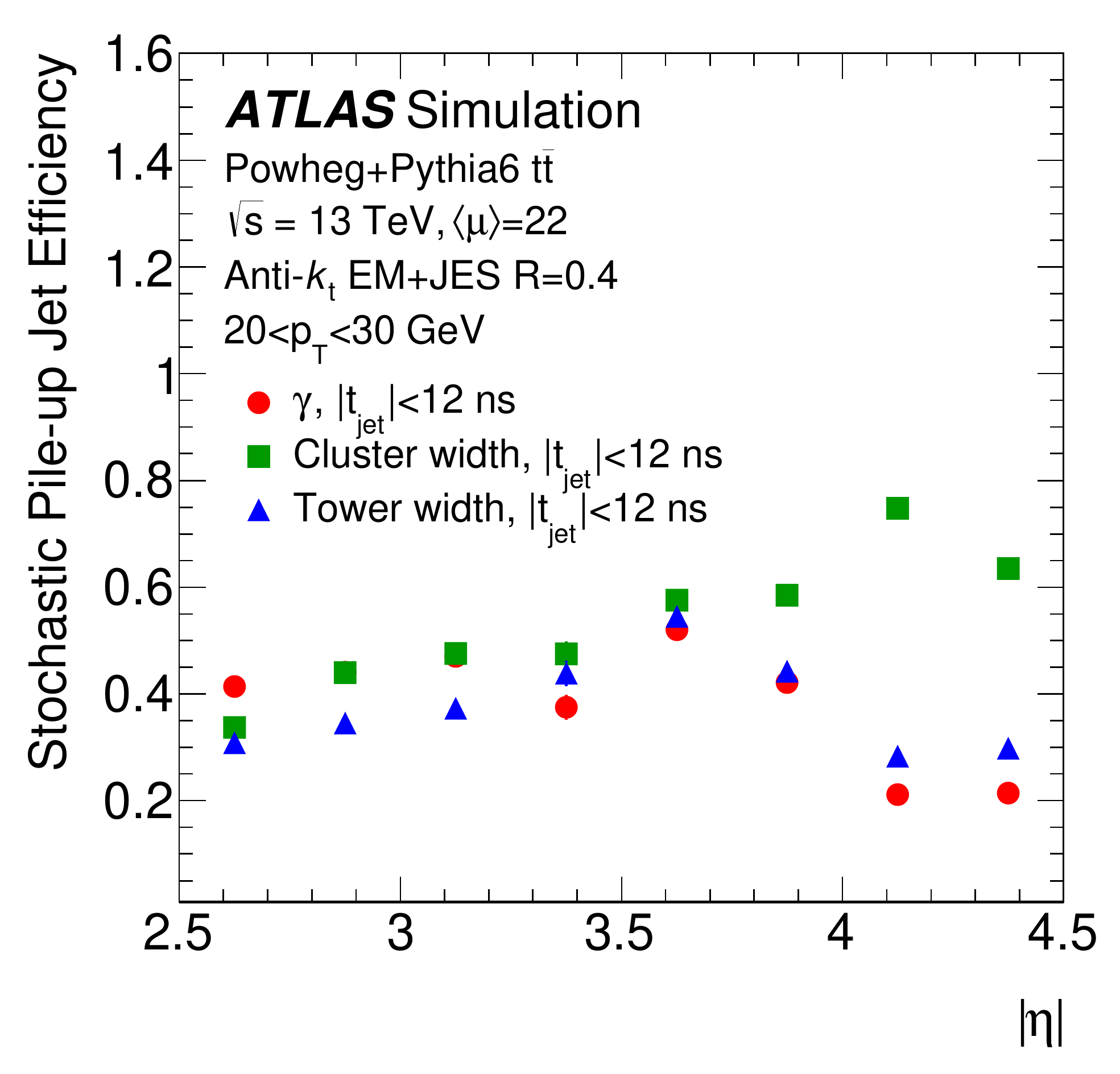}\label{fakeeta}}
\end{center}
\caption{
Hard-scatter jet efficiency as a function of \protect\subref{effnpv} number of reconstructed primary vertices \NPV
and \protect\subref{effeta} pseudorapidity $|\eta|$, as well as
stochastic pile-up jet efficiency as a function of \protect\subref{fakenpv} number of reconstructed primary vertices \NPV and 
\protect\subref{fakeeta} pseudorapidity $|\eta|$ at 90\% efficiency of selecting hard-scatter jets in simulated \ttbar events.}
\label{fig:eff_constCompare}
\end{figure}

The $\gamma$ parameter is a good discriminant for stochastic pile-up jets because it 
provides an estimate of the largest amount 
of \pt in the jet originating from a single vertex. If there is no dominant contribution, 
the \pt distribution does not
feature a prominent core, and therefore $\gamma$ is close to zero.
With this approach, all jets are effectively considered as QCD pile-up jets, and 
$\gamma$ is used to estimate their core \pt. 
Therefore, from this stage, the challenge of pile-up rejection is reduced to the 
identification and
rejection of QCD pile-up jets, which is discussed in the following section.

\section{QCD pile-up jet tagging with topological information}
\label{sec:fjvt}

While it has been shown that pile-up mitigation techniques based on jet shapes
are effective in suppressing stochastic pile-up jets,
such methods do not address QCD pile-up jets that are prevalent in the forward region.
This section describes the development of an effective rejection method 
specifically targeting QCD pile-up jets.

QCD pile-up jets originate from a single \pp interaction where multiple jets can be 
produced.
The total transverse momentum associated with each pile-up interaction is expected to be 
conserved;\footnote{The cross-section of interactions producing high-\pt neutrinos is negligible,
compared to the rate of multijet events.}
therefore all jets and central tracks associated with a given vertex can be exploited 
to identify QCD pile-up jets beyond the tracking coverage of the inner detector.
The principle is clear if the dijet final state alone is considered.
Forward pile-up jets are therefore identified by looking for a pile-up jet 
opposite in $\phi$ in the central region.
The main limitation of this approach is that it only addresses dijet pile-up interactions
in which both jets are reconstructed.

In order to address this challenge, a more comprehensive approach is adopted by considering the total
transverse momentum of tracks and jets associated with each reconstructed vertex independently.
The more general assumption is that the transverse momentum of each pile-up interaction 
should be balanced, and any imbalance would be due to a forward jet from one of the interactions.

In order to properly compute the transverse momentum of each interaction, only QCD pile-up jets
should be considered.
Consequently, the challenge of identifying forward QCD pile-up jets using transverse momentum conservation 
with central pile-up jets requires being able to discriminate between QCD and 
stochastic pile-up jets in the central region.

\subsection{A discriminant for central pile-up jet classification}
\label{sec:diffrpt}

Discrimination between stochastic and QCD pile-up jets
in the central region can be achieved using track and vertex information.
This section describes a new discriminant built for this purpose.

The underlying features of QCD and stochastic pile-up jets are different.
Tracks matched to QCD pile-up jets mostly originate from a vertex PV$_i$ corresponding 
to a pile-up interaction ($i\neq0$), thus yielding $\RpT^i>\RpT^0$ for a given jet. 
Such jets have large values of $\RpT^i$ with respect to the pile-up 
vertex $i$ from which they originated.
Tracks matched to stochastic pile-up jets are not likely to originate from
the same interaction, thus yielding small $\RpT^i$ values with respect to any vertex $i$.
This feature can be exploited to discriminate between these two categories.
For stochastic pile-up jets, the largest $\RpT^i$ value is going to be of similar size as the average $\RpT^i$ value across all vertices,
while a large difference will show for QCD jets, as most tracks originate from the 
same pile-up vertex.

Thus, the difference between the leading and median values of
$\RpT^i$ for a central jet, \diffRpT, can be used for distinguishing QCD pile-up jets 
from stochastic pile-up jets in the central region, as shown in Figure~\ref{fig:diffRpT}.
A minimum \diffRpT requirement can effectively reject stochastic pile-up jets.
In the following a $\diffRpT>0.2$ requirement is applied for central jets with
$\pT < 35 \GeV$.
Above this threshold the fraction of stochastic pile-up jets is negligible, and all pile-up jets are
therefore  assumed to be QCD pile-up jets 
irrespective of their \diffRpT value.
The choice of threshold depends on the pile-up conditions. 
This choice is tuned to be optimal for the collisions considered in this study,
with an average of 13.5 interactions per bunch crossing.

\begin{figure}[htbp]
\begin{center}
\includegraphics[width=0.475\textwidth]{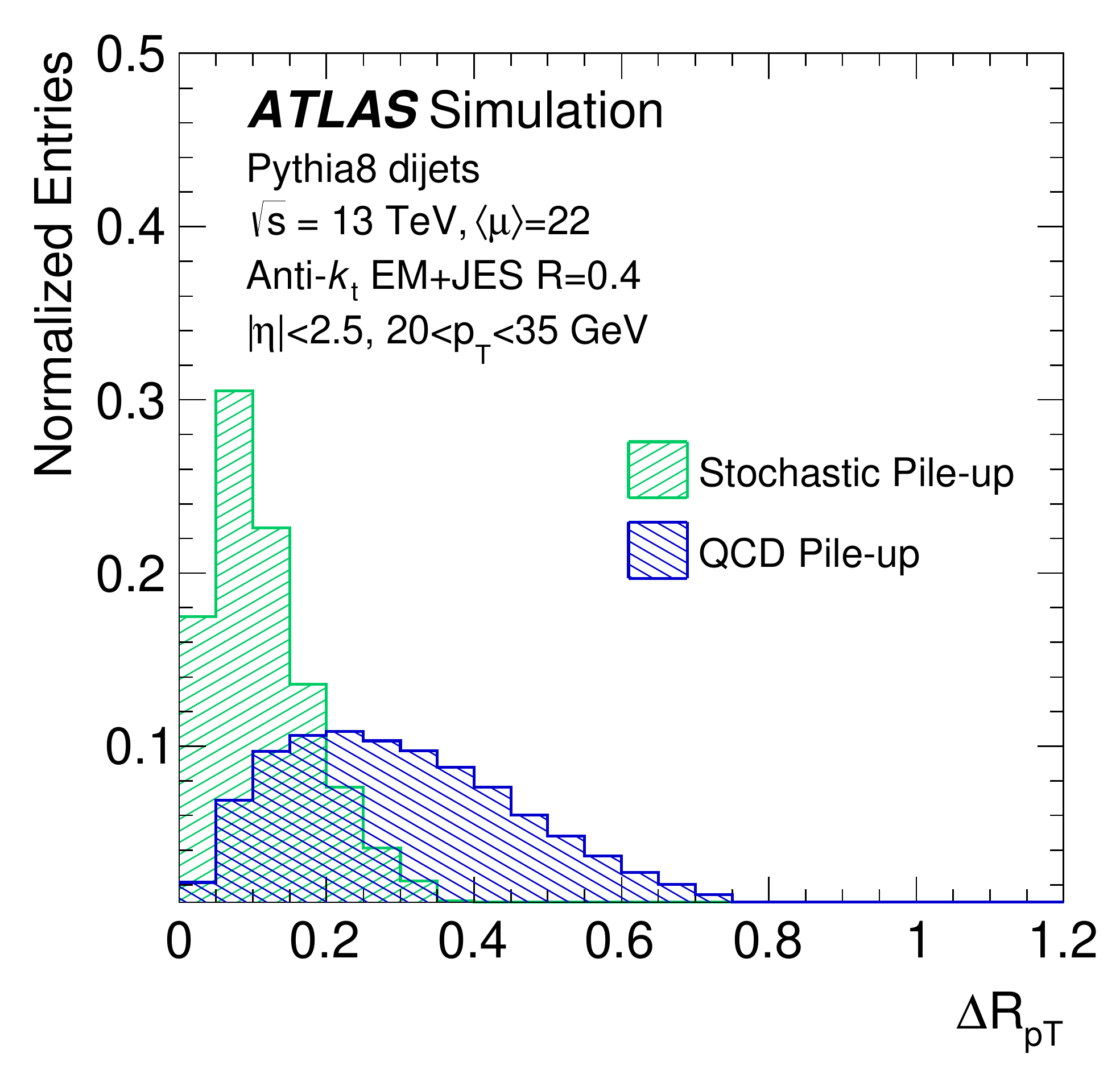}
\end{center}
\caption{Distribution of \diffRpT for stochastic and QCD pile-up jets, 
as observed in dijet events with \PYTHIAeight pile-up simulation.}
\label{fig:diffRpT}
\end{figure}

The total transverse momentum of each vertex is thus computed by averaging, with a vectorial sum, the total
transverse momentum of tracks and central jets assigned to the vertex.
The jet--vertex matching is performed by considering the largest $\RpT^i$ for each jet.
The transverse momentum vector ($\boldsymbol{p}_\mathrm{T}$) of a given forward jet is then compared 
with the total transverse momentum of each vertex
in the event. If there is at least one pile-up vertex in the event with a large total vertex transverse
momentum back-to-back in $\phi$ with respect to the forward jet, the jet itself is likely to have
originated from that vertex.
Figure \ref{fig:fjvtevt} shows an example event, where the $\boldsymbol{p}_\mathrm{T}$ of a forward pile-up jet 
is back-to-back with respect to the
total transverse momentum of the vertex from which it is expected to have originated.

\begin{figure}[htbp]
\begin{center}
\includegraphics[height=150mm]{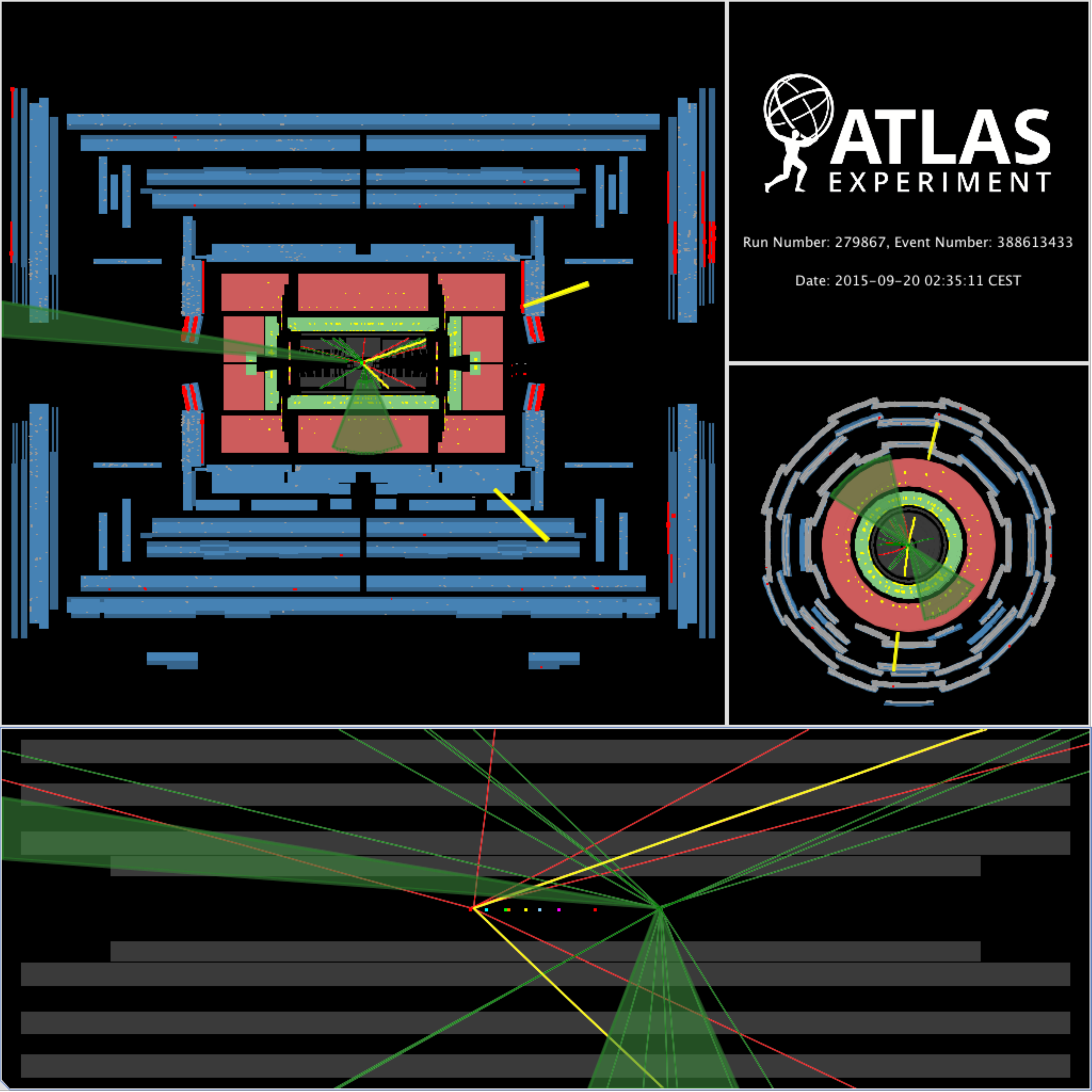}
\end{center}
\caption{
Display of candidate $\Zboson(\to\mu\mu)$ event (muons in yellow) containing two QCD pile-up jets.
Tracks from the primary vertex are in red, those from the pile-up vertex with the highest $\sum p_\mathrm{T}^2$
are in green. The top panel shows a transverse and longitudinal view of the detector, 
while the bottom panel shows the details of the event in the ID in the longitudinal view.
}
\label{fig:fjvtevt}
\end{figure}

\subsection{Forward Jet Vertex Tagging algorithm}

The procedure is referred to as \textit{forward Jet Vertex Tagging} (f\JVT).
The main parameters for the forward JVT algorithm are thus the maximum JVT value,
$\JVT_\mathrm{max}$, to
reject central hard-scatter jets and the minimum \diffRpT requirement to ensure the
selected pile-up jets are QCD pile-up jets.
$\JVT_\mathrm{max}$ is set to 0.14 corresponding to an efficiency of selecting pile-up jets of 
93\% in dijet events.
The minimum \diffRpT requirement defines the operating point in terms of efficiency for 
selecting QCD pile-up jet and contamination from stochastic pile-up jets.
A minimum \diffRpT of 0.2 is required, corresponding to an efficiency of $70\%$ 
for QCD pile-up jets and $20\%$ for stochastic pile-up jets in dijet events.
The selected jets are then assigned to the vertex PV$_i$ corresponding to the highest 
$\RpT^i$ value.
For each pile-up vertex $i$, $i\neq0$, 
the missing transverse momentum $\langle\boldsymbol{p}_{\mathrm{T},i}^\mathrm{miss}\rangle$ is computed as
the weighted vector sum of the jet ($\boldsymbol{p}_\mathrm{T}^\mathrm{jet}$) and track ($\boldsymbol{p}_\mathrm{T}^\mathrm{track}$) transverse momenta:
\begin{equation}
\langle\boldsymbol{p}_{\mathrm{T},i}^\mathrm{miss}\rangle=-\frac{1}{2}\left(\sum_{\mathrm{tracks \in PV}_i}k\boldsymbol{p}_\mathrm{T}^\mathrm{track} + \sum_{\mathrm{jets \in PV}_i}\boldsymbol{p}_\mathrm{T}^\mathrm{jet}\right).
\end{equation}
The factor $k$ accounts for intrinsic differences between the jet and track terms.
The track component does not include the contribution of neutral particles, 
while the jet component is not sensitive to soft emissions significantly below 20 \GeV.
The value $k=2.5$ is chosen as the one that optimizes the overall rejection of 
forward pile-up jets.

The \fJVT discriminant for a given forward jet, with respect to the vertex $i$, 
is then defined as the normalized projection of the missing transverse momentum on 
$\boldsymbol{p}_T^\mathrm{fj}$:
\begin{equation}
\mathrm{fJVT}_i = \frac{\langle\boldsymbol{p}_{\mathrm{T},i}^\mathrm{miss}\rangle\cdot\boldsymbol{p}_\mathrm{T}^\mathrm{fj}}{|\boldsymbol{p}_\mathrm{T}^\mathrm{fj}|^2},
\end{equation}
where $\boldsymbol{p}_\mathrm{T}^\mathrm{fj}$ is the forward jet's transverse momentum.
The motivation for this definition is that the amount of missing transverse momentum in the direction of the
forward jet needed for the jet to be tagged should be proportional to the jet's transverse momentum.
The forward jet is therefore tagged as pile-up if its \fJVT value, defined as 
$\fJVT=\mathrm{max}_i(\fJVT_i)$, is above a threshold.
The choice of threshold determines the pile-up rejection performance.
The \fJVT discriminant tends to have larger values for 
QCD pile-up jets, while the distribution for hard-scatter jets falls steeply,
as shown in Figure~\ref{fig:fjvtbase}.

\begin{figure}[htbp]
\begin{center}
\subfloat[][]{\includegraphics[width=0.475\textwidth]{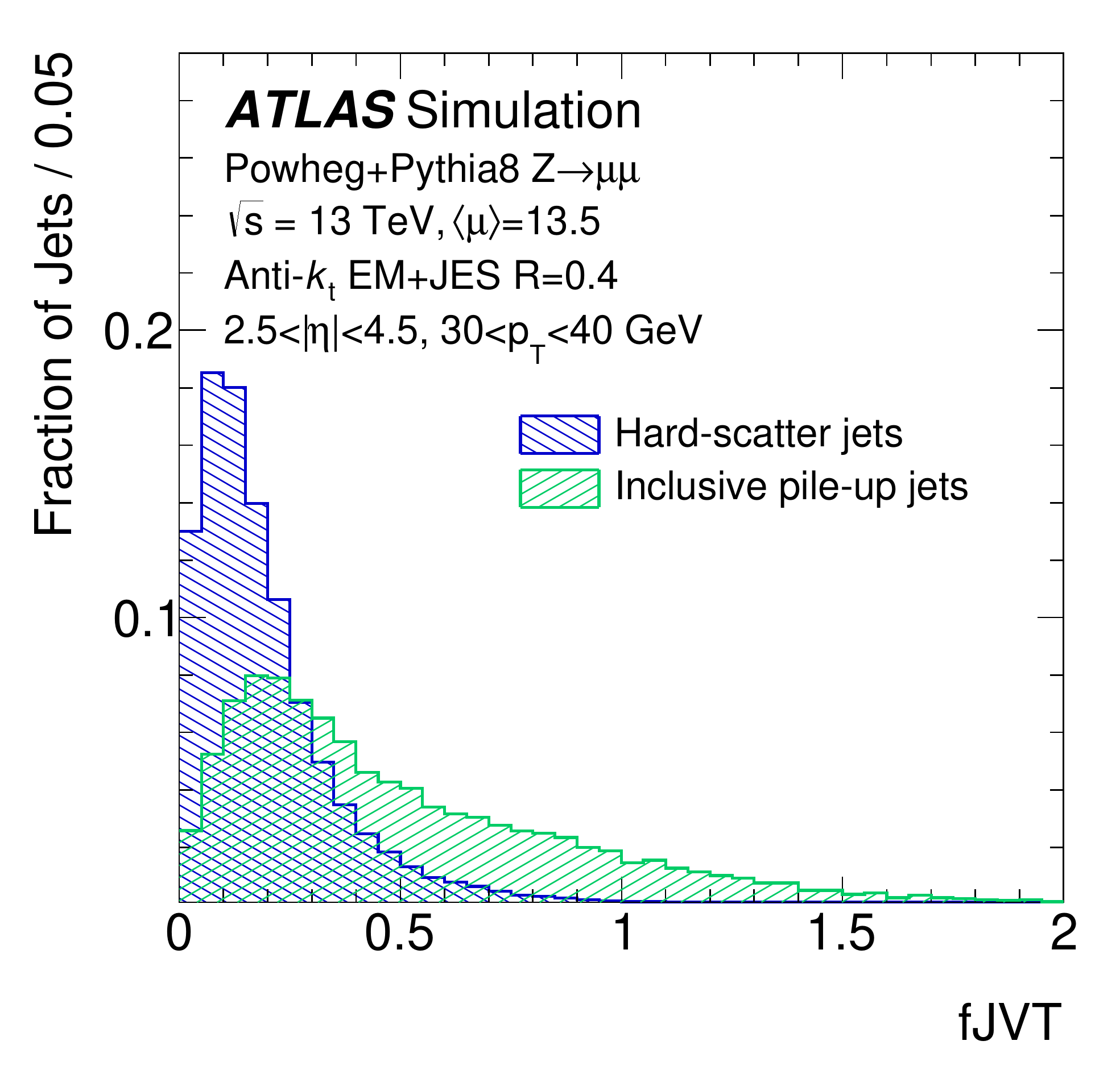}\label{fjvt30}}
\subfloat[][]{\includegraphics[width=0.475\textwidth]{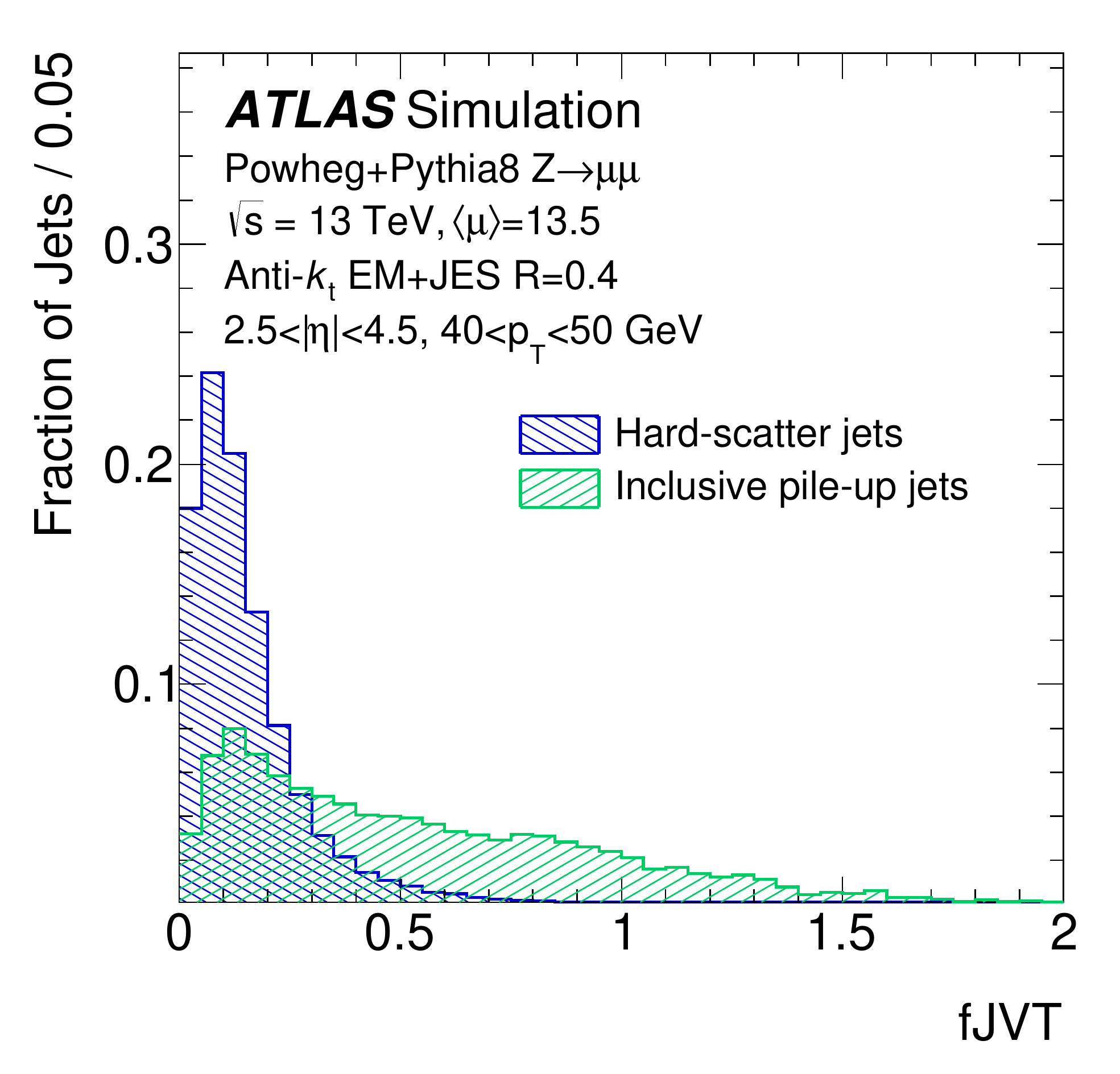}\label{fjvt40}}
\end{center}
\caption{The \fJVT distribution for hard-scatter (blue) and pile-up (green) forward jets 
  in simulated $\Zboson+$jets events with at least one forward jet with \protect\subref{fjvt30} $30<\pt<40$ \GeV 
  or \protect\subref{fjvt40} $40<\pt<50$ \GeV .}
\label{fig:fjvtbase}
\end{figure}

\subsection{Performance}

Figure \ref{fig:fJVTroc} shows the efficiency of selecting forward pile-up jets 
as a function of the efficiency of selecting forward hard-scatter 
jets when varying the maximum \fJVT requirement.

\begin{figure}[htbp]
\begin{center}
\includegraphics[width=0.475\textwidth]{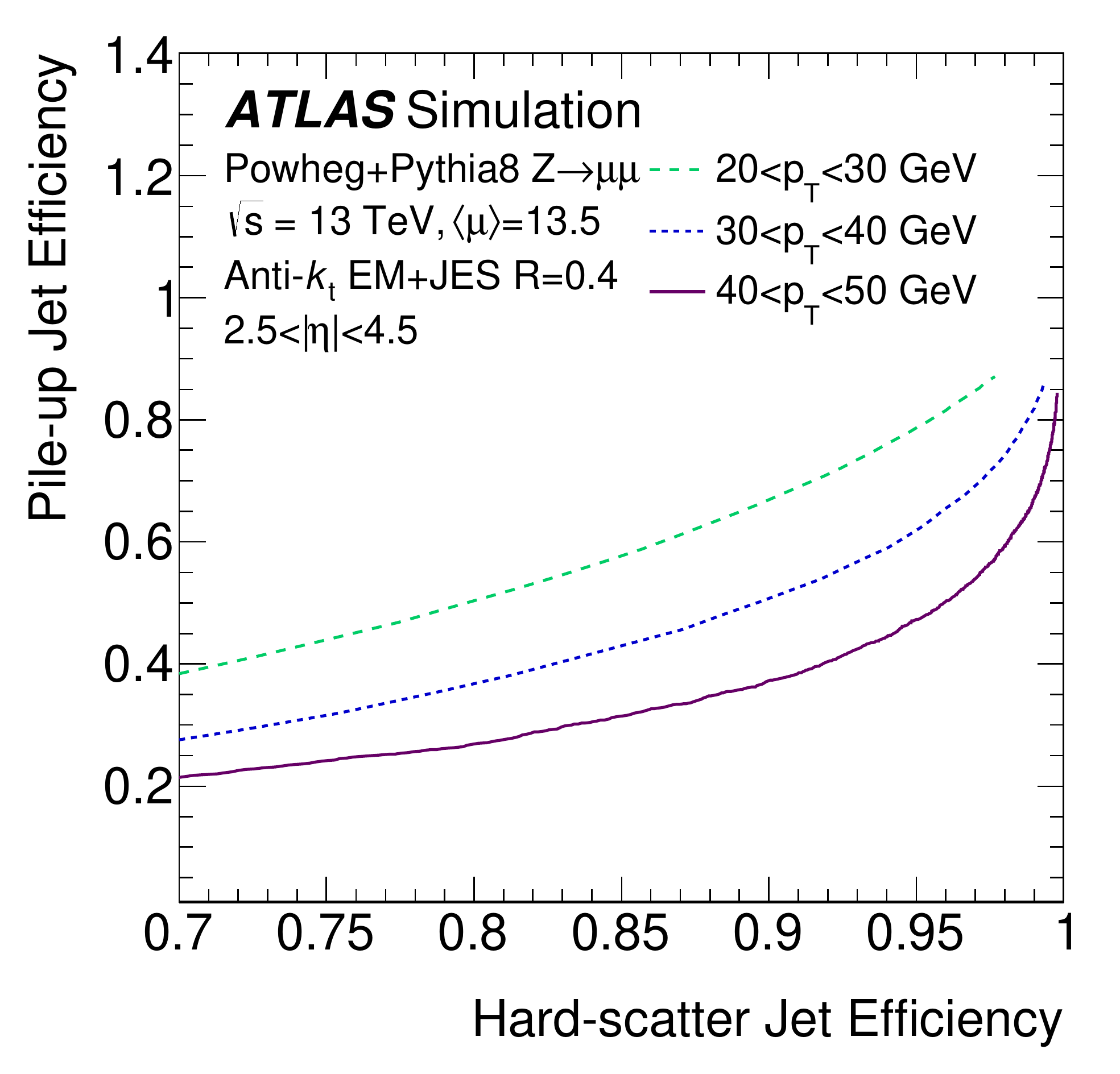}
\end{center}
\caption{
Efficiency for pile-up jets in simulated $\Zboson+$jets events as a function of the efficiency for hard-scatter jets for 
different jet \pt ranges.
}
\label{fig:fJVTroc}
\end{figure}

Using a maximum \fJVT of 0.5 and 0.4 respectively, hard-scatter efficiencies of 
92\% and 85\% are achieved for pile-up efficiencies of 60\% and 50\%, 
considering jets with $20<\pT<50\GeV$.
The dependence of the hard-scatter and pile-up efficiencies on the forward jet \pT is 
shown in Figure~\ref{fig:fJVTeff}.
For low-\pt forward jets, the probability of an upward fluctuation in the \fJVT value 
is more likely, and therefore the efficiency for hard-scatter jets is slightly lower
than for higher-\pt jets.
The hard-scatter efficiency depends on the number of pile-up interactions, 
as shown in Figure~\ref{fig:fJVTeffVsNPV}, as busier pile-up conditions
increase the chance of accidentally matching the hard-scatter jet to a pile-up vertex.
The pile-up efficiency depends on the \pT of the forward jets, due to the \pT-dependence
of the relative numbers of QCD and stochastic pile-up jets.

\begin{figure}[htbp]
\begin{center}
\subfloat[][]{\includegraphics[width=0.475\textwidth]{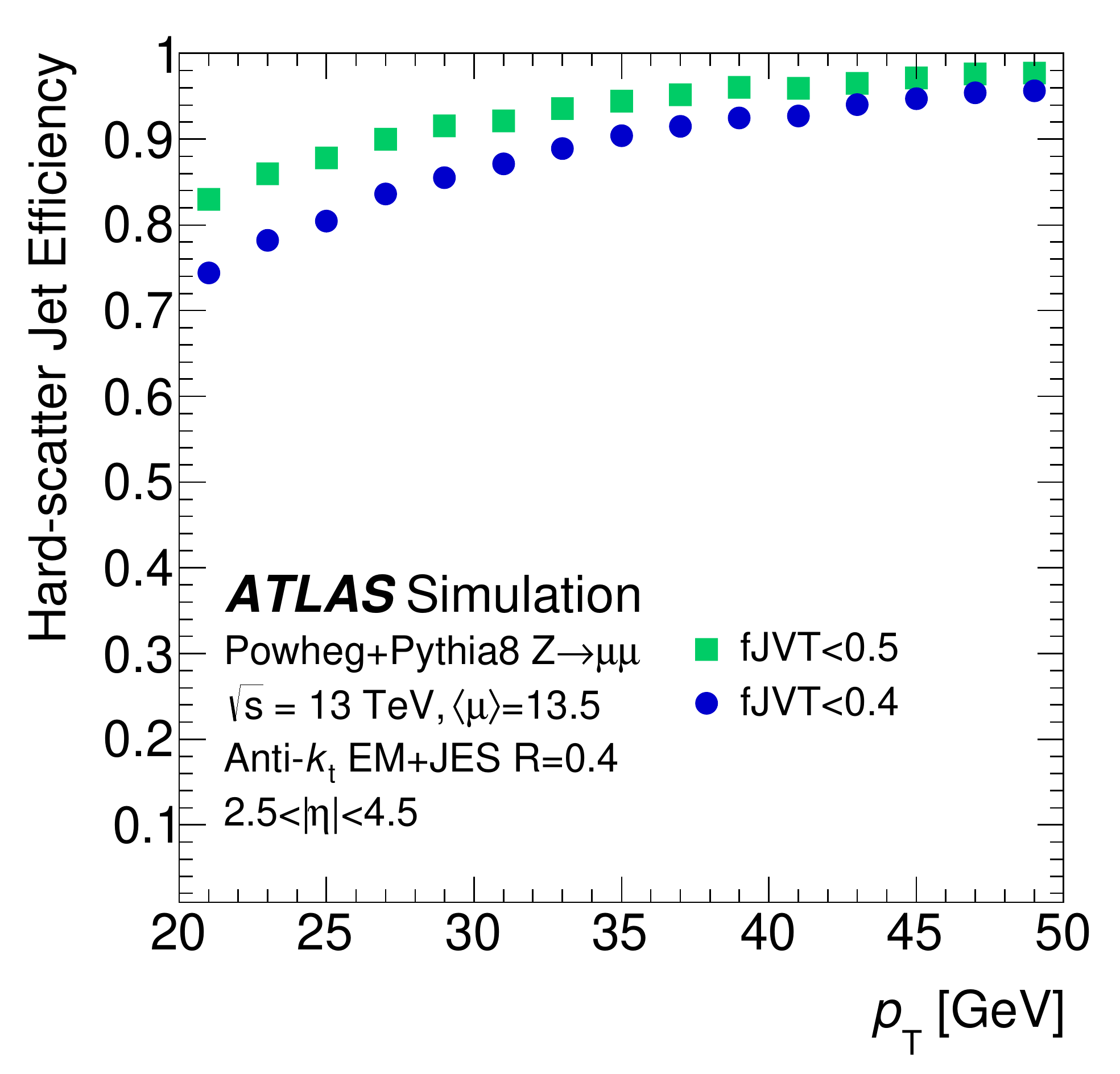}\label{hseff}}
\subfloat[][]{\includegraphics[width=0.475\textwidth]{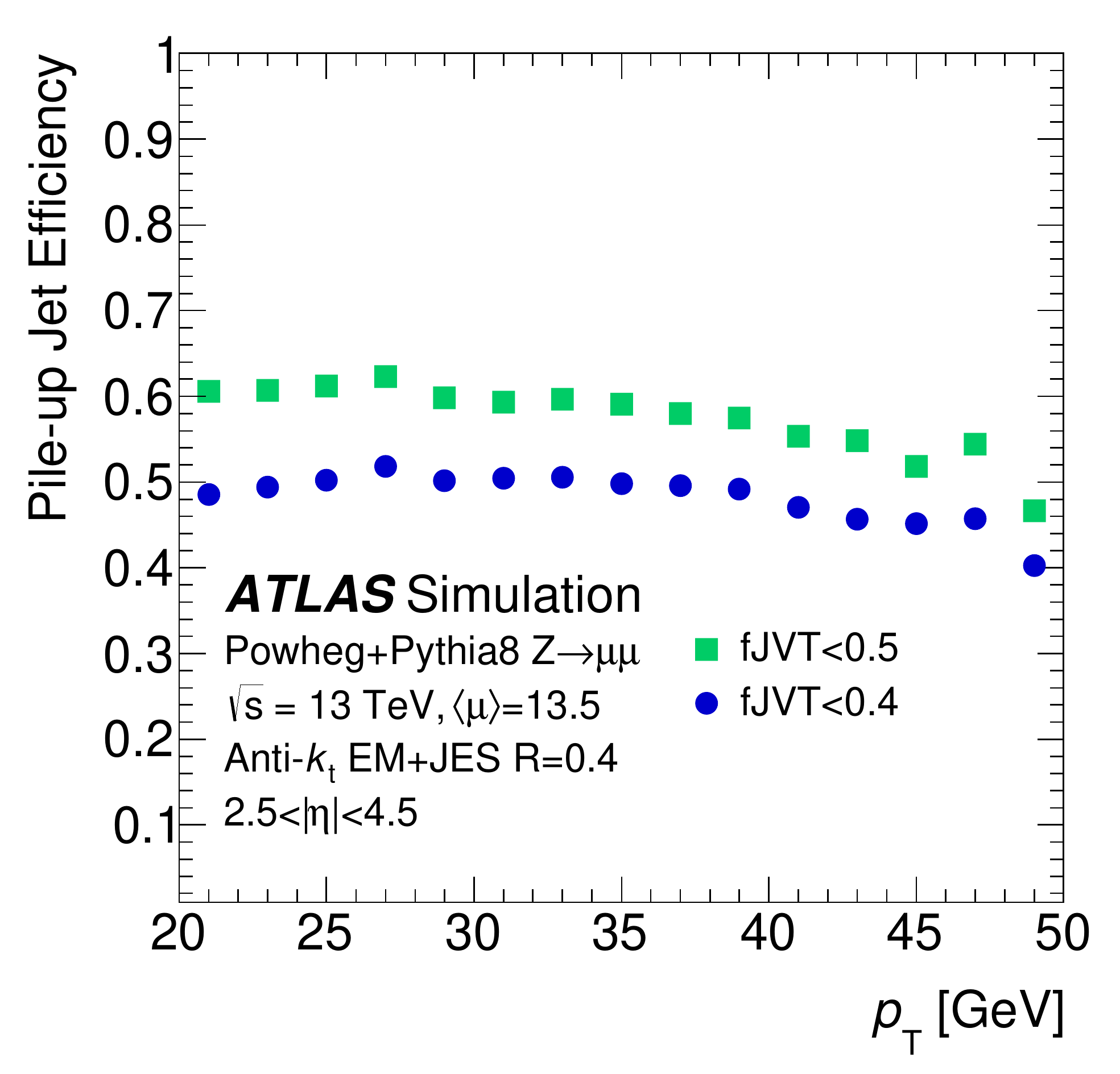}\label{pueff}}
\end{center}
\caption{
Efficiency for \protect\subref{hseff} hard-scatter jets and 
\protect\subref{pueff} pile-up jets as a function of the forward jet \pt in simulated $\Zboson+$jets events.}
\label{fig:fJVTeff}
\end{figure}

\begin{figure}[htbp]
\begin{center}
\subfloat[][]{\includegraphics[width=0.475\textwidth]{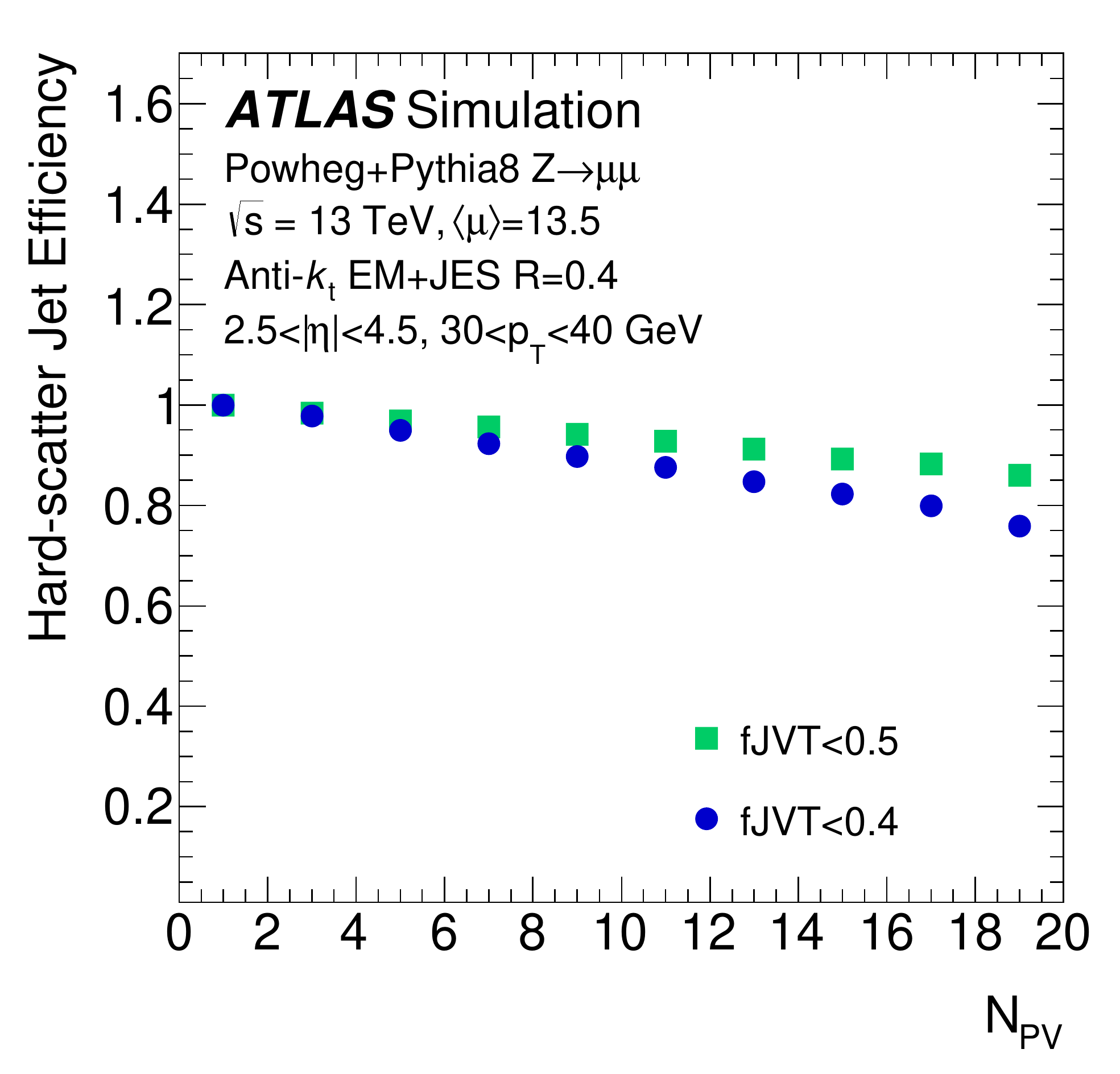}\label{hs30}}\quad
\subfloat[][]{\includegraphics[width=0.475\textwidth]{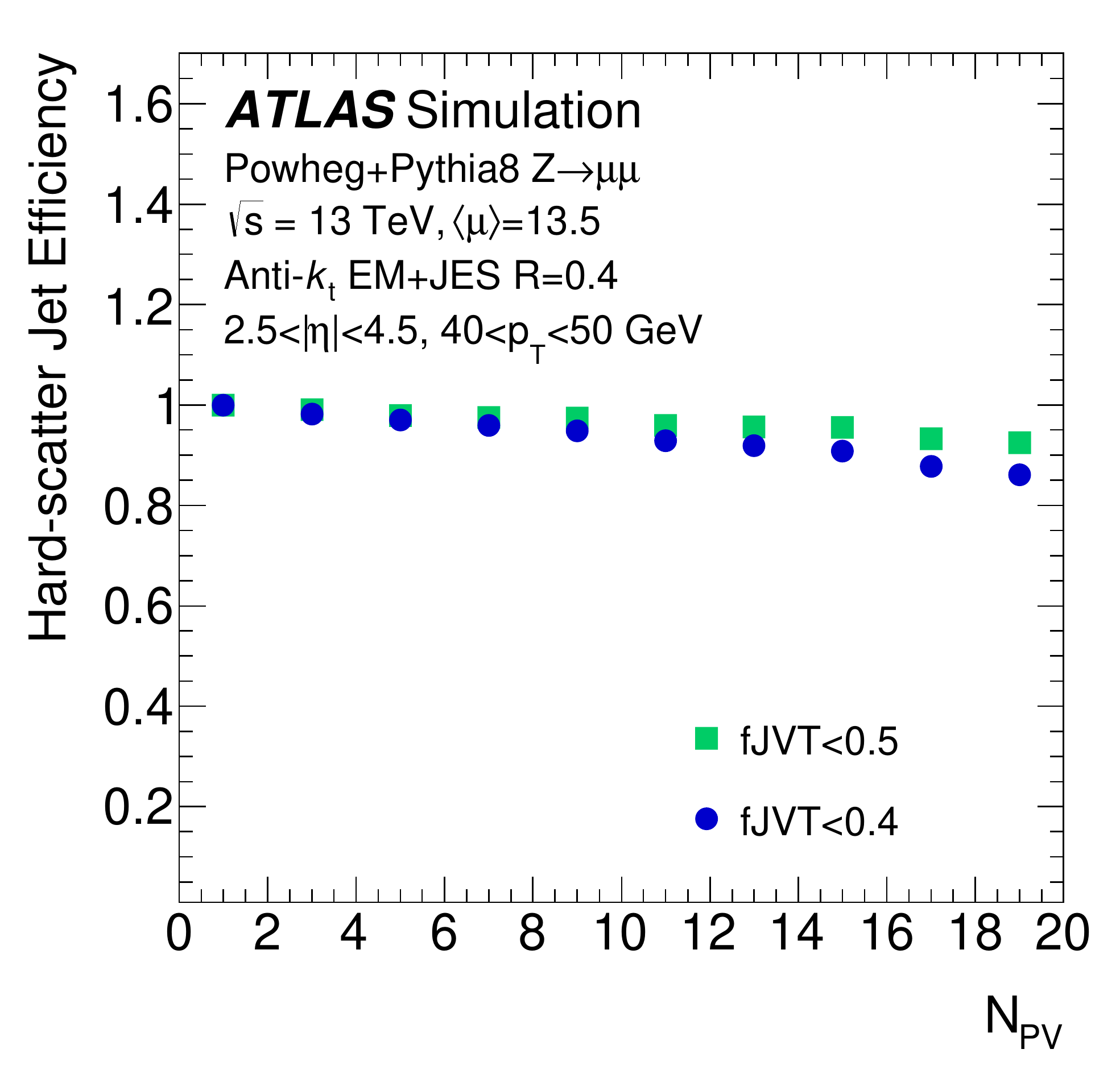}\label{hs40}}\\
\subfloat[][]{\includegraphics[width=0.475\textwidth]{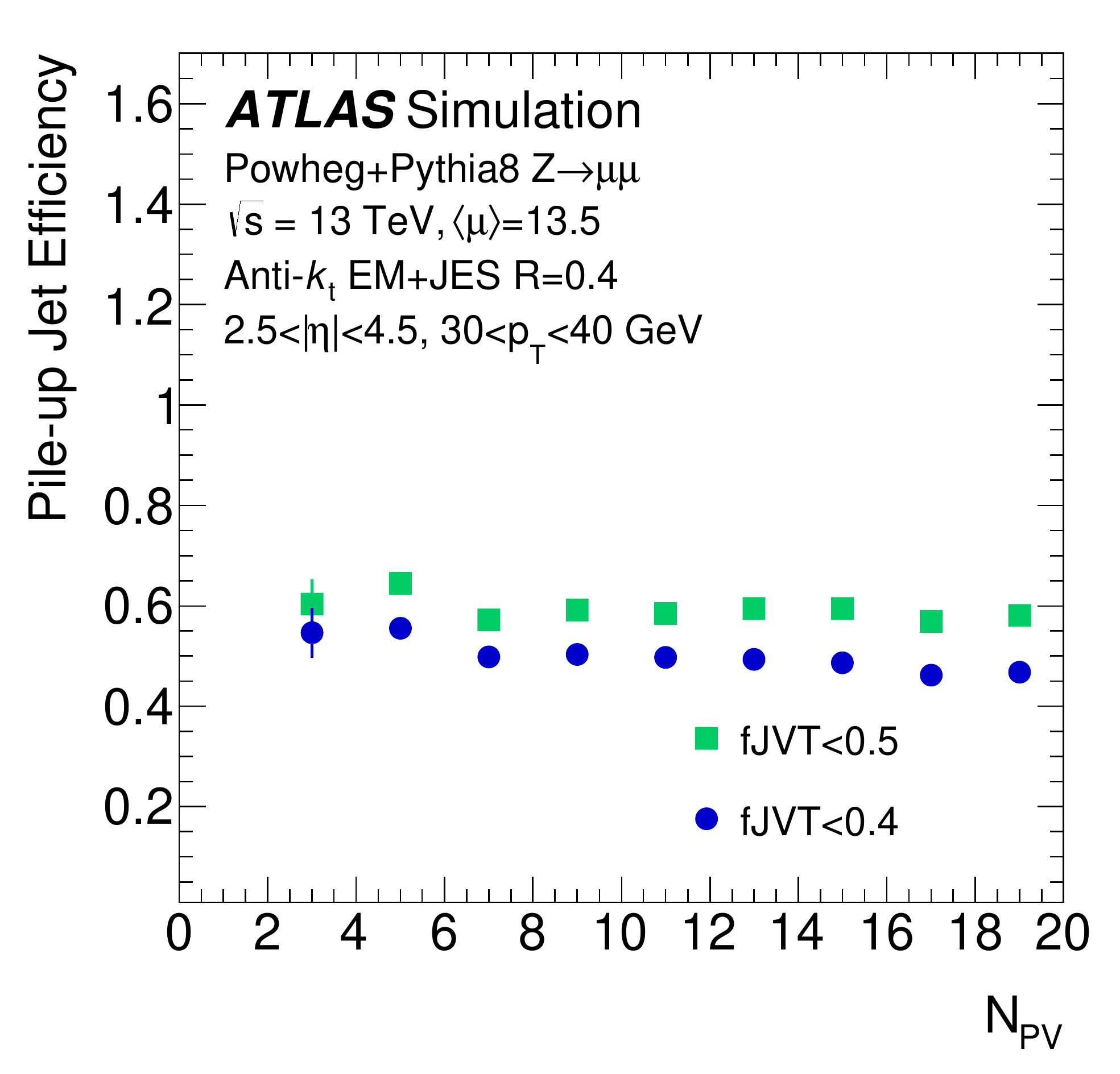}\label{pu30}}\quad
\subfloat[][]{\includegraphics[width=0.475\textwidth]{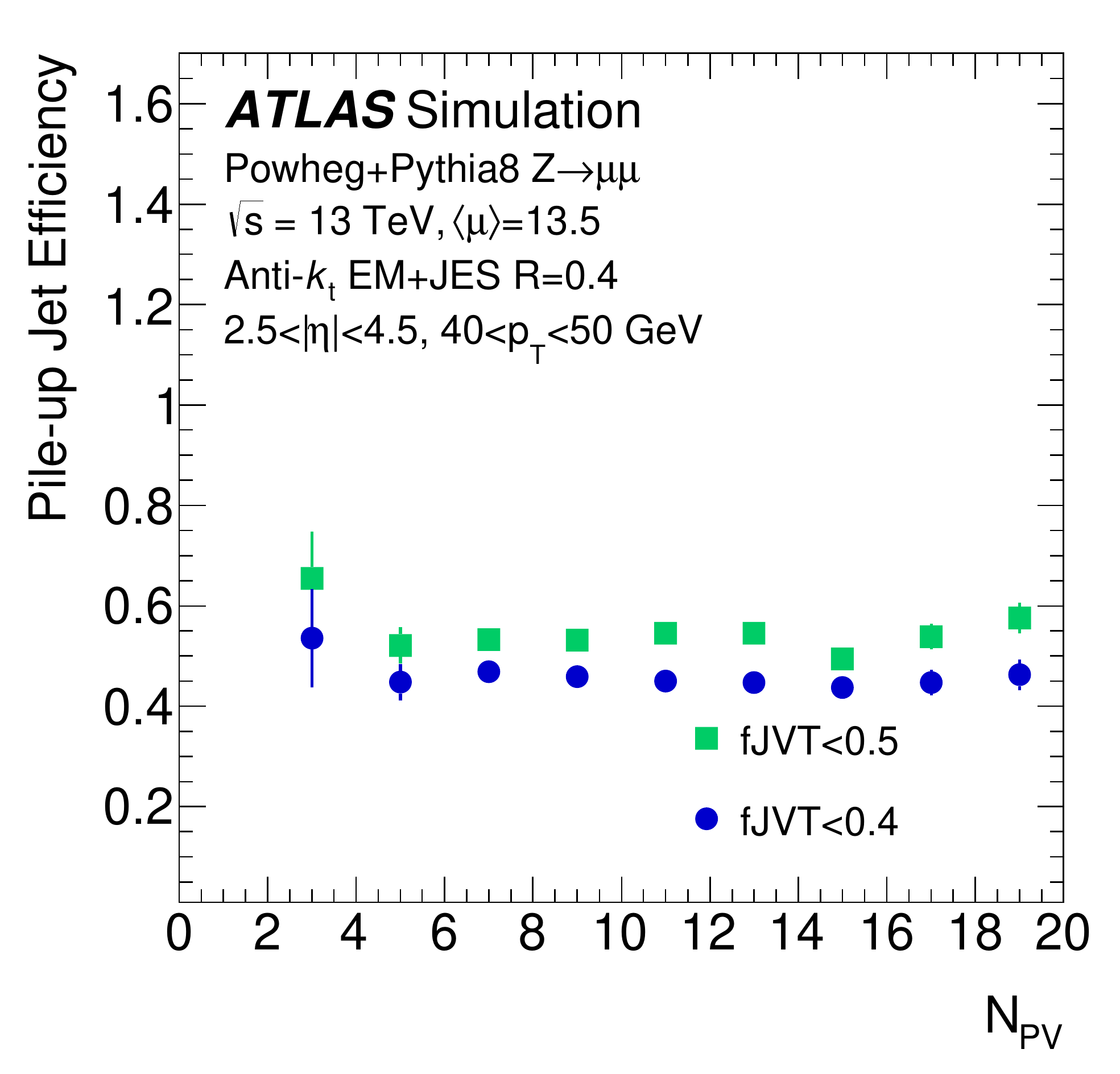}\label{pu40}}
\end{center}
\caption{
Efficiency in simulated $\Zboson+$jets events
as a function of \NPV for hard-scatter forward jets with \protect\subref{hs30} $30\GeV<\pT<40\GeV$ 
and \protect\subref{hs40} $40\GeV<\pT<50\GeV$, and for pile-up forward jets with 
\protect\subref{pu30} $30\GeV<\pT<40\GeV$ \protect\subref{pu40} $40\GeV<\pT<50\GeV$.}
\label{fig:fJVTeffVsNPV}
\end{figure}

\subsection{Efficiency measurements}

The \fJVT efficiency for hard-scatter jets is measured in $\Zboson+\mathrm{jets}$ data events,
exploiting a tag-and-probe procedure similar to that described in 
Ref.~\cite{Aad:2015ina}.

For $Z(\to\mu\mu)+$jets events, selected by single-muon triggers,
two muons of opposite sign and $\pT>25\GeV$ are required,
such that their invariant mass lies between 66 \GeV and 116 \GeV. 
Events are further required to satisfy event and jet quality criteria, 
and a veto on cosmic-ray muons.

Using the leading forward jet recoiling against the \Zboson boson as a probe, 
a signal region of forward hard-scatter jets is defined as the back-to-back region specified by 
$|\Delta \phi (\Zboson, \mathrm{jet})| > 2.8$ rad. 
In order to select a sample pure in forward hard-scatter jets, 
events are required to have no central hard-scatter jets with $\pt>20 \GeV$, identified with \JVT, 
and exactly one forward jet.
The \Zboson boson is required to have $\pT > 20 \GeV$, as events in which the \Zboson boson
has \pT less than the minimum defined jet \pT have a lower hard-scatter purity.
The above selection results in a forward hard-scatter signal region that is greater than  
98\% pure in hard-scatter jets relative to pile-up jets, as estimated in simulation.

The \fJVT distributions for data and simulation in the signal region are compared in Figure 
\ref{fig:basicComparison}.
The data distribution is observed to have fewer jets with high \fJVT than predicted by simulation, 
consistent with an overestimation of the number of pile-up jets, as reported in 
Ref.~\cite{Aad:2015ina}.

\begin{figure}[htbp]
\begin{center}
\subfloat[][]{\includegraphics[width=0.475\textwidth]{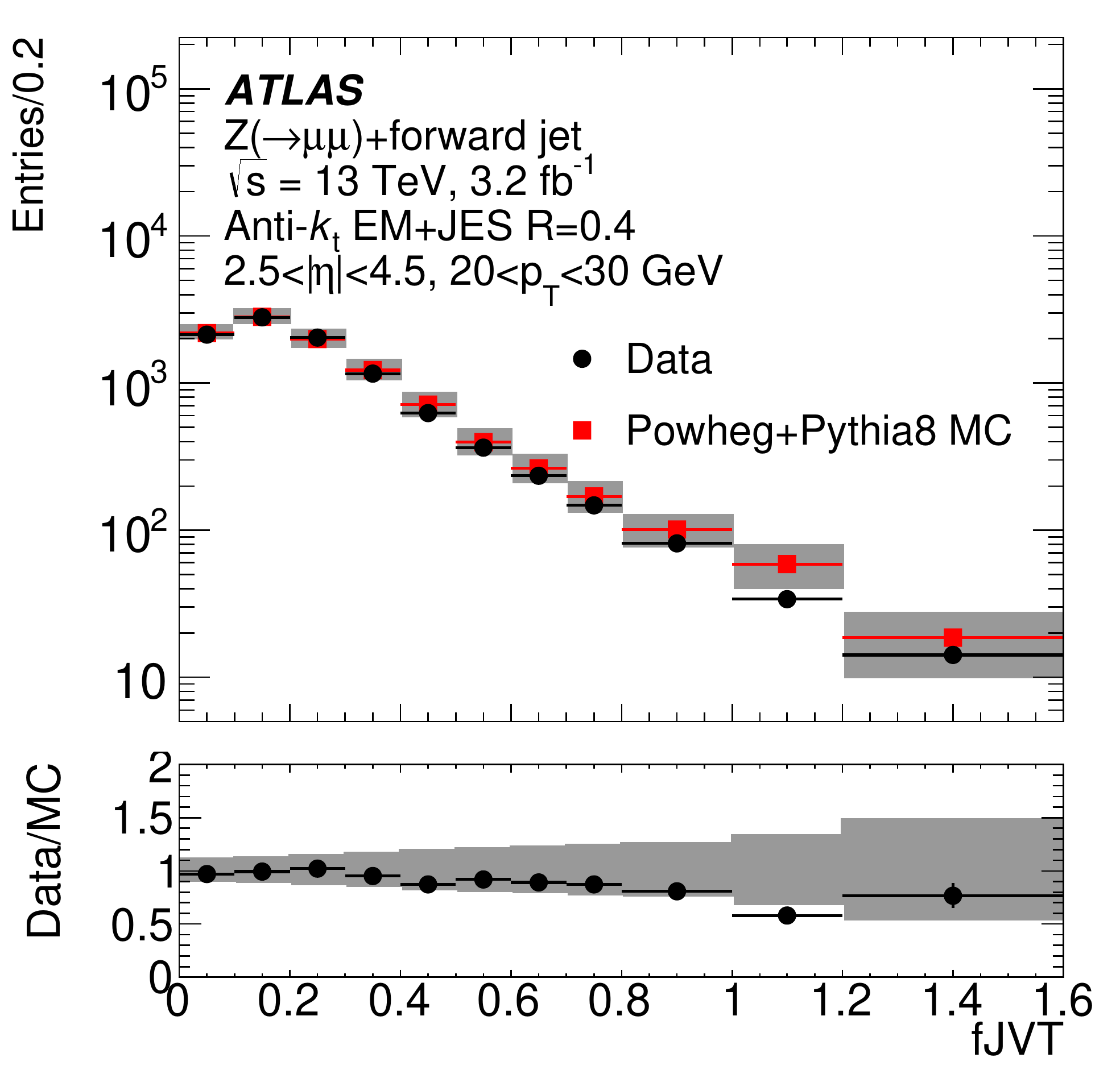}\label{lowpt}}
\subfloat[][]{\includegraphics[width=0.475\textwidth]{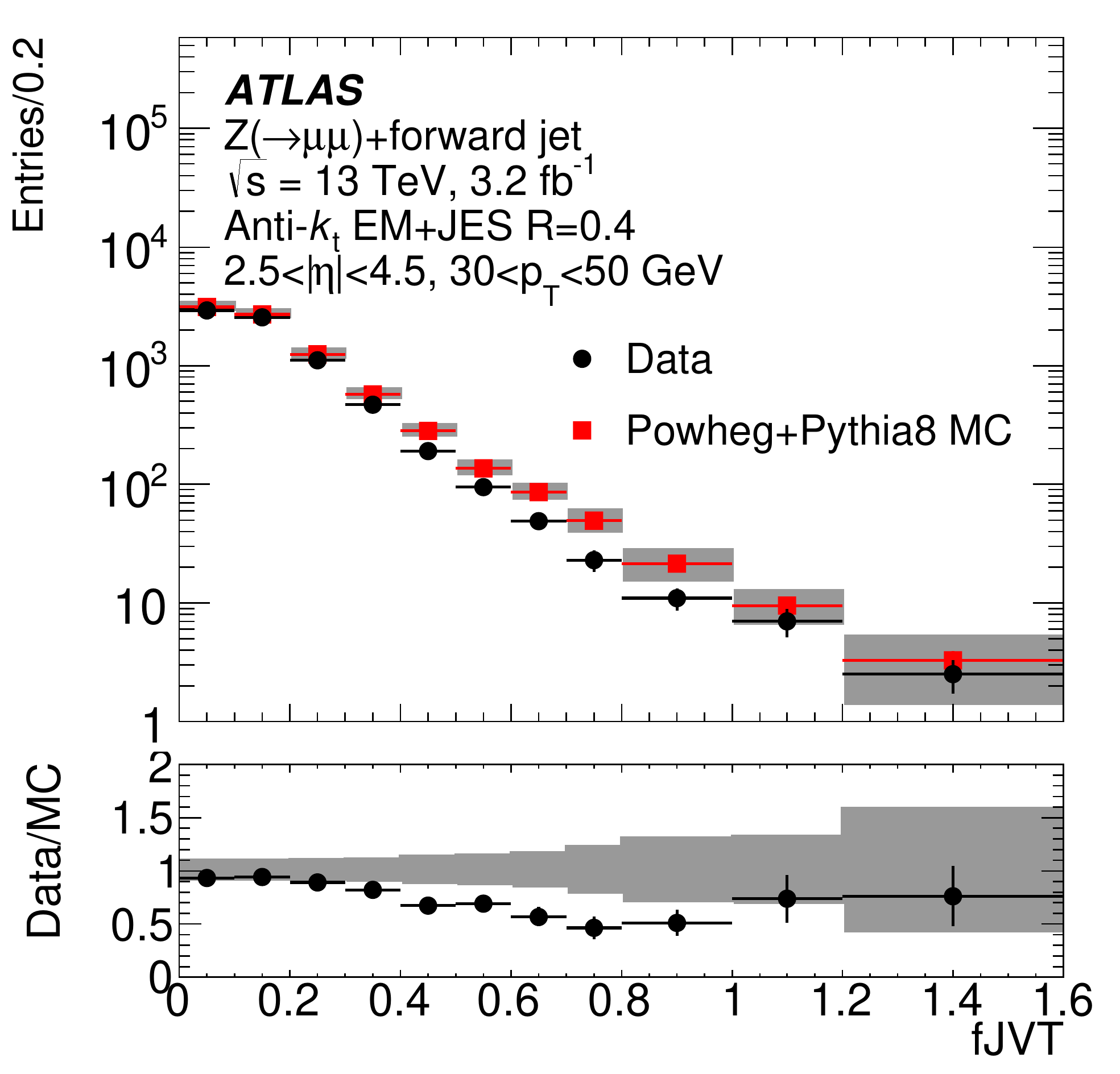}\label{highpt}}
\end{center}
\caption{Distributions of \fJVT for jets with \pT \protect\subref{lowpt} between 20 and 30 \GeV  
and \protect\subref{highpt} between 30 and 50 \GeV for data (black circles) and simulation (red squares).
The lower panels display the ratio of the data to the simulation.
The grey bands account for statistical and systematic uncertainties.}
\label{fig:basicComparison}
\end{figure}

The pile-up jet contamination in the signal region $N_{\mathrm{PU}}^\mathrm{signal}(|\dphiZjet|>2.8~\mathrm{rad})$ is estimated 
in a pile-up-enriched 
control region with $|\dphiZjet|<1.2$ rad, based on the assumption that the $|\dphiZjet|$ 
distribution is uniform for pile-up jets.
The validity of such assumption was verified in simulation.
The pile-up jet rate in data is therefore used to estimate 
the contamination of the signal region as
\begin{multline}
N_{\mathrm{PU}}^\mathrm{signal}(|\dphiZjet|>2.8~\mathrm{rad}) = \\
[N_\mathrm{j}^\mathrm{control}(|\dphiZjet|<1.2~\mathrm{rad}) - N_\mathrm{HS}(|\dphiZjet|<1.2~\mathrm{rad})] \cdot (\pi - 2.8~\mathrm{rad})/1.2~\mathrm{rad},
\end{multline}
where ${N_\mathrm{j}^\mathrm{control}(|\dphiZjet|<1.2~\mathrm{rad})}$ is the number of jets in the data
control region and ${N_\mathrm{HS}(|\dphiZjet|<1.2~\mathrm{rad})}$ is the expected number of hard-scatter jets in the control region,
as predicted in simulation.

The hard-scatter efficiency is therefore measured in the signal region as
\begin{eqnarray}
    \varepsilon = \frac{N_\mathrm{j}^\mathrm{pass} - N_\mathrm{PU}^\mathrm{pass}}{N_\mathrm{j}^\mathrm{signal} - N_{\mathrm{PU}}^\mathrm{signal}},
\end{eqnarray}
where $N_\mathrm{j}^\mathrm{signal}$ and $N_\mathrm{j}^\mathrm{pass}$ denote respectively the overall number 
of jets in the signal region and the number of jets in the signal region satisfying 
the \fJVT requirements. The terms $N_\mathrm{PU}^\mathrm{pass}$ and $N_\mathrm{PU}^\mathrm{signal}$ represent the overall number of pile-up jets in the signal region
and the number of pile-up jets satisfying the \fJVT requirements, respectively, and are both estimated from simulation.
Figure \ref{fig:efficiency} shows the hard-scatter efficiency evaluated in data and 
simulation.
The uncertainties correspond to a 30\% uncertainty in the number of pile-up jets and a 10\% 
uncertainty in the number of hard-scatter jets in the signal region. The uncertainties are
estimated by comparing data and simulation in the pile-up- and hard-scatter-enriched
regions, respectively.
The hard-scatter efficiency is found to be underestimated in simulation.
The level of disagreement is observed to be larger at low jet \pt and high $|\eta|$ and can be as large as about 3\%.
The efficiencies evaluated in this paper are used to define a calibration procedure accounting for this discrepancy.
The uncertainties associated with the calibration and resolution of the jets used to compute \fJVT are estimated in ATLAS analyses
by recomputing \fJVT for each variation reflecting a systematic uncertainty.

\begin{figure}[htbp]
\begin{center}
\subfloat[][]{\includegraphics[width=0.475\textwidth]{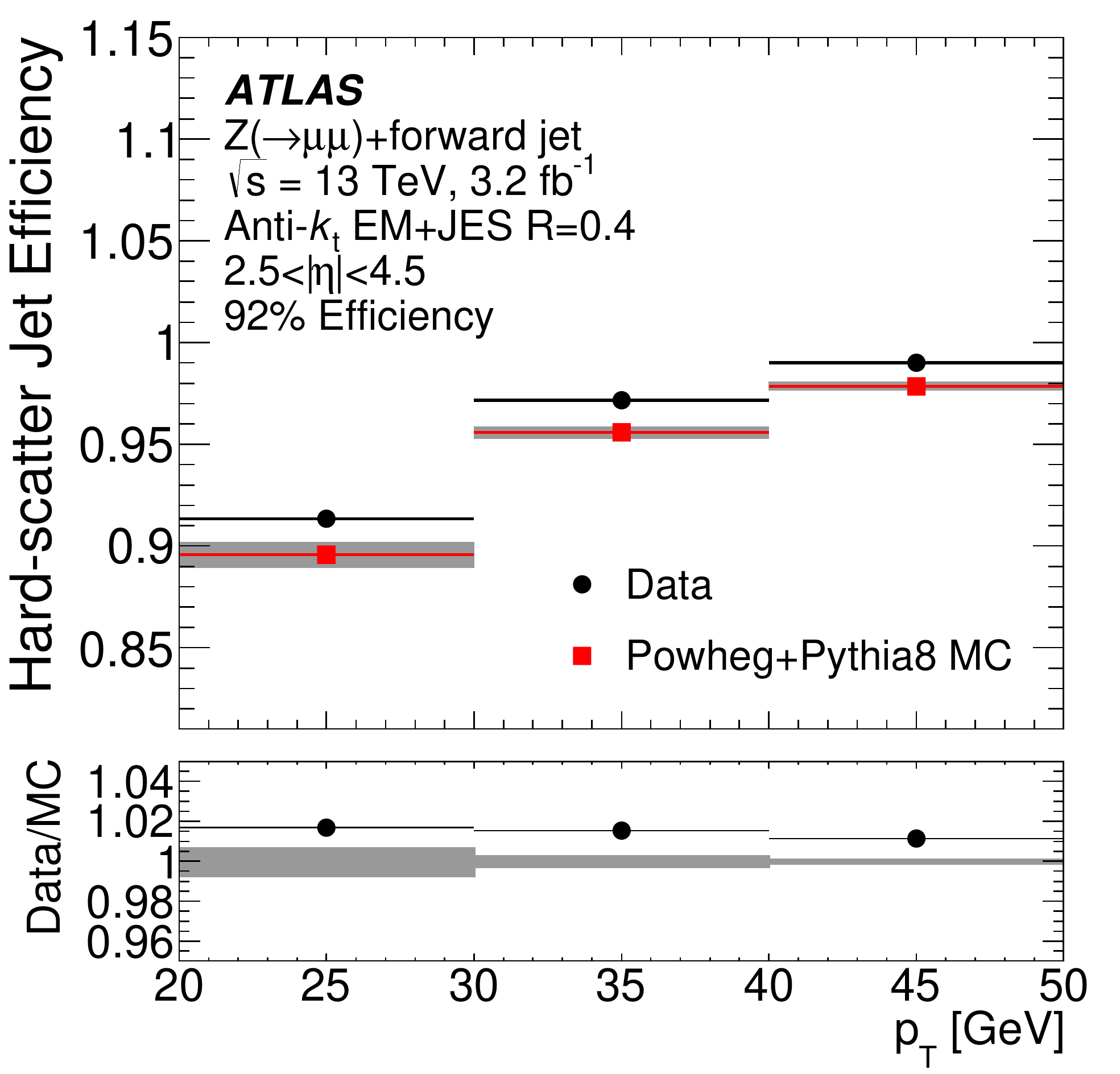}\label{effptloose}}\quad
\subfloat[][]{\includegraphics[width=0.475\textwidth]{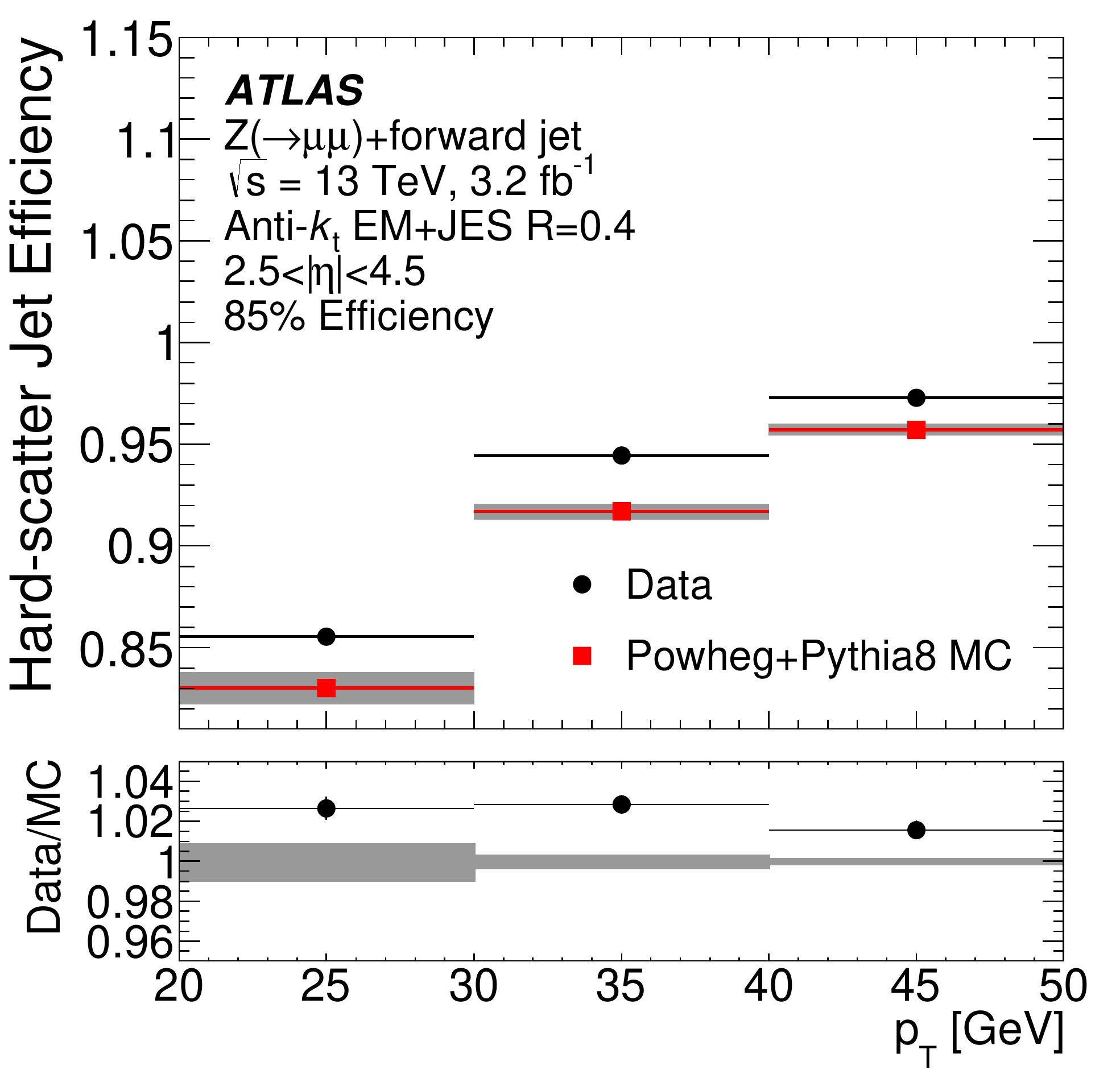}\label{effpttight}}\\
\subfloat[][]{\includegraphics[width=0.475\textwidth]{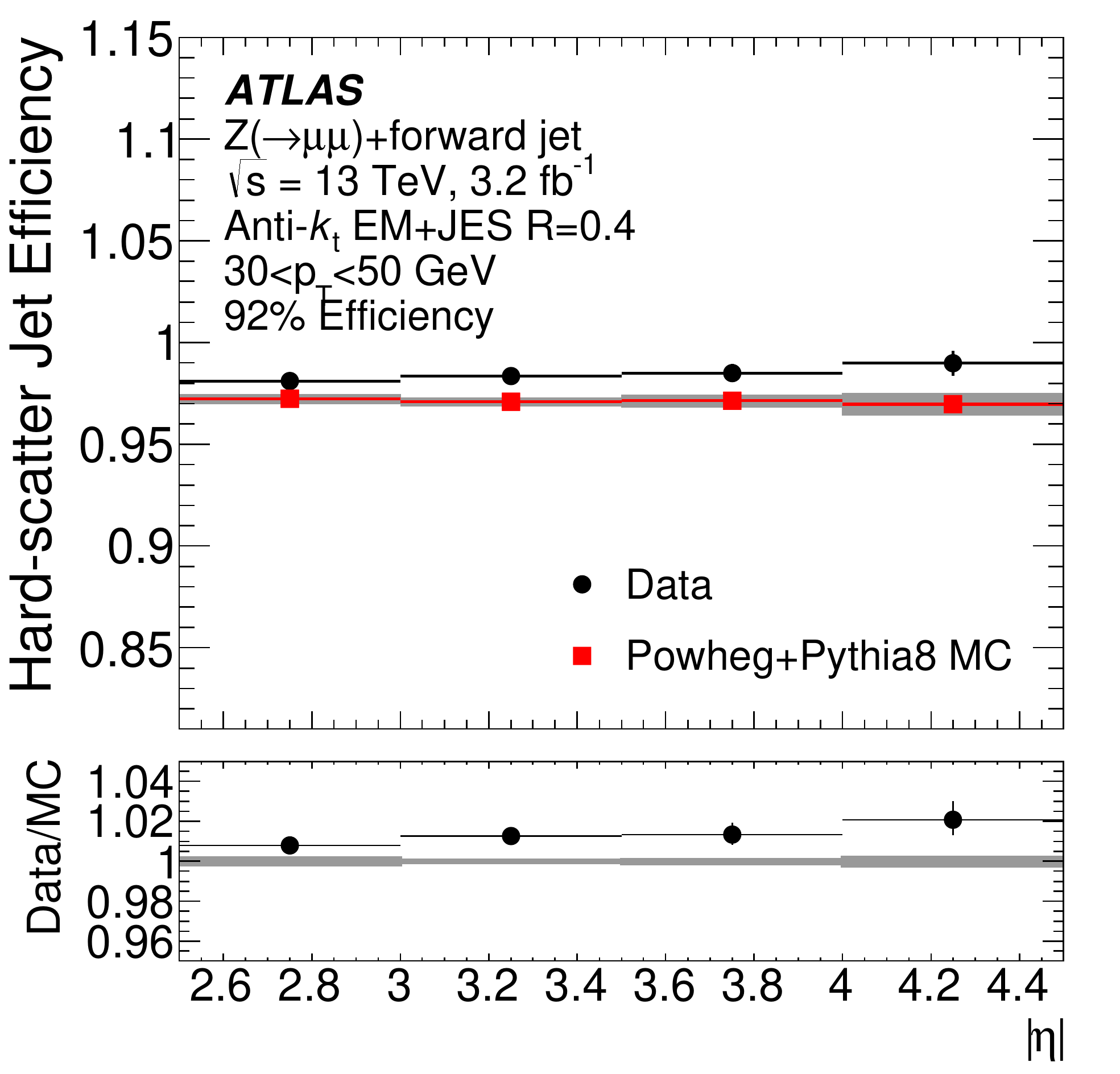}\label{effetaloose}}\quad
\subfloat[][]{\includegraphics[width=0.475\textwidth]{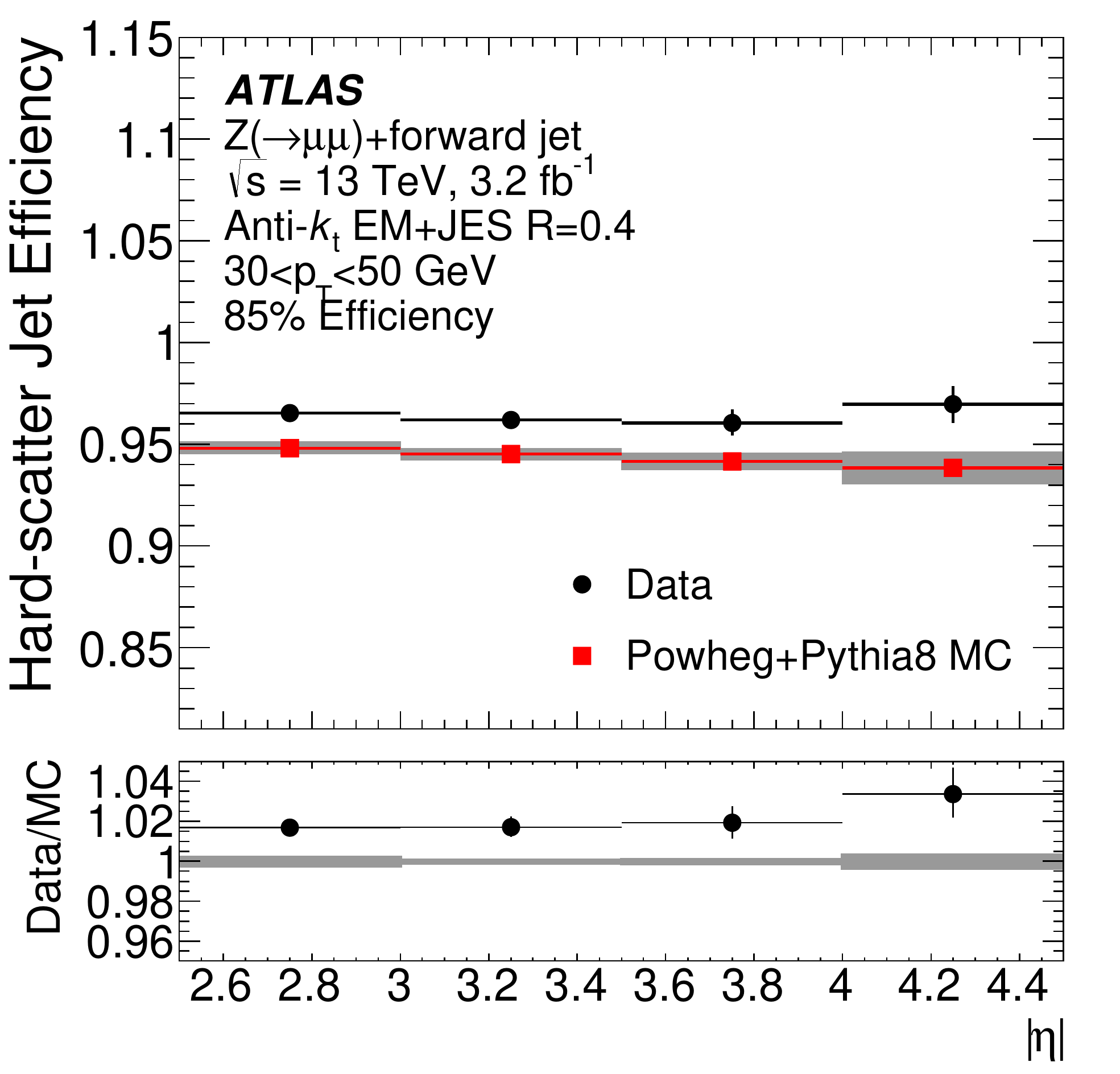}\label{effetatight}}
\end{center}
\caption{Efficiency for hard-scatter jets to pass \fJVT requirements as a function 
of (\protect\subref{effptloose} and \protect\subref{effpttight}) \pT  
and (\protect\subref{effetaloose} and \protect\subref{effetatight}) $|\eta|$  
for the (\protect\subref{effptloose} and \protect\subref{effetaloose}) 92\%
and (\protect\subref{effpttight} and \protect\subref{effetatight}) 85\%
hard-scatter efficiency operating points of the \fJVT discriminant 
in data (black circles) and simulation (red squares).
The lower panels display the ratio of the data to the simulation.
The grey bands account for statistical and systematic
uncertainties.}
\label{fig:efficiency}
\end{figure}

\section{Pile-up jet tagging with shape and topological information}
\label{sec:comb}

The \fJVT and $\gamma$ discriminants correspond to a twofold strategy for pile-up rejection
targeting QCD and stochastic pile-up jets, respectively.
However, as highlighted in Section~\ref{sec:pileup}, this classification is not well defined as
all jets have a stochastic component. Therefore, it is useful to define a coherent strategy
that addresses both the stochastic and QCD nature of pile-up jets at the same time.

The $\gamma$ parameter discussed in Section~\ref{sec:width} provides an estimate of the
\pt in the core of the jet originating from the single interaction contributing 
the largest amount of transverse momentum to the jet.
Therefore, the \fJVT definition can be modified to exploit this estimation by replacing
the jet \pt with $\gamma$, so that
\begin{equation}
\fJVT_{\gamma} = \frac{\langle\boldsymbol{p}_{\mathrm{T},i}^\mathrm{miss}\rangle\cdot\boldsymbol{u}^\mathrm{fj}}{\gamma},
\end{equation}
where $\boldsymbol{u}^\mathrm{fj}$ is the unit vector representing the direction of the forward jet in the 
transverse plane.

Figure \ref{fig:roc_combo_pt} shows the performance of $\fJVT_{\gamma}$ compared with \fJVT
and $\gamma$ independently. The $\fJVT_{\gamma}$ discriminant outperforms the individual 
discriminants over the whole efficiency range. In samples enriched in QCD pile-up jets 
($30<\pt < 50$~\GeV), 
the $\fJVT_{\gamma}$ performance is driven by the topology information, while $\fJVT_{\gamma}$ 
benefits from the shape information for rejecting stochastic pile-up jets.
A multivariate combination of \fJVT and $\gamma$ discriminants was also studied and
found to be similar in performance to $\fJVT_{\gamma}$.

\begin{figure}[htbp]
\begin{center}
\subfloat[][]{\includegraphics[width=0.475\textwidth]{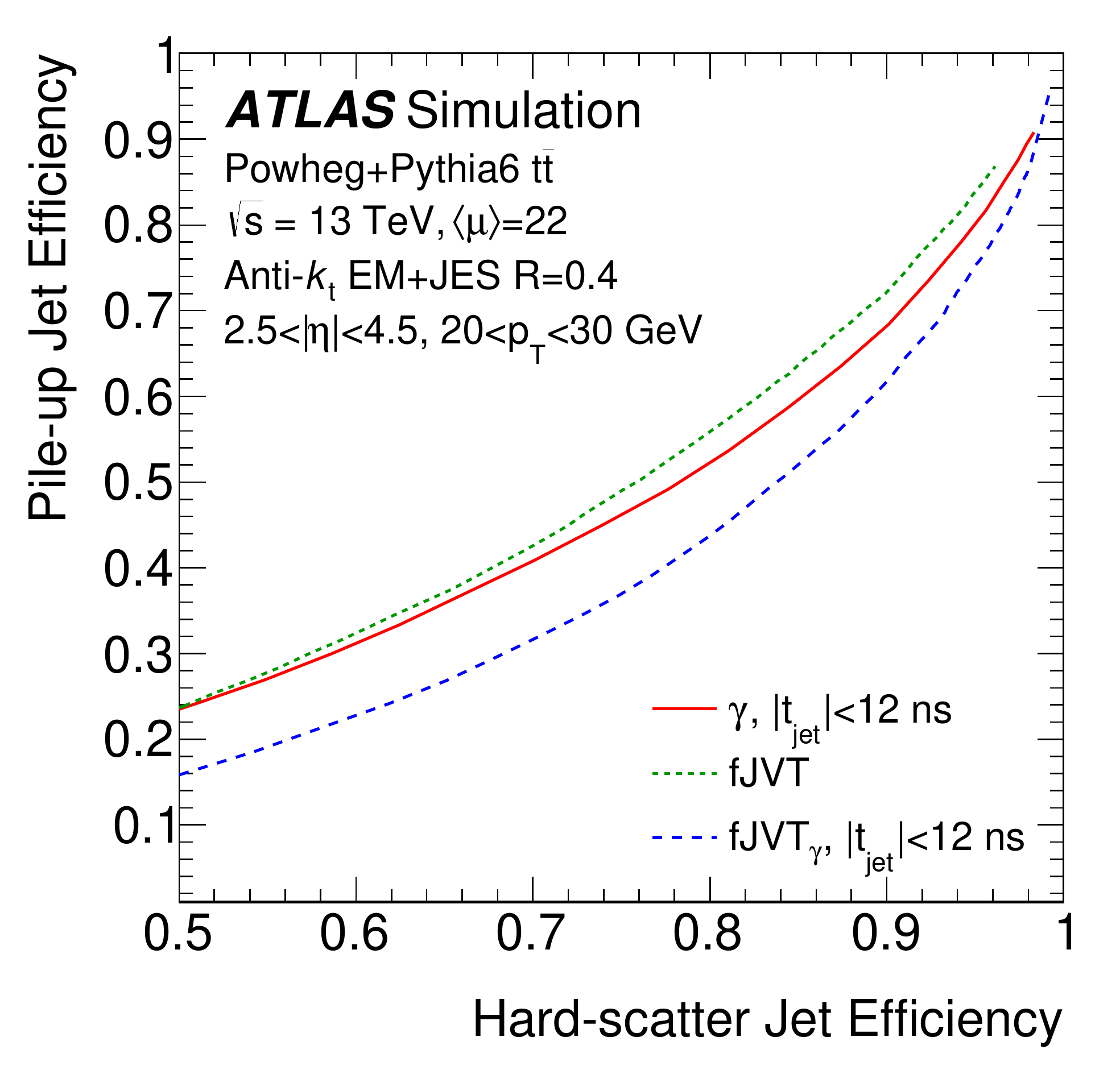}\label{roc2030}}\quad
\subfloat[][]{\includegraphics[width=0.475\textwidth]{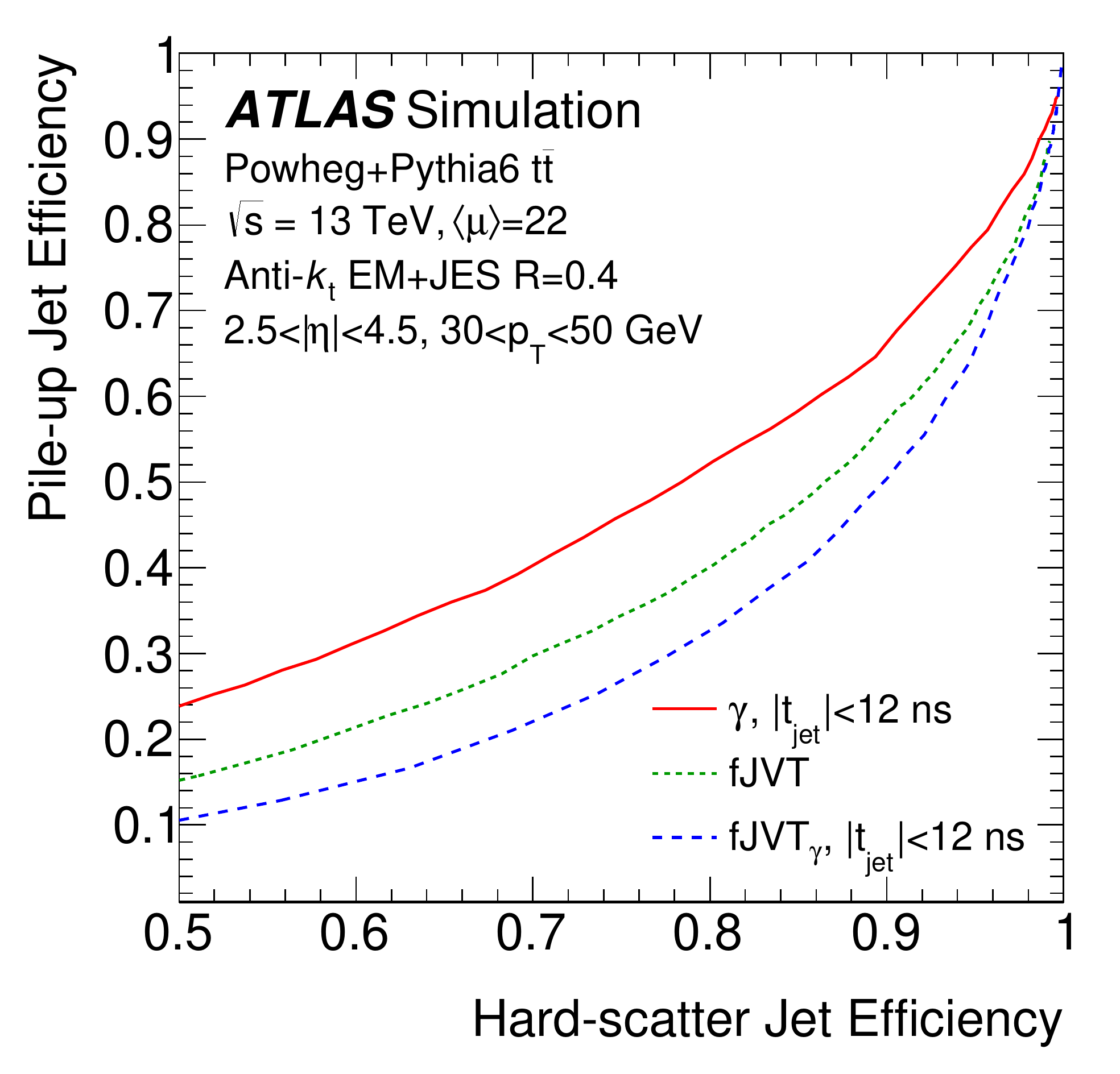}\label{roc3050}}\\
\end{center}
\caption{Efficiency for selecting pile-up jets as a function of the efficiency for 
selecting hard-scatter jets in simulated \ttbar events for \protect\subref{roc2030} jets with $20 \GeV<\pt<30 \GeV$
and \protect\subref{roc3050} jets with $30\GeV < \pt<50 \GeV$.}  
\label{fig:roc_combo_pt}
\end{figure}

\section{Impact on physics of Vector-Boson Fusion}
\label{sec:vbf}

In order to quantify the impact of forward pile-up rejection on a VBF analysis,
the VBF $H\to\tau\tau$ signature is considered, in the case where the $\tau$ decays leptonically.
The pile-up dependence of the signal purity (S/B) is studied in a simplified analysis in the
dilepton channel. Several other channels are used in the analysis of VBF $H\to\tau\tau$ by ATLAS; 
the dilepton channel is chosen for this study by virtue of its simple selection and background composition.
The dominant background in this channel originates from $\Zboson+$jets production,
where the $\Zboson$ boson decays leptonically, either to electrons, muons, or a leptonically decaying $\tau\tau$ pair.
The rate of $\Zboson$ bosons produced in association with two jets satisfying the requirements
targeting the VBF topology is extremely low.
The requirements include large $\Delta\eta$ between the jets and large dijet invariant mass $m_\mathrm{jj}$.
However, background events with forward pile-up jets often have large $\Delta\eta$ and $m_\mathrm{jj}$,
mimicking the VBF topology.
As a consequence, the background acceptance grows almost quadratically with the number of pile-up interactions.
This section illustrates the mitigation of this effect that can be achieved with the pile-up rejection provided 
by $\fJVT_{\gamma}$.

The event selection used for this study was optimized using simulation without pile-up~~\cite{cscbook}:
\begin{itemize}\itemsep0pt
\item The event must contain exactly two opposite-charge same-flavour leptons $\ell^+\ell^-$ (with $\ell=e$,$\mu$) with \pT>15\GeV;
\item The invariant mass of the lepton pair must satisfy $m_{\ell^+\ell^-}<66\GeV$ or $m_{\ell^+\ell^-}>116\GeV$;
\item The magnitude of the missing transverse momentum must be larger than $40\GeV$;
\item The event must contain two jets with $\pT>20\GeV$, one of which has $\pT>40\GeV$. The absolute difference in rapidities $|\eta_{\mathrm{j}_1}-\eta_{\mathrm{j}_2}|$ must exceed 4.4 and the invariant mass of the two jets must exceed 700\GeV.
\item For simulated VBF $H\to\tau\tau$ only, both jets are required to be truth-labelled as hard-scatter jets.
\end{itemize}

The impact of pile-up mitigation is emulated by randomly removing hard-scatter and
pile-up jets to match the performance of a $\fJVT_{\gamma}$ requirement with 85\% overall 
efficiency for hard-scatter jets with $20 < \pt < 50~\GeV$, as estimated in \ttbar simulation with an average $\langle\mu\rangle$ of 13.5.
The efficiencies are estimated as a function of the jet \pt and the average number of interactions per bunch crossing.
\begin{figure}[htbp]
\begin{center}
\subfloat[][]{\includegraphics[width=0.475\textwidth]{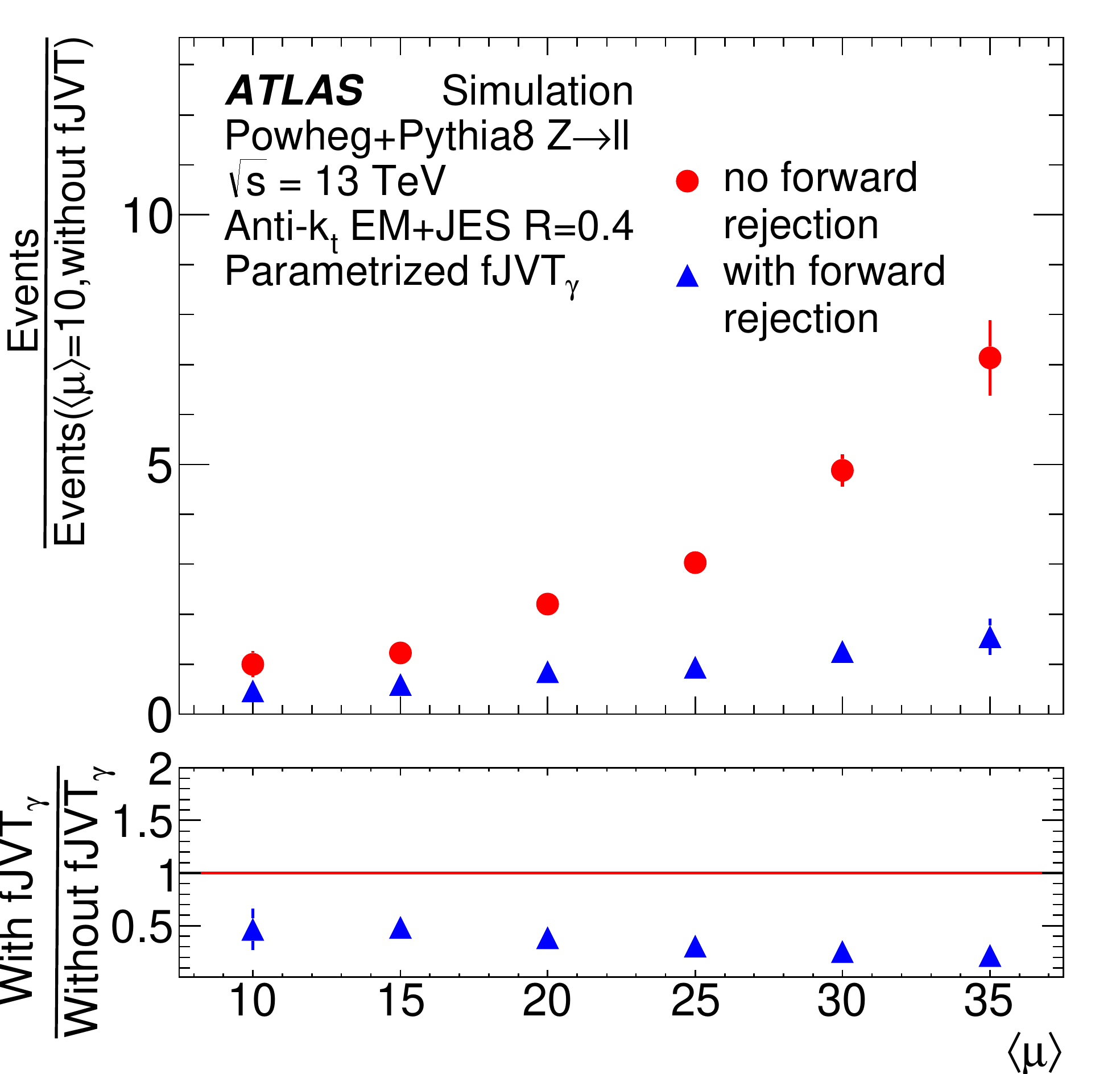}\label{vbfb}}
\subfloat[][]{\includegraphics[width=0.475\textwidth]{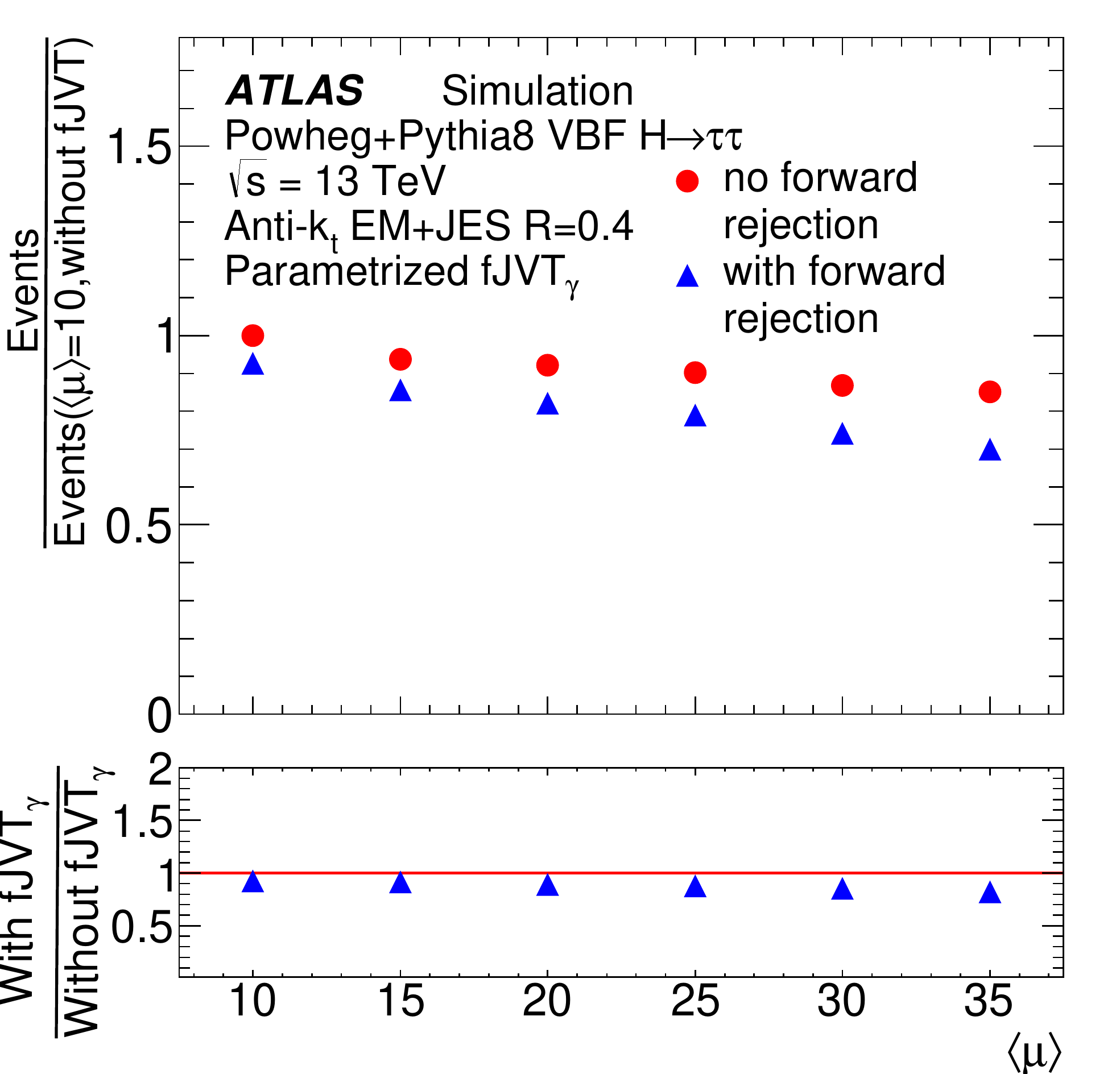}\label{vbfs}}\\
\subfloat[][]{\includegraphics[width=0.475\textwidth]{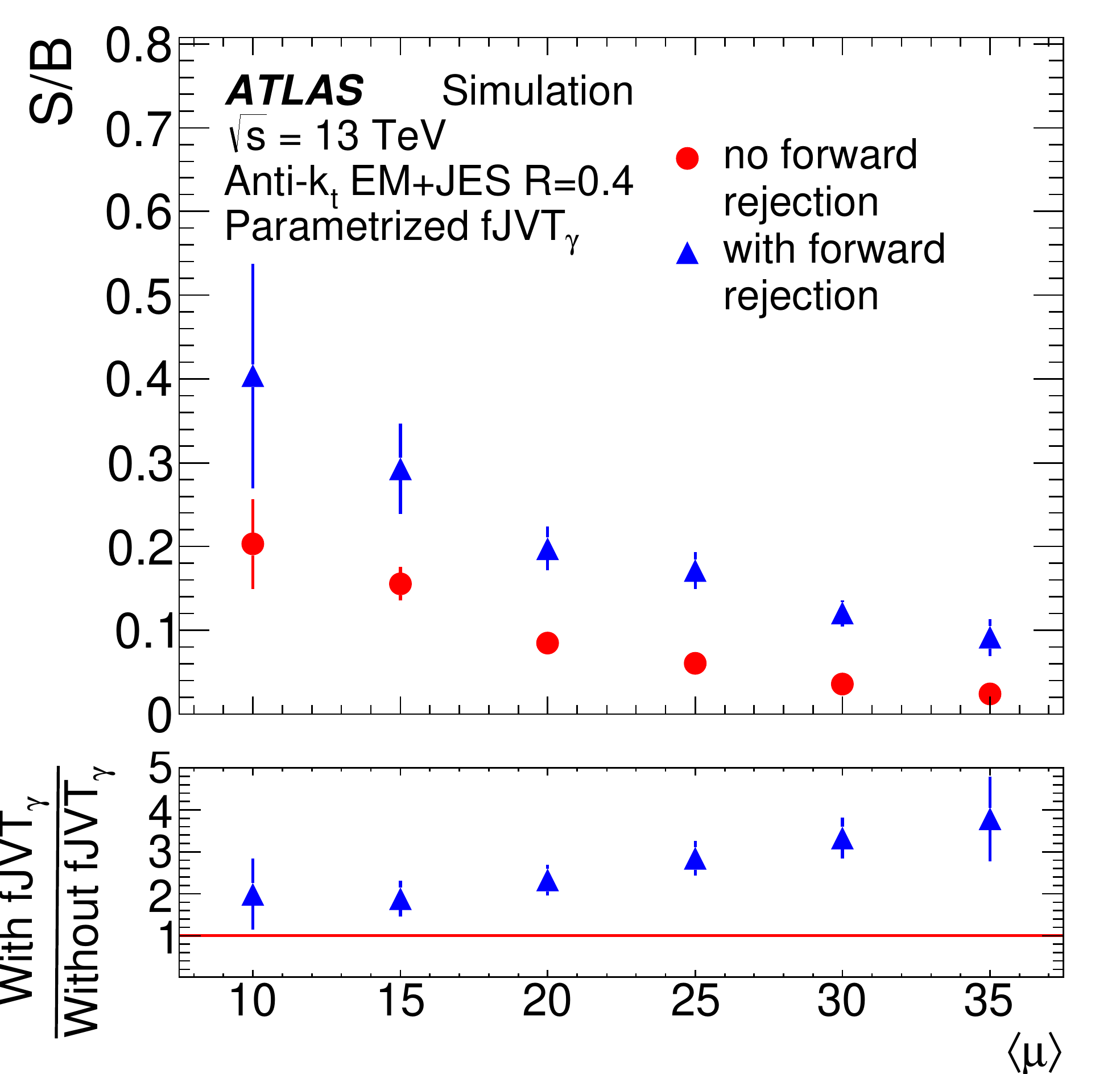}\label{vbfsig}}
\end{center}
\caption{Relative expected yield variation of \protect\subref{vbfb} $\Zboson\to\ell\ell$ and \protect\subref{vbfs} VBF 
$H\to\tau\tau$ events and \protect\subref{vbfsig} signal purity
as a function of the number interactions per bunch crossing ($\langle\mu\rangle$), with
different levels of pile-up rejection using $\fJVT_\gamma$.
The expected signal and background yields at $\langle\mu\rangle=10$ are used as reference.
Parameterized hard-scatter efficiency and pile-up efficiency are used.
The lower panels display the ratio to the reference without pile-up rejection.}
\label{fig:fJVTapp_vbf_paramg}
\end{figure}
Figure~\ref{fig:fJVTapp_vbf_paramg} shows the expected numbers of signal and background events,
as well as the signal purity, as a function of $\langle\mu\rangle$.
When going from $\langle\mu\rangle$ of 10 to 35, the expected number of background events grows by a factor
of seven and the corresponding signal purity drops by a factor of eight, 
indicating that the presence of pile-up jets enhances the background acceptance.
The slight decrease in signal acceptance is due to misidentification of pile-up jets as VBF jets.
The $\fJVT_{\gamma}$ algorithm mitigates the background growth, at the expense of a signal loss
proportional to the hard-scatter jet efficiency.\footnote{Most VBF events are characterized by one
forward jet and one central jet.}
Therefore, the degradation of the purity due to pile-up can be effectively reduced. 
For the specific final state and event selection under consideration, where $\Zboson+$jets production is the dominant background,
this results in about a fourfold improvement in signal purity at $\langle\mu\rangle=35$.

\clearpage

\section{Conclusions}
\label{sec:conclusion}

The presence of multiple $pp$ interactions per bunch crossing at the LHC, 
referred to as pile-up,
results in the reconstruction of additional jets beside the ones from the hard-scatter
interaction.
The ATLAS baseline strategy for identifying and rejecting pile-up jets
relies on matching tracks to jets to determine the $pp$ interaction of origin.
This strategy cannot be applied for jets beyond the tracking coverage of the inner detector.
However, a broad spectrum of physics measurements at the LHC relies on the
reconstruction of jets at high pseudorapidities. 
An example is the measurement of Higgs
boson production through vector-boson fusion. The presence of pile-up jets at high
pseudorapidities reduces the sensitivity for these signatures, by
incorrectly reconstructing these final states in background events.

The techniques presented in this paper allow the identification
and rejection of pile-up jets beyond the tracking coverage of the inner detector.
The strategy to perform such a task is twofold. 
First, the information about the jet shape is used to estimate the
leading contribution to the jet above the stochastic pile-up noise.
Then the topological correlation among particles originating from a pile-up interaction
is exploited to extrapolate the jet vertex tagger, using track and vertex information, 
beyond the tracking coverage of the inner detector to identify and reject pile-up
jets at high pseudorapidities.
When using both shape and topological information, 
approximately 57\% of forward pile-up jets are rejected for a hard-scatter 
efficiency of about 85\% at the pile-up conditions considered in this paper, with an average of 22 pile-up interactions.
In events with 35 pile-up interactions, typical conditions for the LHC operations in the near future,
37\%, 48\%, and 51\% of forward pile-up jets are rejected using, respectively, topological information, 
shape information, and their combination, for the same 85\% hard-scatter efficiency.

A procedure is defined and used to measure the
efficiency of identifying hard-scatter jets in 3.2 \ifb of $pp$ collisions at $\sqrt s=13\TeV$ collected in
2015.
The efficiencies are measured in data and estimated in simulation as a function of the jet
kinematics. Discrepancies of up to approximately 3\% are observed, mainly due to the modelling of pile-up events.

The impact of forward pile-up rejection algorithms presented here
is estimated in a simplified study of Higgs boson production through vector-boson fusion
and decaying into a $\tau\tau$ pair; the signal purity for the baseline
selection under consideration, where $\Zboson+$jets production is the dominant background, 
can be enhanced by a factor of about four for events with 35 pile-up interactions.

\section*{Acknowledgements}

\input{acknowledgements/Acknowledgements}

\printbibliography

\newpage
\input{atlas_authlist}

\end{document}

%% file: fjvt_paper-metadata.tex
\AtlasTitle{Identification and rejection of pile-up jets at high pseudorapidity with the ATLAS detector}

 \usepackage{authblk}

 \usepackage{authblk}
 \author{The ATLAS Collaboration%
 }

\date{\today}

\AtlasRefCode{PERF-2016-06}

 \AtlasJournal{EPJC}

\AtlasAbstract{%
The rejection of forward jets originating from additional proton--proton interactions (pile-up)
is crucial for a variety of physics analyses at the LHC, 
including Standard Model measurements and searches for physics beyond the Standard Model.
The identification of such jets is challenging due to the lack of track and vertex information
in the pseudorapidity range $|\eta|>2.5$.
This paper presents a novel strategy for forward pile-up jet tagging 
that exploits jet shapes and topological jet
correlations in pile-up interactions.
Measurements of the per-jet tagging efficiency are presented
using a data set of $3.2$~\ifb{} of proton--proton collisions at a centre-of-mass energy of
13 \TeV~collected with the ATLAS detector.
The fraction of pile-up jets rejected in the range $2.5<|\eta|<4.5$ is estimated in simulated events with an average of 22
interactions per bunch-crossing.
It increases with jet transverse momentum and, for jets with transverse momentum
between 20 and 50 GeV, it ranges between 49\% and 67\% 
with an efficiency of 85\% for selecting hard-scatter jets.
A case study is performed in Higgs boson production via the vector-boson fusion process,
showing that these techniques mitigate the background growth
due to additional proton--proton interactions,
thus enhancing the reach for such signatures.
}

 \AtlasCoverSupportingNote{Forward pileup jet identification and rejection with the ATLAS detector}{https://cds.cern.ch/record/2144377}
  \AtlasCoverCommentsDeadline{30 March 2017}

 \AtlasCoverAnalysisTeam{Matthew Klein, Francesco Rubbo, Ariel Schwartzman}

 \AtlasCoverEdBoardMember{Juan Terr\'on Cuadrado~(chair), Liza Mijovi\'c, Rainer Stamen}

 \AtlasCoverEgroupEditors{atlas-perf-2016-06-editors@cern.ch}

 \AtlasCoverEgroupEdBoard{atlas-perf-2016-06-editorial-board@cern.ch}

%% file: defines.tex
\newcommand {\ie}{\mbox{i.e.}\xspace}     
\newcommand {\vs}{\mbox{vs.}\xspace}     
\newcommand {\hc}{\mbox{h.c.}\xspace}     
\newcommand {\eg}{\mbox{e.g.}\xspace}     

\newcommand{\bfit}[1]{\textit{\textbf{#1}}}
\newcommand{\Eqref}[1]{Eqn.~\eqref{#1}}
\newcommand{\Figref}[1]{Fig.~\ref{#1}}
\newcommand{\beq}{\begin{equation}}
\newcommand{\eeq}{\end{equation}}
\newcommand{\sign}{\ensuremath{\text{sign}\xspace}}

\renewcommand{\Zboson}{\ensuremath{Z}\xspace}
\newcommand{\Higgs}{\ensuremath{H}\xspace}
\renewcommand{\Wboson}{\ensuremath{W}\xspace}
\newcommand{\proton}{\ensuremath{p}\xspace}
\newcommand{\pq}{\ensuremath{{\mathrm{q}}\xspace}}
\newcommand{\pbbar}{\ensuremath{{\mathrm{\bar{b}}}\xspace}}
\renewcommand{\ttbar}{\ensuremath{{t\bar{t}}}\xspace}
\newcommand{\dijet}{dijet\xspace}

\newcommand{\unit}[1]{\ensuremath{\,\text{#1}}\xspace}
\newcommand{\Unit}[1]{\ensuremath{\,\rm{$#1$}}\xspace}
\newcommand{\Lumino}{20.3\unit{\ifb}}
\renewcommand{\GeV}{\ensuremath{\,\text{Ge\hspace{-.08em}V}}\xspace}
\renewcommand{\TeV}{\ensuremath{\,\text{Te\hspace{-.08em}V}}\xspace}

\newcommand{\eqn}[2]{\begin{subequations}\label{#1} #2 \end{subequations}}

\newcommand*{\metvec}{\ensuremath{\boldsymbol{E}_{\text{T}}^{\text{miss}}}\xspace}
\newcommand*{\metx}{\ensuremath{E_x^{\text{miss}}}\xspace}
\newcommand*{\mety}{\ensuremath{E_y^{\text{miss}}}\xspace}
\newcommand{\MTtwo}{\ensuremath{M_\mathrm{T2}}\xspace}
\newcommand{\JVF}{\ensuremath{\mathrm{JVF}}\xspace}
\newcommand{\JVT}{\ensuremath{\mathrm{JVT}}\xspace}
\newcommand{\fJVT}{\ensuremath{\mathrm{fJVT}}\xspace}
\newcommand{\cJVF}{\ensuremath{\mathrm{corrJVF}}\xspace}
\newcommand{\RpT}{\ensuremath{R_\mathrm{pT}}\xspace}
\newcommand{\ApT}{\ensuremath{A_\mathrm{pT}}\xspace}
\newcommand{\diffRpT}{\ensuremath{\Delta\RpT}\xspace}
\newcommand{\dphiZjet}{\ensuremath{\Delta\phi(\Zboson,\mathrm{jet})}\xspace}
\newcommand{\PURpT}{\ensuremath{R_\mathrm{pT}^{\mathrm{PU}}}\xspace}
\newcommand{\NPV}{\ensuremath{N_\mathrm{PV}}\xspace}
\newcommand{\NPVtruth}{\ensuremath{N_\mathrm{PV}^\mathrm{truth}}\xspace}
\renewcommand{\pT}{\ensuremath{p_{\mathrm{T}}}\xspace}
\newcommand{\pTk}{\ensuremath{p_{\mathrm{T}}^k}\xspace}
\newcommand{\pTi}{\ensuremath{p_{\mathrm{T}}^i}\xspace}
\newcommand{\pTjet}{\ensuremath{p_{\mathrm{T}}^{jet}}\xspace}
\newcommand{\pTcore}{\ensuremath{p_{\mathrm{T}}^{core}}\xspace}
\newcommand{\pTcoreA}{\ensuremath{p_{\mathrm{T}}^{core}-\rho A}\xspace}
\newcommand{\nPUtrk}{\ensuremath{n_\mathrm{trk}^\mathrm{PU}}\xspace}
\newcommand{\DeltaR}{\ensuremath{\Delta R}\xspace}
\newcommand{\cwidth}{\ensuremath{w_\mathrm{cluster}}\xspace}
\newcommand{\twidth}{\ensuremath{w_\mathrm{tower}}\xspace}
\newcommand{\Deltaeta}{\ensuremath{\Delta\eta}\xspace}
\newcommand{\Deltaphi}{\ensuremath{|\Delta\phi|}\xspace}
\newcommand{\Deltaphibare}{\ensuremath{\Delta\phi}\xspace}
\newcommand{\jetHeight}{\ensuremath{H_{jet}}\xspace}
 
\newcommand{\HDECAY}         {\mbox{\small\textsc{HDECAY}}}
\newcommand{\ggWWp}         {\mbox{\small\textsc{GG2WW}}}
\newcommand{\DIPHOX}         {\mbox{\small\textsc{DIPHOX}}}
\newcommand{\AR}         {\mbox{\small\textsc{ARIADNE}}}
\newcommand{\PY}         {\mbox{\small\textsc{PYTHIA}}}
\newcommand{\JSv}        {\mbox{\small\textsc{JETSET\ 7.405}}}
\newcommand{\HWo}        {\mbox{\small\textsc{HERWIG\ 5.8}}}
\newcommand{\HWn}        {\mbox{\small\textsc{HERWIG\ 5.9}}}
\newcommand{\ARv}        {\mbox{\small\textsc{ARIADNE\ 4.05}}}
\newcommand{\PYv}        {\mbox{\small\textsc{PYTHIA\ 5.7}}}
\newcommand{\Sherpa}   {\textsc{Sherpa}\xspace}
\newcommand{\MADGRAPHv}   {\textsc{MadGraph}\ 5v1.1\xspace}
\newcommand{\BRIDGE}   {\textsc{Bridge}\xspace}
\newcommand{\PYTHIAeight}    {\textsc{Pythia8.186}\xspace}
\newcommand{\PYTHIAsix}    {\textsc{Pythia6.428}\xspace}
\newcommand{\DELPHES}   {\textsc{Delphes}\xspace}
\newcommand{\GEANTfour} {{\textsc{Geant4}}\xspace}
\newcommand{\JetPhox}    {\textsc{JetPhox\ 1.3.0}\xspace}
\newcommand{\Powheg}    {\textsc{powheg~box}\xspace}
\newcommand{\Fastjet}   {\textsc{Fastjet\ 2.4.3}\xspace}

\newcommand{\SOFTSUSY}         {\mbox{\small\textsc{SoftSusy}}\xspace}
\newcommand{\PROSPINO}         {\mbox{\small\textsc{Prospino}}\xspace}

\newcommand{\ATLAS}{{\textsc ATLAS}\xspace}

%% file: acknowledgements/Acknowledgements.tex

We thank CERN for the very successful operation of the LHC, as well as the
support staff from our institutions without whom ATLAS could not be
operated efficiently.

We acknowledge the support of ANPCyT, Argentina; YerPhI, Armenia; ARC, Australia; BMWFW and FWF, Austria; ANAS, Azerbaijan; SSTC, Belarus; CNPq and FAPESP, Brazil; NSERC, NRC and CFI, Canada; CERN; CONICYT, Chile; CAS, MOST and NSFC, China; COLCIENCIAS, Colombia; MSMT CR, MPO CR and VSC CR, Czech Republic; DNRF and DNSRC, Denmark; IN2P3-CNRS, CEA-DSM/IRFU, France; SRNSF, Georgia; BMBF, HGF, and MPG, Germany; GSRT, Greece; RGC, Hong Kong SAR, China; ISF, I-CORE and Benoziyo Center, Israel; INFN, Italy; MEXT and JSPS, Japan; CNRST, Morocco; NWO, Netherlands; RCN, Norway; MNiSW and NCN, Poland; FCT, Portugal; MNE/IFA, Romania; MES of Russia and NRC KI, Russian Federation; JINR; MESTD, Serbia; MSSR, Slovakia; ARRS and MIZ\v{S}, Slovenia; DST/NRF, South Africa; MINECO, Spain; SRC and Wallenberg Foundation, Sweden; SERI, SNSF and Cantons of Bern and Geneva, Switzerland; MOST, Taiwan; TAEK, Turkey; STFC, United Kingdom; DOE and NSF, United States of America. In addition, individual groups and members have received support from BCKDF, the Canada Council, CANARIE, CRC, Compute Canada, FQRNT, and the Ontario Innovation Trust, Canada; EPLANET, ERC, ERDF, FP7, Horizon 2020 and Marie Sk{\l}odowska-Curie Actions, European Union; Investissements d'Avenir Labex and Idex, ANR, R{\'e}gion Auvergne and Fondation Partager le Savoir, France; DFG and AvH Foundation, Germany; Herakleitos, Thales and Aristeia programmes co-financed by EU-ESF and the Greek NSRF; BSF, GIF and Minerva, Israel; BRF, Norway; CERCA Programme Generalitat de Catalunya, Generalitat Valenciana, Spain; the Royal Society and Leverhulme Trust, United Kingdom.

The crucial computing support from all WLCG partners is acknowledged gratefully, in particular from CERN, the ATLAS Tier-1 facilities at TRIUMF (Canada), NDGF (Denmark, Norway, Sweden), CC-IN2P3 (France), KIT/GridKA (Germany), INFN-CNAF (Italy), NL-T1 (Netherlands), PIC (Spain), ASGC (Taiwan), RAL (UK) and BNL (USA), the Tier-2 facilities worldwide and large non-WLCG resource providers. Major contributors of computing resources are listed in Ref.~\cite{ATL-GEN-PUB-2016-002}.

%% file: atlas_authlist.tex
\begin{flushleft}
{\Large The ATLAS Collaboration}

\bigskip

M.~Aaboud$^\textrm{\scriptsize 137d}$,
G.~Aad$^\textrm{\scriptsize 88}$,
B.~Abbott$^\textrm{\scriptsize 115}$,
J.~Abdallah$^\textrm{\scriptsize 8}$,
O.~Abdinov$^\textrm{\scriptsize 12}$$^{,*}$,
B.~Abeloos$^\textrm{\scriptsize 119}$,
S.H.~Abidi$^\textrm{\scriptsize 161}$,
O.S.~AbouZeid$^\textrm{\scriptsize 139}$,
N.L.~Abraham$^\textrm{\scriptsize 151}$,
H.~Abramowicz$^\textrm{\scriptsize 155}$
% ...
\bigskip
\\
% ...
$^{8}$ Department of Physics, The University of Texas at Arlington, Arlington TX, United States of America\\
% ...
$^{12}$ Institute of Physics, Azerbaijan Academy of Sciences, Baku, Azerbaijan\\
% ...
$^{88}$ CPPM, Aix-Marseille Universit{\'e} and CNRS/IN2P3, Marseille, France\\
% ...
$^{115}$ Homer L. Dodge Department of Physics and Astronomy, University of Oklahoma, Norman OK, United States of America\\
% ...
$^{119}$ LAL, Univ. Paris-Sud, CNRS/IN2P3, Universit{\'e} Paris-Saclay, Orsay, France\\
% ...
$^{137}$ $^{(a)}$ Facult{\'e} des Sciences Ain Chock, R{\'e}seau Universitaire de Physique des Hautes Energies - Universit{\'e} Hassan II, Casablanca; $^{(b)}$ Centre National de l'Energie des Sciences Techniques Nucleaires, Rabat; $^{(c)}$ Facult{\'e} des Sciences Semlalia, Universit{\'e} Cadi Ayyad, LPHEA-Marrakech; $^{(d)}$ Facult{\'e} des Sciences, Universit{\'e} Mohamed Premier and LPTPM, Oujda; $^{(e)}$ Facult{\'e} des sciences, Universit{\'e} Mohammed V, Rabat, Morocco\\
% ...
$^{139}$ Santa Cruz Institute for Particle Physics, University of California Santa Cruz, Santa Cruz CA, United States of America\\
% ...
$^{151}$ Department of Physics and Astronomy, University of Sussex, Brighton, United Kingdom\\
% ...
$^{155}$ Raymond and Beverly Sackler School of Physics and Astronomy, Tel Aviv University, Tel Aviv, Israel\\
% ...
$^{161}$ Department of Physics, University of Toronto, Toronto ON, Canada\\
% ...
$^{a}$ Also at Department of Physics, King's College London, London, United Kingdom\\
$^{b}$ Also at Institute of Physics, Azerbaijan Academy of Sciences, Baku, Azerbaijan\\
$^{c}$ Also at Novosibirsk State University, Novosibirsk, Russia\\
$^{d}$ Also at TRIUMF, Vancouver BC, Canada\\
$^{e}$ Also at Department of Physics {\&} Astronomy, University of Louisville, Louisville, KY, United States of America\\
% ...
$^{*}$ Deceased
\end{flushleft}
